\documentclass[11pt]{article}
\pdfoutput=1
\usepackage[utf8]{inputenc}
\usepackage{bm}
\usepackage{colortbl} % change \hline color 
\usepackage{mathtools}
\usepackage{cancel}
\usepackage{amsfonts}
\usepackage{amssymb}
\usepackage{graphicx}
\usepackage{lscape} 
\usepackage{relsize}
\usepackage{mathabx}
\usepackage{indentfirst}
\usepackage{algpseudocode,algorithm}
\usepackage{xpatch}%remove the indentation in algorithmic
\makeatletter
\xpatchcmd{\algorithmic}
{\ALG@tlm\z@}{\leftmargin\z@\ALG@tlm\z@}
{}{}
\makeatother
\usepackage{pdfpages}
\usepackage[flushleft]{threeparttable}
\usepackage{float}
\usepackage{array}

\usepackage[justification=centering]{caption} % to use \caption* to get rid of the word table from
\usepackage{breqn}
\DeclareMathAlphabet\mathbfcal{OMS}{cmsy}{b}{n} %for bold mathcal 
\usepackage[toc,page]{appendix}
\usepackage{bbm}
\usepackage{tabularx}
\usepackage{multirow} % useful to use with mutlicolumn which use for notes at end of a table 
\usepackage[bottom]{footmisc} % put footnotes at bottom of pg no matter what
\usepackage[mathscr]{euscript}
\usepackage{pifont} % for check marks and x marks 

\usepackage{makecell} % force a line break in a tabular/table environment 

\usepackage[urlcolor=blue, linkcolor=blue, citecolor=blue,colorlinks=true]{hyperref} 
\usepackage{cleveref}
\usepackage{bookmark}
\usepackage{setspace}   %Allows double spacing with the \doublespacing command
\usepackage[backend=bibtex,style=authoryear, giveninits=true, natbib]{biblatex}
\DeclareFieldFormat{citehyperref}{%
	\DeclareFieldAlias{bibhyperref}{noformat}% Avoid nested links
	\bibhyperref{#1}}

\DeclareFieldFormat{textcitehyperref}{%
	\DeclareFieldAlias{bibhyperref}{noformat}% Avoid nested links
	\bibhyperref{%
		#1%
		\ifbool{cbx:parens}
		{\bibcloseparen\global\boolfalse{cbx:parens}}
		{}}}

\savebibmacro{cite}
\savebibmacro{textcite}

\renewbibmacro*{cite}{%
	\printtext[citehyperref]{%
		\restorebibmacro{cite}%
		\usebibmacro{cite}}}

\renewbibmacro*{textcite}{%
	\ifboolexpr{
		( not test {\iffieldundef{prenote}} and
		test {\ifnumequal{\value{citecount}}{1}} )
		or
		( not test {\iffieldundef{postnote}} and
		test {\ifnumequal{\value{citecount}}{\value{citetotal}}} )
	}
	{\DeclareFieldAlias{textcitehyperref}{noformat}}
	{}%
	\printtext[textcitehyperref]{%
		\restorebibmacro{textcite}%
		\usebibmacro{textcite}}}
\renewbibmacro{in:}{}
\makeatletter
\renewcommand{\@seccntformat}[1]{\csname the#1\endcsname.\quad}
\makeatother
\usepackage{enumerate}

\usepackage[left=2.54cm, right=2.54cm, top=2.54cm, bottom=2.54cm]{geometry}

\newcommand{\bs}[1]{\boldsymbol{#1}}

\usepackage{color}

\newtheorem{assump}{}
\newtheorem{assumpB}{}
\newtheorem{assumpR}{}

\newtheorem{theorem}{Theorem}

\newtheorem{lemma}{Lemma}

\usepackage{sectsty}
\usepackage{titlesec}
\usepackage{amsmath}
\newcommand\numberthis{\addtocounter{equation}{1}\tag{\theequation}}
\usepackage{authblk}
\usepackage[T1]{fontenc}
\usepackage{times}

\def\keywords{\vspace{.5em}
	{\textit{Keywords}:\,\relax%
}}

\titleformat{\paragraph}
{\normalfont\normalsize\bfseries}{\theparagraph}{1em}{}
\titlespacing*{\paragraph}
{0pt}{3.25ex plus 1ex minus .2ex}{1.5ex plus .2ex}
\providecommand{\keywords}[1]{\textbf{\textit{Index terms---}} #1}
\providecommand{\JEL}[1]{\textit{JEL Classifications:} #1}

\title{\Large  Inferential Theory for Granular Instrumental Variables in High Dimensions\footnote{We are grateful to Xu Cheng, Xavier Gabaix, Gloria Gonzalez-Rivera, Bruce Hansen, Jean Helwege, Bo Honor\'e, Guido Imbens, Ralph Koijen, Andrew Patton, Markus Pelger, Hashem Pesaran, Ekaterina Seregina, Ruoyao Shi, Aman Ullah, Tiemen Woutersen, Qiankun Zhou, seminar participants at UC Riverside and JSM 2021 for helpful comments and suggestions. We thank Michael Bates for providing access to UCR's High Performance Computing Center to carry out simulations efficiently.}}

\author[]{ \small Saman Banafti\textsuperscript{\textdagger} }
\author[]{ \small Tae-Hwy Lee\textsuperscript{\ddag} }
\affil[]{\small\textsuperscript{\textdagger}Amazon; \textsuperscript{\ddag}University of California, Riverside. \\ Email: \texttt{\href{mailto:sbanafti@amazon.com}{sbanafti@amazon.com}\color{black}},
\texttt{\href{mailto:taelee@ucr.edu}{taelee@ucr.edu}\color{black}}.} 
\date{\small \today}

\begin{document}
\begin{refsegment}
	\begin{titlepage}
\clearpage\maketitle
\thispagestyle{empty}
%\pubMonth{Month}
%\pubYear{Year}
%\pubVolume{Vol}
%\pubIssue{Issue}
%\JEL{}
\begin{abstract}
	\setlength{\parindent}{0.0cm}
	\singlespacing
	The Granular Instrumental Variables (GIV) methodology exploits panels with factor error structures to construct instruments to estimate structural time series models with endogeneity even after controlling for latent factors. We extend the GIV methodology in several dimensions. First, we extend the identification procedure to a large $N$ and large $T$ framework, which depends on the asymptotic Herfindahl index of the size distribution of $N$ cross-sectional units. Second, we treat both the factors and loadings as unknown and show that the sampling error in the estimated instrument and factors is negligible when considering the limiting distribution of the structural parameters. Third, we show that the sampling error in the high-dimensional precision matrix is negligible in our estimation algorithm. Fourth, we overidentify the structural parameters with additional constructed instruments, which leads to efficiency gains. Monte Carlo evidence is presented to support our asymptotic theory and application to the global crude oil market leads to new results. 
\end{abstract}\thispagestyle{empty}
 \bigskip

\JEL{C26, C36, C38, C46, C55}
\end{titlepage}

\section{Introduction}\label{intro}
\setlength{\parindent}{0.5cm}

In the absence of randomized control trials, \textit{finding} valid and strong instruments to circumvent unobserved confounders is a very challenging task. The Granular Instrumental Variables, hereafter GIV, methodology that \citet{gabaix2020granular} propose, establishes a systematic way to \textit{construct} instruments from suitably weighted idiosyncratic shocks, from observational datasets and use them as instruments for aggregate endogenous variables. 

\textbf{\large Constructing instruments.} There are some existing methodologies which seek to eliminate the need to find an instrument. A leading example is the \citet{arellano1991some} framework in the context of estimating the speed of adjustment or state dependence parameters using dynamic panel data models with fixed effects, in which higher order lags of the dependent variable serve as instruments for the included lags of the dependent variable. The \citet{bartik1991benefits} methodology (aka shift-share estimators) where instruments are constructed from identities involving the (endogenous) explanatory variable whose shift component is interacted with shares. The \citet{rigobon2003identification} setting exploits the existence of structural breaks in the conditional heteroskedasticity regime, which is common place in many applications of interest. This allows one to bring a system with less equations than unknowns to a just identified system with as many equations as unknowns. The \citet{bai2010instrumental} methodology lays out a panel simultaneous equations model (similar to the model analyzed in this paper) where the estimated (strong) factors can be used as instrumental variables under certain conditions. We will return to the \citet{bai2010instrumental} methodology when we overidentify the structural parameters of interest as it is inspired by their framework. The vast methodological refinements cited within the papers referenced above are not listed here for brevity. 

\textbf{\large Microeconomic (granular) origins of aggregate fluctuations.} How can idiosyncratic shocks be relevant for endogenous aggregate variables? The literature on "granularity" traces back to historic debates in macroeconomics; no attempt to fully catalog this debate is made here, rather a concise summary is offered. \citet{long1983real} demonstrate that in a multisector stochastic neoclassical growth model, sectoral shocks (as opposed to aggregate shocks) can potentially lead to GDP fluctuations. Intuitively, complex production processes form sectoral linkages which in turn provide a transmission mechanism of shocks across sectors. Subsequently, \citet{horvath2000sectoral} and \citet{dupor1999aggregation} debate whether sectoral shocks decay according to $\frac{1}{\sqrt{N}}$ as the central limit theorem would suggest. \citet{gabaix2011granular} provides an initial theoretical solution to the debate by showing that when the firm size distribution is heavy tailed, the central limit theorem does not apply and sectoral volatility decays much slower than $\frac{1}{\sqrt{N}}$. \citet{gabaix2011granular} coins this mechanism as the so-called "granular" hypothesis, in which the economy is composed of incompressible grains as opposed to infinitesimally small micro units. \citet{acemoglu2012network} formulates a network approach to demonstrate that sectoral idiosyncratic shocks generate non-negligible aggregate volatility when there exists sufficient asymmetry in the input-output relationships. \citet{pesaran2020econometric} build off of the theoretical approach of \citet{acemoglu2012network} and develop econometric theory to measure the degree of network dominance and in their application they find some evidence of sector-specific shock propagation albeit not overwhelmingly strong for the US input-output accounts data over the period 1972-2002. More empirical evidence for such propagation mechanism is presented in  \citet{gatti2005new}, \citet{canals2007trade}, \citet{koren2007volatility}, \citet{blank2009shocks}, \citet{malevergne2009professor},  \citet{yan2011role}, \citet{gabaix2011granular}, \citet{carvalho2013great}, \citet{schiaffi2013granularity},   \citet{acemoglu2017microeconomic}, \citet{jannati2017geographic} and \citet{lera2017quantification}. 

\textbf{\large GIV, Gabaix and Koijen (2021).} In an econometric framework, GK illustrate that when the market under consideration is sufficiently concentrated, then one can use the collection of idiosyncratic shocks to individual micro units, at each time period $t$, as an instrument for endogenous aggregate variables. The instrumental relevance follows heuristically from the paragraphs above. The exogeneity condition, as in any instrumental variables procedure, requires assumptions on unobserved random variables. However it should be noted that the exogeneity condition exploited in this framework is a relatively mild assumption that is often made in factor models (e.g. \citet{bai2002determining}) for identification purposes. The insight and contribution of GK opens the doors to a wide possibility of ways in which one can continue building on the promising new GIV methodology.  

\textbf{\large Contributions of this paper.} Our contributions to the GIV methodology are primarily focused on the underlying econometric issues. First, we naturally extend GK's identification procedure to a large $N$ and large $T$ framework (GK formally introduced GIV for a fixed $N$ and large $T$) by establishing and restricting the asymptotic behavior of the Herfindahl index for large $N$ markets as a function of the tail index of the size distribution. Given the large $N$ and large $T$ framework, we treat both the factors and loadings as unknown and allow the idiosyncratic error term to be weakly cross-sectionally correlated.\footnote{GK treat the factor loadings as known and extract the factors via period-by-period cross-sectional regressions. While they advocate extraction of latent factors via principal components analysis when loadings are unknown, they abstract away from the corresponding sampling error. We will show that the sampling error is indeed negligible.} As such, from our preliminary stage, we extract not only the estimated factors but also the estimated loadings via principal components analysis, PCA hereafter, or depending on the generality of the model ($k_x \neq 0$ in our notation from Section \ref{Model}), we use the iterative OLS-PCA method of \citet{bai2009panel}. Second, we show that the sampling error in the estimated instrument and estimated factors is negligible when considering the limiting distribution of the structural parameters of interest; that is, the estimator is robust to the latent factor structure. Moreover, the exogeneity requirement for one of the structural parameters generally depends on a potentially high dimensional precision matrix (the inverse of the covariance matrix). Third, we show that the sampling error in the high dimensional precision matrix is negligible in our iterative estimation algorithm for said structural parameter. Fourth, we overidentify the structural parameters which leads to efficiency gains. This leads to new and improved results in our empirical application of GIV to the global crude oil markets. Monte Carlo evidence is presented to confirm the finite sample behavior of our estimators are well approximated by the asymptotic distributions. We label our refinement to the GIV methodology as \textit{Feasible Granular Instrumental Variables} or FGIV for short. Finally, an empirical application of the estimation methods to estimate demand and supply elasticities of the global crude oil markets are presented to demonstrate the estimation procedures.

\textbf{\large Notation.} We distinguish vectors and matrices from scalars by making an object bold. Let $\{X_{it}, i=1,\dots,N; t=1,\dots,T\}$ be a double index process of random variables where $N$ denotes the number of cross-sectional units and $T$ denotes the number of time periods. We frequently stack across $i$, in which we obtain $\underset{N \times 1}{\bs{X}_{\cdot t}}:= \begin{pmatrix} X_{1t}& \dots &X_{Nt} \end{pmatrix}'$. Similarly, if we stack across $t$ we obtain $\underset{T \times 1}{\bs{X}_{i \cdot}} :=  \begin{pmatrix} X_{i1} & \dots & X_{iT}\end{pmatrix}'$. When $\bs{X}_{it}$ is itself a vector, say of dimension $k$, then we obtain a matrix when we stack across $i$ or $t$, e.g. $\underset{N \times k}{\bs{X}_{\cdot t}}$ or $\underset{T \times k}{\bs{X}_{ i \cdot}}$. Define $X_{\bs{w}t}$ as the cross-sectionally weighted average of $X_{it}$, that is $X_{\bs{w}t}:= \bs{w}'X_{\cdot t}=\sum^N_{i=1} w_i X_{it}$. Common weights, $\underset{N \times 1}{\bs{w}}=(w_i)$, used frequently throughout the paper are (1) the precision weights, $\bs{E}:=  \frac{\bs{\Sigma}^{-1}_u \bs{\iota}}{\bs{\iota}'\bs{\Sigma}^{-1}_u\bs{\iota}}$ where $\underset{N \times N}{\bs{\Sigma}_u}:=  \mathbbm{E}(\bs{u}_{\cdot t}\bs{u}_{\cdot t}')$ is the covariance matrix of the idiosyncratic error term, $u_{it}$, $\underset{N \times 1}{\bs{\iota}}$ is a vector of ones and (2) the share weights, which we simply refer to as size weights, $\bs{S}:=  \begin{pmatrix} S_1 & \dots & S_N \end{pmatrix}'$. Let $\widetilde{X}_{it}=X_{it}-\bar{X}_t$, where $\bar{X}_t=\frac{1}{N}\sum_{i=1}^N X_{it}$, denote a cross-sectionally demeaned variable. Unless otherwise specified, we denote the $L^2$-norm as $||\cdot||$ or sometimes explicitly as $||\cdot||_2$, the $L^1$-norm as $|| \cdot ||_1$ and the Frobenius norm as $||\cdot||_F$; if another norm is used, it will be explicitly noted. Given a square matrix $\bs{A}$, let $\gamma_{max}(\bs{A})$ denote the maximum eigenvalue of $\bs{A}$. Joint convergence of $N$ and $T$ will be denoted as $(N,T) \overset{j}\rightarrow \infty$ without any restriction on the relative rates; whenever restrictions on relative rates of convergence are imposed, it will be explicitly noted. The expression $\overset{p}{\rightarrow}$ denotes convergence in probability while $\overset{d}{\rightarrow}$ denotes convergence in distribution. The equation $\bs{y} = \mathcal{O}_p(\bs{x})$ states that the vector of random variables $\bs{y}$ is at most of order $\bs{x}$ in probability. The equation $a=\Theta_p(b)$ states that $a$ is stochastically bounded by $b$ and $b$ is stochastically bounded by $a$, hence $a$ and $b$ rise jointly proportionally. 
\section{Model}\label{Model}
A general formulation of the model examined in this paper is given in the following panel simultaneous equations model with factor error structure
\begin{align*}
\bs{y}_{it} &=\bs{B}\bs{x}_{it} + \bs{C}\bs{a}_t  + \bs{v}_{it}, \\
\bs{v}_{it} &=\bs{\Lambda}_i'\bs{F}_t + \bs{u}_{it},
\end{align*}
where $\bs{y}_{it}=\begin{pmatrix} y_{1,it} & \dots &  y_{G,it} \end{pmatrix}'$ is a $G \times 1$ vector of dependent variables, $\bs{x}_{it}=\begin{pmatrix} x_{1,it} & \dots & x_{k_x,it} \end{pmatrix}'$ is a $k_x \times 1$ vector of strictly exogenous variables (which can be arbitrarily correlated with the common factors, $\bs{F}_t$, and/or the loadings, $\bs{\Lambda}_i$), $\bs{a}_t= \begin{pmatrix} a_{1,t} & \dots & a_{k_a,t} \end{pmatrix}'$ is a $k_a  \times 1$ vector of potentially endogenous aggregate variables, $\bs{v}_{it}$ is a $G  \times 1$ vector of composite error terms which admit a low-rank plus sparse (factor structure) error decomposition, where $\bs{\Lambda}_i$ is an $r \times G$ matrix of latent factor loadings and $\bs{F}_t$ is an $r \times 1$ vector of latent factors. 

In our exposition, we focus on the canonical setting of estimating the supply and demand elasticities in the global crude oil market, so we set the dimension of $G=2$ for supply and demand variables respectively. We take $k_x=0$ for ease of exposition but we present a general estimation algorithm for when $k_x \neq 0$. Moreover, we assume that only one of the $G=2$ variables has a panel structure, whereas the other variable is an aggregate time series. The main results extend relatively naturally to the case where both variables have a panel model. That is, $\bs{y}_{it}=\begin{pmatrix} d_{t} & y_{it} \end{pmatrix}'$ where $d_t$ is the log change of aggregate crude oil consumption and $y_{it}$ is the log change of country $i$'s crude oil production, $a_t=p_t$, with $k_a =1$, is the log change of real crude oil price (where we deflate the nominal oil price with the U.S. general price in3dex).\footnote{One may wonder why $p_{t}$ is not disaggregated; in fact the $p_t$ we use can be considered as the weighted average of country specific real oil prices (in changes). As shown in \citet{mohaddes2016country}, for a proper global analysis, deflating the nominal oil price in U.S. dollars by the U.S. price index is generally theoretically invalid unless the law of one price holds universally. Namely, let $P_{it}$ denote the general price index faced by country $i$, $E_{it}$ denotes country $i$'s exchange rate measured as units of country $i$'s currency per U.S. dollar, $p_{it}$ denote country specific log of real oil prices and $\tilde{p}_t$ denotes nominal oil prices in U.S. dollars, if $E_{it}P_{US,t}=P_{it} \,\, \forall i$; then it follows that $\sum_{i=1}^N w_i p_{it}=\tilde{p}_t + \sum_{i=1}^N w_i \text{log}(E_{it}/P_{it})=\tilde{p}_t + \sum_{i=1}^N w_i \text{log}(1/P_{US,t})=\tilde{p}_t-p_{US,t} :=  p_t$. As it turns out, $p_t=\tilde{p}_t-p_{US,t}$ is an appropriate approximation as documented in \citet{mohaddes2016country} for their long run analysis, in the sense that it respects the long-run equilibrium relationships. We assume it is an appropriate approximation for our short-run analysis.} Given our stylizations the coefficient matrix $\bs{C}$ and composite error, $\bs{v}_{it}$ becomes
\begin{align*}
\bs{C}&=\begin{pmatrix}
\phi^d & 0 \\ 0 & \phi^s
\end{pmatrix}, \\ 
\bs{v}_{it}&=
\bs{\Lambda}_i'\bs{F}_t + \bs{u}_{it}=\begin{pmatrix}
1 & 0 \\  0 & \bs{\lambda}_i
\end{pmatrix}\begin{pmatrix}
\varepsilon_t \\ \bs{\eta}_t
\end{pmatrix} + \begin{pmatrix}
0\\u_{it}
\end{pmatrix},
\end{align*}
where the coefficients $\phi^d$ and $\phi^s$ denote the crude oil demand and supply elasticities, respectively, and $\bs{\eta}_t, \bs{\lambda}_i$ are $r \times 1$ vectors of latent factors and latent loadings, respectively. Our stylized simultaneous equations model takes the simple form 
\setcounter{equation}{0}
\begin{align}
d_{t}  &= \phi^d p_t  + \varepsilon_{t}\\
y_{it} &= \phi^s p_t + \bs{\lambda}_i'\bs{\eta}_t + u_{it}.\label{eq:1.2}
\end{align}
The global market clearing condition is given by $y_{St}=d_t$, where  $y_{St}:=\bs{S}'\bs{y}_{\cdot t} =\sum_{i=1}^N S_{i} y_{it} $, $\bs{S}$ is the $N \times 1$ vector of shares that are normalized such that $\sum_{i=1}^N S_i =1$ and $i$ and $t$ take the values $i=1,\dots,N$ and $t=1,\dots,T$, respectively.\footnote{As oil is a storable good, one could easily allow oil prices to adjust to the gap between supply and demand, e.g. as in \citet{mohaddes2016country}. This introduces more complex notations without adding any substance to the main points of the paper.} Making use of the global market clearing condition we see that
\begin{align}
p_t = \frac{1}{\phi^d - \phi^s} \left(u_{St} +\bs{\lambda}_S' \bs{\eta}_t -\varepsilon_t \right), \label{eqprice}
\end{align}
which makes the simultaneity clear, e.g., that prices are composed of size-weighted idiosyncratic shocks, aggregate supply shocks and the demand shock. The objective of the GIV methodology is to extract the idiosyncratic shocks and use them as instruments for price. 

 \textbf{\large Demand estimation in the case of uniform loadings ($\bs{\lambda}_i = \bs{\lambda} \,\forall \, i$).} To momentarily fix ideas, it is helpful to consider a major simplification when constructing the instrument. Suppose that the loadings are uniform, $\bs{\lambda}_i=\bs{\lambda} \, \forall i$. Then, the instrument$, z_t$, can be formed as 
 \begin{align*}
 z_t &= y_{St}-\frac{1}{N}\sum_{i=1}^N y_{it}=(\phi^dp_t + \bs{\lambda}'\bs{\eta}_t + u_{St}) - (\phi^d p_t + \bs{\lambda}'\bs{\eta}_t + \frac{1}{N}\sum_{i=1}^N u_{it}),  \\
 &=u_{St} - \frac{1}{N}\sum_{i=1}^N u_{it} := u_{\Gamma t}, \numberthis  \label{toy}
 \end{align*}
 where $\bs{\Gamma} := \bs{S}- \bs{\iota}/N$ is an $N \times 1$ random vector such that $\bs{\iota}'\bs{\Gamma}=\sum_{i=1}^N \Gamma_i=0$, by construction. $\Gamma_i$ is random because we assume the shares follow a fat-tailed distribution, see \ref{A4}. Identification and estimation of demand by GIV requires that 
 \begin{align}
 \mathbbm{E}(z_t\varepsilon_t)=\mathbbm{E}\left(\sum_{i=1}^N\Gamma_i\mathbbm{E}(u_{it}\varepsilon_t|\bs{\Gamma})\right)=0. \label{id1}
 \end{align}
 \eqref{id1} is our exogeneity condition and \eqref{eqprice} gives $\mathbbm{E}(z_tp_t)\neq 0$, relevance. A sufficient condition for the moment condition in \eqref{id1} to be zero is $\mathbbm{E}(u_{it}\varepsilon_t|\bs{\Gamma})=0$, which effectively requires that conditional on size, $u_{it}$ and $\varepsilon_t$ are uncorrelated. Given relevance, exogeneity implies the following demand elasticity estimator $\widehat{\phi}^d=\frac{\sum_td_tz_t}{\sum_t p_tz_t}$. Intuitively, $z_t$ places larger weights on the idiosyncratic shocks to larger oil producers, these granular shocks will shift the supply curve while keeping the aggregate demand curve fixed since demand responds to these shocks only through their affects on prices. This allows for consistent estimation of the demand elasticity. The uniform loadings assumption in this case tremendously facilitate the analysis. Uniform loadings allow one to construct the instrument, as in \eqref{toy}, from observables. \color{black} In practice, uniform loadings are quite restrictive and we subsequently relax this assumption. However, before moving on to the general case, we also illustrate supply estimation under simplifying assumptions to fix ideas. \color{black}
 
 \textbf{\large Supply estimation in the case of uniform loadings and $u_{it} \,\,\bs{ i.i.d.}$} Continuing on with the uniform loadings case, remarkably, GK show that one can use the \textit{same} instrument, $z_t$, to \textit{also} estimate the supply elasticity using a cross-sectionally aggregated supply equation. Now, GK further assume that $u_{it}$ are $i.i.d.$, $\mathbbm{E}(\bs{u}_{\cdot t}\bs{u}_{\cdot t}'):= \bs{\Sigma}_u=\sigma^2_u \bs{I}_N$, where $\bs{u}_{\cdot t}:=\begin{pmatrix} u_{1t} & \dots & u_{Nt}\end{pmatrix}'$ and $\bs{I}_N$ is the identity matrix and define the $N \times 1$ precision weight vector $\bs{E} :=  \frac{\bs{\Sigma}^{-1}_u\bs{\iota}}{\bs{\iota}'\bs{\Sigma}_u^{-1}\bs{\iota}}$ which reduces to $\bs{\iota}/N$ when $u_{it}$ are $i.i.d.$ across $i$. Aggregation of the supply equation is performed using the vector $\bs{E}$, we have that $y_{Et} =\phi^sp_t + \bs{\lambda}'\bs{\eta}_t + u_{Et}.$ Identification and estimation of supply by GIV requires that the instrument satisfies exogeneity with respect to the \textit{composite} error term\footnote{\color{black}In the general case to follow, we estimate the factors and thus only exploit $\mathbbm{E}(u_{Et}z_t)=0$ to estimate $\phi^s$.\color{red}}
 \begin{align}
 \mathbbm{E}((\bs{\lambda}'\bs{\eta}_t+u_{Et})z_t)=0. \label{exogs}
 \end{align}
 The first term in \eqref{exogs} has similar interpretation as in \eqref{id1}, i.e., size-weighted idiosyncratic supply shocks are uncorrelated with the aggregate supply component, $\bs{\lambda}'\bs{\eta}_t$. Miraculously, the second term is exactly zero
 \begin{align*}
 \mathbbm{E}(u_{Et}z_{t})&=\mathbbm{E}(\bs{E}'\bs{u}_{\cdot t}\bs{u}_{\cdot t}'\bs{\Gamma})=\mathbbm{E}(\bs{E}'\mathbbm{E}(\bs{u}_{\cdot t}\bs{u}_{\cdot t}'|\bs{\Gamma})\bs{\Gamma})=\frac{\sigma^2_u}{N}\mathbbm{E}(\bs{\iota}'\bs{\Gamma}) =0.\text{\footnotemark} \numberthis \label{babymagic}
 \end{align*}
 \footnotetext{One may wonder then why this particular form of $\bs{\Gamma}$ was selected. Appealing to Proposition 3 in GK, they establish that $\bs{\Gamma}=\bs{S}-\bs{E},$ for this example, turns out to be the optimal weight vector, amongst the class of weights which sum-to-zero. $\bs{\Gamma}$ is optimal in the sense that it minimizes the asymptotic variance of the structural parameters.} 
% however, \eqref{babymagic} also reveals that this moment condition may be asymptotically valid (for large $N$) even if $\bs{\iota}'\bs{\Gamma}$ deviates from zero, say due to measurement error in the shares. In our framework, we can show that $\bs{\Gamma}=\Theta(1)$ and hence it will be required that $\bs{\iota}'\bs{\Gamma}=0$.\footnote{When $\bs{\iota}'\bs{\Gamma} \neq 0$, this would technically mean that $z_t$ is not a valid instrument in the conventional sense for the supply elasticity, in any finite sample. However, a suitable bias-corrected pooled estimator was developed in \citet{bai2010instrumental}, which can still achieve asymptotic normality in a similar setting. The analog of the bias-corrected pooled estimator for the supply elasticity is not pursued in this paper.} 
The moment condition \eqref{babymagic} is zero due to  independence of $\Gamma_i$ and $u_{it}$ by assumption and the sum-to-zero property of $\bs{\Gamma}$. For identification with large $N$, we assume size to follow a power law in tail (see \ref{A4}), thus $\Gamma_i$ is stochastic and assumed to be independent of $u_{it}$.\footnote{For a fixed $N$, it is not required to assume independence of $\Gamma_i$ and $u_{it}$ because $\Gamma_i$ can be treated as constant and \eqref{babymagic} is zero solely by virtue of the fact that $\bs{\iota}'\bs{\Gamma}=0$.} So again, we have $\mathbbm{E}(z_tp_t)\neq 0$ and for this simplified example, we avoid the need to estimate the factor structure since (i) due to uniform loadings, $z_t$ is constructed from observables and (ii) $z_t$ is uncorrelated with the \textit{composite} error term. If either of (i) or (ii) fails to hold, estimation of the factor structure becomes a preliminary step, as in our general procedure. Nevertheless, \eqref{exogs} leads to the following simple supply elasticity estimator $\widehat{\phi}^s=\frac{\sum_ty_{Et}z_t}{\sum_tp_t z_t}$. The intuition here is that, again, $z_t$ places larger weights on the idiosyncratic shocks to larger oil producers, these granular shocks keep the simple average (or more generally precision-weighted, i.e., weighted heavily towards more stable oil producers) supply curve fixed. That is, on average, precision-weighted supply responds to these granular shocks only through their effects on prices (due to $\mathbbm{E}((\bs{\lambda}'\bs{\eta}_t+u_{Et})z_t)=0$) and at the same time since smaller oil producers take as given price changes caused by these granular shocks, it will shift their \color{black}supply \color{black} curves which enables consistent estimation of the supply elasticity. 
 
 \textbf{\large \color{black}Discussion.\color{black}} In the case of uniform loadings and $u_{it}$ $i.i.d.$, the vector $\bs{E}$ and the instrument are constructed from observables, the large sample properties of $\widehat{\phi}^s$ and $\widehat{\phi}^d$ only entail fixed $N$, large $T$ asymptotics for which GK have laid out. In general, however, the cross-section will need to be exploited to estimate $\bs{E}$ since one can not know if $u_{it}$ are $i.i.d.$ across $i$. Indeed, the factors typically take care of a substantial portion of the cross-sectional correlations but it is prudent to allow for cross correlations in $u_{it}$ since the exogeneity condition for estimation of the supply elasticity heavily exploits the structure of $\bs{\Sigma}_u$. Therefore, it will be important to generally allow for some weak cross correlations in $\bs{\Sigma}_u$, which our algorithm accommodates, as discussed in Section \ref{Fgiv} and Section \ref{FS}. 
 
   \color{black}Moreover, \color{black} although homogeneous loadings was only an abstraction to illustrate the instrument, GK advocate the use of $y_{\Gamma t}=y_{St}-\frac{1}{N}\sum_i y_{it}$ in practice even when the loadings are not uniform. In the general heterogeneous loadings case, their instrument becomes 
 \begin{align}
 Z_t &:=  y_{\Gamma t} =u_{\Gamma t} + \bs{\lambda}_{\Gamma}'\bs{\eta}_t. \label{ZGK}
 \end{align} 
 They label this instrument with a capital case convention, to distinguish it because it is no longer solely composed of weighted idiosyncratic shocks, $u_{\Gamma t}$, as the $\bs{\lambda}_{\Gamma}'\bs{\eta}_t$ term is contaminating the instrument. However, this clever formulation is possible because they advocate estimation of the factors in practice, which they augment to their structural equations, thereby controlling for the second term which can potentially make their moment conditions different from zero. 
 \section{Feasible Granular Instrumental Variables}\label{Fgiv}
 Homogeneous loadings are overly restrictive but relaxing this can be easily accommodated in practice via PCA or iterative OLS-PCA methods, e.g., \citet{bai2003inferential} or \citet{bai2009panel} in a preliminary stage to construct an estimate of the instrument.\footnote{For our theory, we assume a balanced panel. However, in the case of unbalanced panels with data missing at random (which is beyond the scope of this paper) one can instead use the \citet{bai2015unbalanced} method or \citet{bai2021matrix} method to estimate the factor structure and the instrument. \color{black}In the more realistic case where data are not missing at random, one can use the methods developed in \citet{xiong2019large}. Remark?\color{black}} Although in GK's asymptotic theory they assume homogeneous loadings and that the instrument is exogenous with respect to the composite error, which circumvents the need to estimate the factor structure, they indeed advocate augmenting their structural equations with estimated factors either via period-by-period cross sectional regressions when the loadings are known or via PCA in the case of non-parametric (unknown) loadings.  GK abstract away from the sampling error in suggesting the use of augmented factors, which only vanishes for both large $N$ and $T$. \citet{bai2006confidence} and \citet{greenaway2012asymptotic} have developed the asymptotic distribution for structural parameters in factor augmented regressions in time series and panel models respectively. In this paper, a variant of their corresponding result is established in showing the sampling error from estimating the high dimensional precision matrix, the factors, as well as the instrument is negligible in the asymptotic distribution of the structural parameters. 

 \textbf{\large The general heterogeneous loadings case and $u_{it}$ non-$\bs{i.i.d.}$}
 Now we formulate the estimation approach in the general case, which makes much heavier use of the cross-section. When we cross-sectionally demean the supply equation and stack across $i$ we obtain (recall $\widetilde{\bs{X}}$ denotes a generic demeaned variate)
 \begin{align}
 \bs{\widetilde{y}}_{\cdot t}=\bs{\widetilde{\Lambda}}\bs{\eta}_t+\bs{\widetilde{u}}_{\cdot t},\label{runPCA}
 \end{align}
 which is estimable with vanilla PCA when the factor structure is strong.\footnote{Strong factors in the sense that $\bs{\Lambda}'\bs{\Lambda}/N \overset{p}{\rightarrow} \bs{\Sigma}_{\Lambda}>0$; thus we assume the factors are strong/pervasive in the sense that a significant fraction of cross-sectional units are affected by their presence. Consistent estimation of weak factors is beyond the scope of this paper, see for example \citet{onatski2012asymptotics},  \citet{bailey2016exponent} or \citet{freyaldenhoven2021factor} for suitable conditions for which it is possible. Even when estimable, their convergence rates are slower relative to estimates of strong factors, e.g., see \citet{bai2021approximate}. This will generally require modifications to the limiting distributions we derive in this paper.} Letting $\bs{Q}=(\bs{I}_N-\bs{\widetilde{\Lambda}(\widetilde{\Lambda}'\widetilde{\Lambda})^{-1}\widetilde{\Lambda}}')$, then $\bs{Q}\bs{\widetilde{y}}_{\cdot t}=\bs{Q}\bs{\widetilde{u}}_{\cdot t}$,
 completely purges the process of the common factors through the loading space. Premultiplying the share weights gives the instrument
 \begin{align}
 z_t& :=\bs{S}'\bs{Q}\bs{\widetilde{y}}_{\cdot t},\label{GIV}\\
 &=\bs{S}'\bs{Q}\bs{\widetilde{u}}_{\cdot t} :=  \bs{\Gamma}'\bs{\widetilde{u}}_{\cdot t},
 \end{align}
 where $\bs{\Gamma}:=  \bs{QS}$ is unknown because $\bs{Q}$ is unknown, but $\bs{Q}$ is easily estimated from data. Once we have $\bs{\widehat{Q}}$, \color{black}which just replaces $\widetilde{\Lambda}$ with $\widehat{\widetilde{\Lambda}}$ \color{black}, we form $\widehat{z}_t=\bs{S}'\bs{\widehat{Q}}\bs{\widetilde{y}}_{\cdot t}$ from observables. Importantly, when $\bs{\lambda}_i=\bs{\lambda} \, \forall i$, then $\bs{\Gamma}=(\bs{I}_N-\bs{\widetilde{\Lambda}(\widetilde{\Lambda}'\widetilde{\Lambda})^{-1}\widetilde{\Lambda}}')\bs{S}=\bs{S}-\bs{\iota} /N$ as in the previous case with homogenous loadings. This gives rise to a more general demand elasticity estimator
 \begin{align}
 \widehat{\phi}^d=\widehat{\phi}^d(\bs{\widehat{z}})&=\frac{\sum_td_t\widehat{z}_t}{\sum_t p_t\widehat{z}_t}. \label{justidentifiedGIV}
 \end{align}
In Section \ref{suppest}, we show that the demand elasticity can be estimated \textit{as if} the infeasible instrument, $z_t$, is used. 

In the case of the supply elasticity, the estimator will additionally depend on the estimated (potentially high dimensional) precision matrix. That is, $\widehat{\phi}^s=\widehat{\phi}^s(\bs{\widehat{z}},\bs{\widehat{\Sigma}}^{-1}_u)$. This creates the need to jointly estimate $\bs{\widehat{\Sigma}}^{-1}_u$ to form $\bs{\widehat{E}}$ in order to aggregate the panel to estimate $\widehat{\phi}^s$. We propose a simple iterative procedure and show that the supply elasticity can be estimated \textit{as if} the infeasible precision matrix, $\bs{\Sigma}_u^{-1}$, and instrument, $z_t$, were used. More specifically, let $y_{Et} =\phi^s p_t +\bs{\lambda}_E'\bs{\eta}_t +u_{Et} :=\bs{f}_t' \bs{\theta}^s  + u_{Et},$ where $\bs{\theta}^s=\begin{pmatrix}\phi^s & \bs{\lambda}_E'\end{pmatrix}'$ and $\bs{f}_t=\begin{pmatrix}p_t & \bs{\eta}'_t\end{pmatrix}'$ are $(1+r) \times 1$ vectors. The remarkable result $\mathbbm{E}(z_t u_{Et})=0$, shown in \eqref{babymagic} for the previous simple example with homogeneous loadings, continues to hold in this setting as well, with $z_t=\bs{S}'\bs{Q}\bs{\tilde{y}_{\cdot t}}=\bs{S}'\bs{Q}\bs{\tilde{u}_{\cdot t}}$ and $\bs{\Gamma}=\bs{Q}\bs{S}$ (recall that $\bs{\iota}'\bs{\Gamma}=0$) 
\begin{align*}
\mathbbm{E}(u_{Et}z_t )&=\mathbbm{E}\left(\bs{E}'\bs{u}_{\cdot t} \bs{\tilde{u}}_{\cdot t}'\bs{\Gamma}\right)= \mathbbm{E}(\bs{E}'\bs{u}_{\cdot t}(\bs{u}_{\cdot t}-\bar{u}_{t}\bs{\iota})'\bs{\Gamma})=\mathbbm{E}(\bs{E}'\bs{u}_{\cdot t}\bs{u}_{\cdot t}'\bs{\Gamma}) -\mathbbm{E}(\bs{E}'\bs{u}_{\cdot t} \bar{u}_{t}\bs{\iota}'\bs{\Gamma})\\
&=\mathbbm{E}(\bs{E}'\,\mathbbm{E}(\bs{u}_{\cdot t}\bs{u}_{\cdot t}'|\bs{\Gamma})\bs{\Gamma})-0=\frac{1}{\bs{\iota}'\bs{\Sigma}_u^{-1}\bs{\iota}}\mathbbm{E}\left(\bs{\iota}'\bs{\Gamma}\right) =0. \numberthis \label{magic}
\end{align*} 
So we have that (where the estimated factors self-instrument)
\begin{align}
\mathbbm{E}\left[ \begin{pmatrix}
z_t  \\ \bs{\eta}_t
\end{pmatrix}\cdot u_{Et}\right]&=\mathbbm{E}\left[ \begin{pmatrix}
z_t \\ \bs{\eta}_t
\end{pmatrix}\cdot \left(y_{Et}-\phi^s p_t -\bs{\lambda}_E'\bs{\eta}_t \right) \right]=\bs{0}. 
\end{align}
However, given our interest lies in inference for $\phi^s$, it is useful to stack over $t$, $\bs{y}_{E} = \bs{p} \,\phi^s + \bs{\eta} \,\bs{\lambda}_E + \bs{u}_{E},$ where $\bs{y}_{E}, \bs{p},$ and $\bs{u}_{E}$ are $T \times1$ vectors and $\bs{y}_{\widehat{E}}$ is the feasible counterpart of $\bs{y}_{E}$. Let $\bs{M}_{\widehat{\eta}}=(\bs{I}_T-\bs{\widehat{\eta}(\widehat{\eta}'\widehat{\eta})^{-1}\widehat{\eta}}')$, then it follows from standard partitioned regression results that
\begin{align} 
\widehat{\phi}^s =\widehat{\phi}^s(\bs{\widehat{z}},\bs{\widehat{\Sigma}}_u^{-1})&= \dfrac{\bs{\widehat{z}}^{\,'}\,\bs{M}_{\widehat{\eta}}\,\bs{y}_{\widehat{E}}}{\bs{\widehat{z}}^{\,'} \,\bs{M}_{\widehat{\eta}}\,\bs{p}}.\label{iterphid}
\end{align}
As $\bs{\widehat{\Sigma}}_u^{-1}$ depends on $\widehat{\phi}^s$, \eqref{iterphid} generally requires an iterative estimation procedure. To that end, note that if $\phi^s$ were known, $y_{it}-p_t\phi^s=\bs{\lambda}_i'\bs{\eta}_t+u_{it}$ follows an approximate factor structure. Thus, a covariance estimator, $\bs{\widehat{\Sigma}}_u$, for the idiosyncratic part can be obtained following \citet{fan2013large} by applying thresholding to the eigenvalue decomposition, $\frac{1}{T}\sum_{t=1}^T(\bs{y}_{\cdot t}-\bs{\iota}p_t\phi^s)(\bs{y}_{\cdot t}-\bs{\iota}p_t\phi^s)'=\sum_{i=1}^N \gamma_i \bs{\xi}_i\bs{\xi}_i',$ where $\gamma_i$ and $\bs{\xi}_i$ are the eigenvalues (sorted in decreasing order) and corresponding eigenvectors, respectively. More specifically, if $\phi^s$ were known, we have
\begin{align}
\bs{\widehat{\Sigma}}_{(\bs{y}_{\cdot t}-\bs{\iota}p_t\phi^s)} :=\sum_{i=1}^r \widehat{\gamma}_i \bs{\widehat{\xi}}_i\bs{\widehat{\xi}}_i' + \bs{\widehat{\Sigma}_u}^{\mathbfcal{T}}, \label{POET}
\end{align}
where $\bs{\widehat{\Sigma}}_u^{\mathbfcal{{T}}}=\sum_{i=r+1}^N \widehat{\gamma}_i \bs{\widehat{\xi}}_i\bs{\widehat{\xi}}_i'=(\widehat{\sigma}^{\mathcal{T}}_{u,ij})_{N \times N}$,
\begin{align}
\widehat{\sigma}^{\mathcal{T}}_{u,ij}=\begin{cases} \widehat{\sigma}_{u,ii}, &i=j, \\
h_{ij}(\widehat{\sigma}_{u,ij}), & i \neq j,
\end{cases} \label{POETu}
\end{align}
and $h_{ij}(\cdot)$ is a generalized shrinkage function of \citet{antoniadis2001regularization}.\footnote{Examples of $h_{ij}(\cdot)$ include hard thresholding $h_{ij}(x)=x\mathbbm{1}(|x|\geq \tau_{ij})$ and soft thresholding $h_{ij}(x)=\text{sgn}(x)(|x| - \tau_{ij})_{+}$. The entry dependent threshold, $\tau_{ij}>0$, can be defined as $C\omega_T \sqrt{\widehat{\alpha}_{ij}}$, where $\widehat{\alpha}_{ij}=\frac{1}{T}\sum_{t=1}^T(\widehat{u}_{it}\widehat{u}_{jt}-\widehat{\sigma}_{u,ij})^2$, $\widehat{\sigma}_{u,ij}=\frac{1}{T}\sum_{t=1}^T\widehat{u}_{it}\widehat{u}_{jt}$ and $\widehat{u}_{it}=y_{it}-\phi^sp_t-\bs{\widehat{\lambda}}_i'\bs{\widehat{\eta}}_t$ for some predetermined decreasing sequence $\omega_T>0$ and $C>0$. The choice of $C$ can be data driven; \citet{fan2013large} choose $C$ through multifold cross-validation to maintain positive definiteness of $\bs{\widehat{\Sigma}}_u^{\mathbfcal{{T}}}(C)$. In our algorithm below, we make use of the R package for POET, written by the authors \citet{fan2013large}.} Of course, $\phi^s$ can not be known as it requires an estimate of $\bs{\Sigma}_u^{-1}$.  Thus, we now address joint estimation of $\phi^s$ and $\bs{\Sigma}_u^{-1}$ in what follows and subsequently establish that the sampling error in $\bs{\widehat{E}}$ is negligible given some regularity conditions. The iterative procedure is summarized in Algorithm \ref{alg:alg1} presented below.
\begin{center}
	\begin{algorithm}[!htbp]
		\renewcommand{\thealgorithm}{1}
		\caption{\textbf{FGIV for $\phi^s$ (when $k_x=0$):}}
		\label{alg:alg1}
		\begin{algorithmic}[1]
							\begin{itemize}
								\item \textit{Step 1:} Run PCA on \eqref{runPCA} and obtain $\widehat{z}_t=\bs{S}'\bs{\widehat{Q}}\bs{\widetilde{y}}_{\cdot t}$ as the sample counterpart of \eqref{GIV}.
								\item  \textit{Step 2:} Initialize $\bs{\widehat{\Sigma}}_u^{-1}=\bs{I}_N$.
								\item  \textit{Step 3:} Obtain  $\bs{y}_{\widehat{E}}(\bs{\widehat{\Sigma}}_u^{-1})$ and $\widehat{\phi}^s(\bs{\widehat{z}},\bs{\widehat{\Sigma}}_{u}^{-1})$ as in \eqref{iterphid}. 				
								\item \textit{Step 4:} Update $\bs{\widehat{\Sigma}}_u^{-1}$ by inverting $\bs{\widehat{\Sigma}}_u^{\mathbfcal{T}}$ defined in \eqref{POETu}, $\bs{y}_{\widehat{E}}(\bs{\widehat{\Sigma}}_u^{-1})$ and $\widehat{\phi}^s(\bs{\widehat{z}},\bs{\widehat{\Sigma}}_{u}^{-1})$.
								\item \textit{Step 5:} Iterate \textit{Step 3} and \textit{Step 4} until convergence.
							\end{itemize} 					
		\end{algorithmic}
	\end{algorithm}
%\footnotetext{Thresholding in \textit{Step 3} is necessary when $N>T$. If $N<T$, then Step 3 can be simplified such that thresholding is not applied since in this case $\text{rank}(\bs{\Sigma_u})=\text{min}\{N,T\}=N$ and thus $\bs{\Sigma_u}$ is invertible.}
\end{center}
When $r$ is unknown, one can augment \textit{Step 1} and estimate $r$ using a procedure as in \citet{bai2002determining}, \citet{onatski2010determining} or \citet{ahn2013eigenvalue}; we use the $ER$ and $GR$ methods of \citet{ahn2013eigenvalue} (hereafter AH). For more details of the $ER$ and $GR$ methods, see Section \ref{rhat} of the Supplementary Appendix. 

 \textbf{\large FGIV algorithm accommodating cross-section specific covariates.} When $k_x \neq 0$ then the demeaning transformation from \eqref{runPCA} results in
$\bs{\widetilde{y}}_{\cdot t}=\bs{\widetilde{\Lambda}}\bs{\eta}_t+\bs{\widetilde{x}}_{\cdot t}\bs{\beta}+\bs{\widetilde{u}}_{\cdot t},$ where $\bs{\widetilde{x}}_{\cdot t}$ is an $N \times k_x$ matrix, which leaves $\underset{k_x \times 1}{\bs{\beta}}$ as an additional parameter to estimate. $\bs{\beta}$ can be easily estimated by adapting the procedure of \citet{bai2017inferences}, which is generalizing \citet{bai2009panel}, to handle endogeneity of prices even after controlling for latent common factors. More specifically,
\begin{align}
\bs{\beta}(\bs{\widetilde{\Lambda}},\bs{\eta}_t,\bs{\Sigma}_u^{-1})&=\left(\sum_{t=1}^T\bs{\widetilde{x}}_{\cdot t}' \bs{\Sigma}_u^{-1} \bs{\widetilde{x}}_{\cdot t} \right)^{-1}  \sum_{t=1}^T\bs{\widetilde{x}}_{\cdot t}'\bs{\Sigma}_u^{-1}(\bs{\widetilde{y}}_{\cdot t}-\bs{\widetilde{\Lambda}}\bs{\eta}_t),\label{beta} \\ 
(\bs{\widetilde{y}}_{\cdot t}-\bs{\widetilde{x}}_{\cdot t}\bs{\beta})&=\bs{\widetilde{\Lambda}}\bs{\eta}_t+\bs{\widetilde{u}}_{\cdot t}, \label{betaPCA}
\end{align} 
since \eqref{betaPCA} follows a factor structure, the $T \times r$ factor matrix, $\bs{\eta}(\bs{\beta},\bs{\Sigma}_u^{-1})$, can be estimated using the principal components estimator whose columns are the eigenvectors corresponding to the largest $r$ eigenvalues of the $T \times T$ matrix $(\bs{\widetilde{y}}_{\cdot \cdot}-\bs{\widetilde{x}}_{\cdot \cdot}(\bs{\beta}))\bs{\Sigma}_u^{-1}(\bs{\widetilde{y}}_{\cdot \cdot}-\bs{\widetilde{x}}_{\cdot \cdot}(\bs{\beta}))'$, where the $T \times N$ matrix $\bs{\widetilde{x}}_{\cdot \cdot}(\bs{\beta}):=  \begin{pmatrix}\bs{\widetilde{x}}_{1 \cdot}\bs{\beta} & \dots & \bs{\widetilde{x}}_{N \cdot}\bs{\beta} \end{pmatrix}$ and $\bs{\widetilde{\Lambda}}(\bs{\beta},\bs{\Sigma}_u^{-1})=\frac{1}{T}\sum_{t=1}^T(\bs{\widetilde{y}}_{\cdot t}-\bs{\widetilde{x}}_{\cdot t}\bs{\beta})\bs{\eta}_t' (\bs{\beta},\bs{\Sigma}_u^{-1}).$ 
Thus, to deal with general (strictly exogenous) covariates, $\bs{x}_{it}$, Algorithm \ref{alg:alg2} can be applied.
\begin{center}
	\begin{algorithm}[!htbp]
		\renewcommand{\thealgorithm}{2}
		\caption{\textbf{FGIV for $\phi^s$ (when $k_x\neq0$):}}
		\label{alg:alg2}
		\begin{algorithmic}[2]
			\begin{itemize}
				\item \textit{Step 1:} Initialize $\bs{\widehat{\beta}}=\bs{0}$, $\bs{\widehat{\Sigma}}_{u}^{-1}=\bs{I}_N$. 
				\item \textit{Step 2:} Run PCA on \eqref{betaPCA} to obtain $\bs{\widehat{\eta}}_t(\bs{\widehat{\beta}},\bs{\widehat{\Sigma}}_u^{-1})$ and $\bs{\widehat{\widetilde{\Lambda}}}(\bs{\widehat{\beta}},\bs{\widehat{\Sigma}}_u^{-1})$ as explained above. 
				\item \textit{Step 3:} Update $\bs{\widehat{\beta}}$ as the sample counterpart of \eqref{beta}. 
				\item \textit{Step 4:} Obtain $\widehat{z}_t=\bs{S}'\bs{\widehat{Q}}(\bs{\widetilde{y}}_{\cdot t}-\bs{\widetilde{x}}_{\cdot t}\bs{\widehat{\beta}})$.
				\item  \textit{Step 5:} Initialize $\bs{y}_{\widehat{E}}(\bs{\widehat{\Sigma}}_u^{-1})$ and $\widehat{\phi}^s(\bs{\widehat{z}},\bs{\widehat{\Sigma}}_{u}^{-1})=\left(\bs{\widehat{z}}^{\,'} \,\bs{M}_{\widehat{\eta}}\,\bs{p}\right)^{-1} \bs{\widehat{z}}^{\,'}\,\bs{M}_{\widehat{\eta}}\,(\bs{y}_{\widehat{E}}-\bs{x}_{i \cdot}\bs{\beta})$. 				
				\item \textit{Step 6:} Update $\bs{\widehat{\Sigma}}_u^{-1}$ by inverting $\bs{\widehat{\Sigma}}_u^{\mathbfcal{T}}$ defined in \eqref{POETu}, where $\widehat{\gamma}_i$ and $\bs{\widehat{\xi}}_i$ are the eigenvalues and eigenvectors (sorted in decreasing order) corresponding to the sample analog of \newline  $\frac{1}{T}\sum_{t=1}^T(\bs{y}_{\cdot t}-\bs{\iota}p_t\phi^s-\bs{x}_{\cdot t}\bs{\beta})(\bs{y}_{\cdot t}-\bs{\iota}p_t\phi^s-\bs{x}_{\cdot t}\bs{\beta})'$ respectively.
				\item \textit{Step 7:}  Iterate \textit{Step 2} through \textit{Step 6} until convergence.
			\end{itemize} 						
		\end{algorithmic}
	\end{algorithm}
%	\footnotetext{Thresholding in \textit{Step 6} is necessary when $N>T$. If $N<T$, then Step 6 can be simplified such that thresholding is not applied since in this case $\text{rank}(\bs{\Sigma_u})=\text{min}\{N,T\}=N$ and thus $\bs{\Sigma_u}$ is invertible.}
\end{center}
When $r$ is unknown, one can augment \textit{Step 2} and iteratively estimate $r$ using the $ER$ and $GR$ methods of \citet{ahn2013eigenvalue}. 

The main takeaway is that when both $(N,T)$ are large, one can generalize the GIV estimators proposed by GK along different dimensions; here we accommodate latent heterogeneous loadings, latent factors and latent precision matrix (e.g., $u_{it}$ can be weakly cross-correlated and heteroskedastic). As mentioned earlier, we call the proposed estimators of the elasticities in \eqref{justidentifiedGIV}, Algorithm \ref{alg:alg1} and Algorithm \ref{alg:alg2} as FGIV estimators.

\begin{assumpR} \label{pseudoinstruments}
	 In principle, the theory for the estimators proposed in this paper allows for $N \gg T$. This case is relevant in many empirical settings (e.g., empirical industrial organization and finance). However, it may be beneficial to avoid estimating the precision matrix for cases where $N \ll T$ (e.g., empirical macro). But, as \eqref{babymagic} and \eqref{magic} show, to have a valid instrument for which the moment equation is \textit{exactly} zero, we must specify $\bs{\Sigma}_u$ correctly. This is the primary motivation for estimating the general precision matrix in Algorithms \ref{alg:alg1} and \ref{alg:alg2}. In order to avoid estimating the precision matrix, we must assume (potentially erroneously) $u_{it}$ are cross-sectionally independent. We now analyze the consequences of making this assumption when in fact $u_{it}$ are cross-sectionally correlated. Suppose we erroneously assume cross-sectional independence, then the vector $\bs{E}$ reduces to $\bs{\iota}/N$ and we end up with the following moment equation
	\begin{align*}
	\mathbbm{E}(u_{Et}z_t )&=\mathbbm{E}\left(\bs{E}'\bs{u}_{\cdot t} \bs{\tilde{u}}_{\cdot t}'\bs{\Gamma}\right)=\mathbbm{E}(\bs{E}'\bs{u}_{\cdot t}(\bs{u}_{\cdot t}-\bar{u}_{t}\bs{\iota})'\bs{\Gamma}) \\
	&=\mathbbm{E}(\bs{E}'\mathbbm{E}(\bs{u}_{\cdot t}\bs{u}_{\cdot t}'|\bs{\Gamma})\bs{\Gamma}) -\mathbbm{E}(\bs{E}'\bs{u}_{\cdot t} \bs{\iota}'\bs{\Gamma})\bar{u}_{t} =\frac{1}{N}\mathbbm{E}(\bs{\iota}'\bs{\Sigma}_u\bs{\Gamma})-0= o(1). \label{misspecifiedmagic}\numberthis
	\end{align*} 
	Hence, $z_t$ is not a valid instrument in the traditional sense because we allow $\mathbbm{E}(u_{Et}z_t )\neq 0$ for any given sample. Nevertheless, this moment converges to zero for large $N$. Indeed, the moment satisfies $\mathbbm{E}(u_{Et}z_t )=o\left(1\right)$ under our regularity assumptions, and thus, $z_t$ is asymptotically a valid instrument.\footnote{It can be shown that $\bs{\iota}'\bs{\Sigma}_u\bs{\Gamma} =\bs{\iota}'\bs{\Sigma}_u\bs{QS}  \leq \bs{\iota}'\bs{\Sigma}_u\bs{S} \gamma_{max}(\bs{Q}) =\sum_{i,j}\sigma_{u,ij} S_j \leq \left(\sum_{i,j} \sigma^2_{u,ij} \right)^{1/2} ||\bs{S}||^2_2 = || \bs{\Sigma}_u ||_{F} || \bs{S} ||^2_2 \leq  || \bs{\Sigma}_u ||_{1} || \bs{S} ||^2_2 \leq \mathcal{O}(m_N) \Theta_p(1) = o(N)\Theta_p(1)=o(N)$, where $m_N$ is defined in \ref{A3a} and $m_N=o(N)$.} This insight reveals that this moment is approaching zero, hence it may prove to be beneficial to aggregate the panel, $y_{it}$, using weights $\bs{\iota}/N$ regardless of the covariance structure. The immediate implication is that $\widehat{\phi}^s=\widehat{\phi}^s(\bs{\widehat{z}},\bs{I}_N)$, so there is no need for an algorithmic estimation procedure, the simple analytical formula for the supply elasticity estimator with potentially misspecified covariance structure for $u_{it}$ is given by 
	$\widehat{\phi}^s(\bs{\widehat{z}},\bs{I}_N) = \frac{\bs{\widehat{z}}^{\,'}\,\bs{M}_{\widehat{\eta}}\,\bs{\bar{y}}}{\bs{\widehat{z}}^{\,'} \,\bs{M}_{\widehat{\eta}}\,\bs{p}},$ where $\bs{\bar{y}}$ stacks $\bar{y}_t = \frac{1}{N}\sum_{i=1}^N y_{it}$ for each $t=1,\dots, T$; this estimator is essentially \textit{Step 2} and \textit{Step 3} of Algorithm \ref{alg:alg1}. Asymptotically, it holds that $\widehat{\phi}^s(\bs{\widehat{z}},\bs{\widehat{\Sigma}}_u^{-1}) = \widehat{\phi}^s(\bs{\widehat{z}},\bs{I}_N) + o_p(1).$ However, regarding performance in finite samples, when $u_{it}$ are not $i.i.d.$ and when $N \ll T$, it is not clear ex-ante if $\widehat{\phi}^s(\bs{\widehat{z}},\bs{I}_N)$ will outperform $\widehat{\phi}^s(\bs{\widehat{z}},\bs{\widehat{\Sigma}}_u^{-1})$. When $N \gg T$ one would expect ex-ante that $\widehat{\phi}^s(\bs{\widehat{z}},\bs{I}_N)$ will be less efficient than $\widehat{\phi}^s(\bs{\widehat{z}},\bs{\widehat{\Sigma}}_u^{-1})$ since the former is not optimally weighting the observations, whereas the latter is. When $u_{it}$ are indeed $i.i.d.$ we would expect $\widehat{\phi}^s(\bs{\widehat{z}},\bs{I}_N)$ to perform better.\footnote{In unreported simulations where $N \ll T$  and $u_{it}$ are non-i.i.d., we find that	$\widehat{\phi}^s(\bs{\widehat{z}},\bs{\widehat{\Sigma}}_u^{-1})$ typically has a smaller bias than $\widehat{\phi}^s(\bs{\widehat{z}},\bs{I}_N)$ (in absolute terms, the bias of both estimators are very small) but with a slightly larger variance.} 
\end{assumpR}

\section{Efficient GMM Estimation: Factor-Augmented FGIV}\label{FS} 
We now proceed to overidentify the elasticities, which yields overidentified FGIV estimators. We will refer to the overidentified FGIV estimators simply as efficient GMM estimators and the just identified FGIV estimators simply as FGIV estimators. It will be of interest to practitioners to see if overidentification is possible for the supply and demand equations. In this section, we show that the system is indeed overidentified to varying degrees for the supply and demand equations. 

\textbf{\large Demand.} It is common practice to assume uncorrelated aggregate supply and aggregate demand shocks, that is $\mathbbm{E}(\bs{\eta}_t \varepsilon_t)=0$. When we are willing to entertain this, then our supply factors, estimated via principal components, serve as valid instruments in estimation of the demand elasticity, rendering an overidentified parameter. In fact, the theory for using principal components as instruments was laid out in \citet{bai2010instrumental} under strong instrument asymptotics, as well as \citet{kapetanios2010factor} under many/weak instrument asymptotics. In the remainder of this section, we let the GIV be denoted as $z_{t,GIV}:= z_t$ to distinguish it from the full instrument vector we introduce with upper case conventions. Our full instrument matrix for the demand equation is $\underset{T \times (1+r)}{\bs{Z}_d} := \begin{pmatrix} \bs{z}_{GIV} & \bs{\eta} \end{pmatrix}$ with $\mathbbm{E}(\bs{Z}_{dt}\varepsilon_t)=0$; $\bs{Z}_{dt}$ simply augments factors to be used as instruments. Making use of the $(1 + r) \times 1$ dimensional moment condition, the efficient GMM demand elasticity estimator is defined as 
\begin{align}
\widehat{\phi}^d_{GMM} &= \underset{\phi^d}{\text{argmin}} \,\frac{\bs{\varepsilon}'\bs{Z}_d}{T} \,\bs{W}_d \, \frac{\bs{Z}_d'\bs{\varepsilon}}{T}, \nonumber \\
&=\left(\bs{p}'\,\bs{\widehat{Z}}_d \,\bs{\widehat{\Omega}}_d^{-1} \, \bs{\widehat{Z}}^{\,'}_d \bs{p}\right)^{-1}\bs{p}'\,\bs{\widehat{Z}}_d \,\bs{\widehat{\Omega}}_d^{-1}\, \bs{\widehat{Z}}^{\,'}_d \bs{d}, \label{gmmdemand}
\end{align}  
where $\underset{ (1+r) \times (1+r)}{\bs{W}_d}$ is an arbitrary positive definite weight matrix, but is optimally set as $\bs{\widehat{W}}_d=\bs{\widehat{\Omega}}_d^{-1}$, where $\bs{\widehat{\Omega}}_d=\frac{1}{T}\sum_{t=1}^T \bs{\widehat{Z}}_{dt}\bs{\widehat{Z}}_{dt}'(d_t-p_t\widehat{\phi}^d_{2SLS})^2$. It is clear that \eqref{gmmdemand} nests the FGIV estimator for the demand elasticity as a special case. In this sense, $\widehat{\phi}^d_{GMM}$ will be robust to scenarios where $ z_t$ is weaker. 

\textbf{\large Supply.} In the same vein, the supply elasticity can \textit{always} be overidentified given our identifying assumptions because $\mathbbm{E}(\varepsilon_t u_{Et})=0$ and thus $\varepsilon_t$ can serve as an additional instrument. To estimate the entire parameter vector for the supply equation, let $\underset{T \times (2+r)}{\bs{Z}_s}:=\begin{pmatrix} \bs{z}_{GIV} & \bs{\varepsilon} & \bs{\eta} \end{pmatrix}$, where the augmented factors self-instrument as they are part of the supply equation. Then $\bs{y}_E=\bs{f}\bs{\theta}^s +\bs{u}_E$ and recall $\bs{\theta}^s=\begin{pmatrix}\phi^s & \bs{\lambda}_E'\end{pmatrix}'$ and $\bs{f}_t=\begin{pmatrix}p_t & \bs{\eta}'_t\end{pmatrix}'$ are $(1+r) \times 1$ vectors and the matrix $\bs{f}$ is $T \times (1+r)$, which stacks $\bs{f}_t$. We have $\mathbbm{E}(\bs{Z}_{st} \bs{u}_{Et})=0$; hence, making use of the $(2+r)\times 1$ dimensional moment conditions, the efficient GMM supply elasticity estimator is defined as  
\begin{align}
\bs{\widehat{\theta}}^s_{GMM}&= \underset{\bs{\theta}^s}{\text{argmin}} \,\frac{\bs{u}_E'\bs{Z}_s}{T} \,\bs{W}_s\, \frac{\bs{Z}_s'\bs{u}_E}{T}, \nonumber \\
&=\left(\bs{\widehat{f}}'\bs{\widehat{Z}}_s \,\bs{\widehat{\Omega}}_s^{-1}\,\bs{\widehat{Z}}^{\,'}_s\bs{\widehat{f}} \right)^{-1}\bs{\widehat{f}}'\bs{\widehat{Z}}_s \,\bs{\widehat{\Omega}}_s^{-1}\,\bs{\widehat{Z}}^{\,'}_s \bs{y}_{\widehat{E}}. \label{gmmsupply}
\end{align}
where $\underset{ (2+r) \times (2+r)}{\bs{W}_s}$ is an arbitrary positive definite weight matrix, but is also optimally set as $\bs{\widehat{W}}_s=\bs{\widehat{\Omega}}_s^{-1}$, where $\bs{\widehat{\Omega}}_s=\frac{1}{T}\sum_{t=1}^T \bs{\widehat{Z}}_{st}\bs{\widehat{Z}}_{st}'(y_{\widehat{E}t}-\widehat{\bs{f}}'_t\bs{\widehat{\theta}}^s_{GMM})^2$.\footnote{In the case of the demand elasticity estimator in \eqref{gmmdemand} we use 2SLS residuals to construct $\bs{\widehat{\Omega}}_d$. However, we implement \eqref{gmmsupply} via Algorithm \ref{alg:alg4} which, by iteration, renders the residuals used to construct $\bs{\widehat{\Omega}}_s$ to be GMM residuals.} It is clear that \eqref{gmmsupply} nests the FGIV estimator for the supply equation as a special case. As in the just identified case in \eqref{iterphid}, $\widehat{\bs{\theta}}^s_{GMM}$ in \eqref{gmmsupply} depends on $\widehat{\bs{\Sigma}}_u^{-1}$, hence, will generally require an iterative estimation procedure. Algorithm \ref{alg:alg4} below generalizes Algorithm \ref{alg:alg1} by extending the joint estimation of the supply elasticity estimator and the precision matrix to the overidentified case for when $k_x=0$. In view of Algorithm \ref{alg:alg2}, Algorithm \ref{alg:alg4} can be further extended to the case when $k_x > 0$, but we omit the details for brevity. 
\begin{center}
	\begin{algorithm}[!htbp]
		\renewcommand{\thealgorithm}{3}
		\caption{\textbf{Efficient GMM for $\phi^s$ (when $k_x=0$):}}
		\label{alg:alg4}
		\begin{algorithmic}[1]
			\begin{itemize}
				\item \textit{Step 1:} Run PCA on \eqref{runPCA} and obtain $\widehat{z}_t=\bs{S}'\bs{\widehat{Q}}\bs{\widetilde{y}}_{\cdot t}$ as the sample counterpart of \eqref{GIV}.
				\item  \textit{Step 2:} Initialize $\bs{\widehat{\Sigma}}_u^{-1}=\bs{I}_N$.
				\item  \textit{Step 3:} Estimate \eqref{gmmdemand} to obtain $\bs{\widehat{\varepsilon}}$, initialize $\bs{\widehat{W}}_s=(\bs{\widehat{Z}}_s'\bs{\widehat{Z}}_s)^{-1}$ and obtain $\bs{\widehat{\theta}}_{2SLS}^s(\bs{\widehat{Z}}_s,\bs{\widehat{\Sigma}}_{u}^{-1})$.
				\item  \textit{Step 4:} Obtain  $\bs{y}_{\widehat{E}}(\bs{\widehat{\Sigma}}_u^{-1})$. 		
				\item \textit{Step 5:} Update $\bs{\widehat{W}}_s = \left(\frac{1}{T}\sum_{t=1}^T\bs{\widehat{Z}}_{st} \bs{\widehat{Z}}'_{st} \widehat{u}^2_{\widehat{E}t}\right)^{-1}$, where $\widehat{u}_{\widehat{E}t}=y_{\widehat{E}t}-\bs{\widehat{\theta}}_{GMM}^s(\bs{\widehat{Z}}_s,\bs{\widehat{\Sigma}}_{u}^{-1})'\bs{\widehat{f}}_t$ and construct $\bs{\widehat{\theta}}_{GMM}^s(\bs{\widehat{Z}}_s,\bs{\widehat{\Sigma}}_{u}^{-1})$ as the sample counterpart of \eqref{gmmsupply}.
				\item \textit{Step 6:} Update $\bs{\widehat{\Sigma}}_u^{-1}$ by inverting $\bs{\widehat{\Sigma}}_u^{\mathbfcal{T}}$ defined in \eqref{POETu}. 
				\item \textit{Step 7:} Iterate \textit{Step 4} through \textit{Step 6} until convergence.
			\end{itemize} 						
		\end{algorithmic}
	\end{algorithm}
\end{center}  
In addition to efficiency gains, the efficient GMM estimators exhibit superior finite sample properties and are also robust to the GIV itself being a weak instrument. We illustrate these points in greater detail in \ref{FStheory} and Section \ref{weakGIV}.

The intuition for the overidentified estimators can be seen from observing the reduced form equation for (equilibrium) prices, $p_t=\frac{1}{\phi^d - \phi^s} \left(u_{St} +\bs{\lambda}_S' \bs{\eta}_t -\varepsilon_t \right)$. Clearly $\mathbbm{E}(p_t\bs{\eta}_t) \neq 0$ and $\mathbbm{E}(p_t\varepsilon_t) \neq 0$ and so instrumental relevancy is established. Thus, we are effectively back to the classical approach of finding exogenous supply shifters, in this case $\varepsilon_t$, to estimate the supply elasticity and finding exogenous demand shifters, in this case $\bs{\eta}_t$, to estimate the demand elasticity. With the exception that these shifters, $\bs{\eta}_t$ and $\varepsilon_t$ are unobserved. In what follows, we show that estimating $\bs{\eta}_t$ and $\varepsilon_t$ has a negligible effect on the limiting distributions of the estimators of demand and supply elasticities, respectively. 

\section{Assumptions}
Below we lay out the assumptions needed to derive our main results. \ref{A1}, \ref{A2} and \ref{A3a} are standard in the literature; see, for example, \citet{bai2003inferential}, \citet{fan2013large} and \citet{bai2017inferences}, \color{black}but are relevant for a thorough understanding of the subsequent theorems\color{black}. Whereas, \ref{A4} parts ii.) and iii.) are new so we provide more details. 
\begin{assump}[Factor Error Structure]\label{A1}
	The composite error term in \eqref{eq:1.2} is assumed to admit an (approximate) factor structure representation $v_{it} :=  \bs{\lambda}'_i \bs{\eta}_t + u_{it},$ %\footnote{Although cross sectional dependence is more reasonable than independence, one can test for cross sectional dependence for pure first-order dynamic panel data models by implementing CD tests as in \citet{pesaran2015testing}; however, tests for dynamic models and additional covariates, like the model in this paper, remains an open problem in the literature.}
where $\bs{\eta}_t=\begin{pmatrix}\eta_{1t} & \dots & \eta_{rt}\end{pmatrix}'$ is an $r \times 1$ vector of latent common factors and $\bs{\lambda}_i=\begin{pmatrix} \lambda_{1i} & \dots & \lambda_{ri}\end{pmatrix}'$ is an $r \times 1$ vector of latent factor loadings. We assume the factors are pervasive in the sense that $\bs{\Lambda}'\bs{\Lambda}/N$ converges to some $r \times r$ positive definite matrix. 
\end{assump}   
\begin{assump}\label{A2} \textbf{\emph{(Strict Stationarity, Exponential Tails \& Strong Mixing)}} \newline
	$\textbf{(A2i.)} \,\, \{\bs{\eta}_t, u_{i t}, \varepsilon_t\}_{t \geq 1}$ is strictly stationary and each with a zero mean. \newline 
	$\textbf{(A2ii.)} \,\,\exists \,\, c_1, c_2 > 0$ with $\gamma_{\emph{min}}(\bs{\Sigma}_u) > c_2$, $\underset{j \leq N}{\emph{max}} || \gamma_j || < c_1$, $c_2 < \gamma_{\emph{min}}(\mathbbm{cov}(\bs{\eta}_t)) \leq \gamma_{\emph{max}}(\mathbbm{cov}(\bs{\eta}_t)) < c_1.$\newline 
	$\textbf{(A2iii.)} \,\,$Exponential tail: $\exists \,\, r_1, r_2 > 0$ and $b_1, b_2 >0$, such that for any $s >0$, $i \leq N$ and $j \leq r$, $\mathbbm{P}(|u_{it}|>s) \leq \emph{exp}(-(s/b_1)^{r_1})),$ and $\mathbbm{P}(|\eta_{t,j}|>s) \leq \emph{exp}(-(s/b_2)^{r_2})).$ \newline 
	$\textbf{(A2iv.)} \,\,$Strong Mixing: $\exists \,\, r_3, \,C > 0$ $\forall \,\, T >0, \,\, r_1^{-1} + r_{2}^{-1} + r_{3}^{-1} > 1$, $\underset{A \in \mathcal{F}_{-\infty}^0, \,\, B \in \mathcal{F}_{T}^{\infty}}{\emph{sup}} |\mathbbm{P}(A)\mathbbm{P}(B)-\mathbbm{P}(AB)| < \emph{exp}(-CT^{r_3}),$ where $\mathcal{F}_{-\infty}^0$ and $\mathcal{F}_{T}^{\infty}$ denote the $\sigma$-algebras generated by $\{(\bs{\eta}_t, u_{it}, \varepsilon_t) : t<0\}$ and $\{(\bs{\eta}_t, u_{it}, \varepsilon_t) : t>T\}$ respectively.
%	\color{red} Remark referring to Acemoglu that states heavy-tailed shares + exponential tails in the errors is observationally equivalent to heavy-tailed errors. Also, new assumption 4 implies mixing here so can remove mixing here in A2\color{black}
\end{assump}

%\begin{assumpR}\label{R1}
%	The first statement rules out weak factors by assuring convergence of $\Lambda'\Lambda/N$ to a p.d. matrix, thus we assume the factors are strong/pervasive in the sense that a significant fraction of the units are affected by their prescense. This identifies the factors and effectively separates them from the idiosyncratic term. Hence why the factors are estimable via PCA.\footnote{Consistent estimation of weak factors is problematic, see for example \citet{onatski2012asymptotics} or \citet{bailey2016exponent} for suitable conditions for which it is possible. Even when estimable, their convergence rates are slower relative to estimates of strong factors, which will generally require modifications of the limiting distributions. This would be an interesting extension to consider for this paper.} Further, the second statement allows for $\eta_t$ to be dynamic itself, but that the dynamics do not enter our equation for the dependent variable. 
%\end{assumpR} 

%\edef\oldassump{\the\numexpr\value{assump}+1}
%\setcounter{assump}{0}
%\renewcommand{\theassump}{\oldassump.\alph{assump}}
\begin{assump}[Sparsity on $\bs{\Sigma}_u$]\label{A3a} Let $\bs{\Sigma}_u =(\sigma_{u,ij})$, for some $q \in [0,1/2),$ define
\begin{align}
m_N=\underset{i \leq N}{\emph{max}}\sum_{j=1}^N |\sigma_{u,ij}|^q.
\end{align}
We require that there is $q \in [0,1/2)$ such that $m_N \omega_{N,T}^{1-q}=o(1)$, where $\omega_{N,T}=\sqrt{\frac{\emph{log}(N)}{T}} +\frac{1}{\sqrt{N}}$.
\newline 
\end{assump}
%\begin{assumpR}\label{R2}
%	\ref{A2} and \ref{A3a} imply $v_{it}$ follows the so-called approximate factor model of \citet{arbitrage1983mean}, with the exception that we allow $\alpha_u >0$, whereas the Chamberlain model considered $\alpha_u=0$. \ref{A2} and \ref{A3a} are standard in the literature on large $N$ and $T$ factor estimation, e.g. \citet{bai2002determining}, \citet{bai2003inferential} \& \citet{bai2009panel}, with the exception that we impose weak stationarity. In the aforementioned papers, time heteroskedasticity is allowed. However, as we will aggregate along the cross-sectional dimension to arrive at a time series to estimate the short-run demand elasticity, we will impose weak stationarity. 
%\end{assumpR} 

%\begin{assumpR} \label{R3}
%	Assumption 3 is effectively stating that the idiosyncratic shocks are fully diversifiable for large-$N$. However, this will ultimately not be the case in this model, which will be made explicit after the following assumption, in \ref{R5}.
%\end{assumpR} 
\setcounter{assumpB}{3}
\begin{assumpB}\label{A4}\textbf{\emph{(Identification by GIV)}}\newline 
	$\textbf{(A4i.)} \,\,\,	\mathbbm{E} (z_t u_{Et})=\mathbbm{E} (z_t \varepsilon_t) =\mathbbm{E} (\bs{Z}_{st} u_{Et})=\mathbbm{E} (\bs{Z}_{dt} \varepsilon_t) = 0$. \newline
	\textbf{(A4ii.)} The sizes $\mathscr{S}_1,\dots,\mathscr{S}_N$ are drawn $i.i.d.$ from an arbitrary distribution for which the tail of the size distribution (i.e. above some threshold) follows a power law, with tail index, $\mu>0$
	\begin{align*}
	\mathbbm{P}(\mathscr{S}>s) = cs^{-{\mu}}. \nonumber 
	\end{align*} 
	The tail index $\mu$ determines the probability of observing extreme values. We assume that $\mathscr{S}_i$ is independent of $u_{it}$.  \newline
%	\color{red}\textbf{(A4ii.)} Heavy-tailed cross-sectional size distribution: The sizes $\mathscr{S}_{1t},\dots,\mathscr{S}_{Nt}$ are drawn $i.i.d.$ over $i$ from an arbitrary distribution for which the tail of the size distribution (i.e. above some threshold) follows a power law, with tail index, $\mu \in (0,1)$
%	\begin{align*}
%	\mathbbm{P}(\mathscr{S}>s) = cs^{-{\mu}}. \nonumber 
%	\end{align*} 
%	The tail index $\mu$ determines the probability of observing extreme values. We assume that $\mathscr{S}_{i}$ is independent of $u_{js} \,\, \forall i,j,s$ . 	\color{black} \newline
	\textbf{(A4iii.)} Suppose the sizes are ordered in decreasing fashion as such:
	$\mathscr{S}_{(1)} \geq \mathscr{S}_{(2)}  \geq \dots \geq \mathscr{S}_{(N-1)} \geq \mathscr{S}_{(N)} ,$ and we partition the cross-section as, $\mathcal{N}_{dominant}:= \{1,\dots,N_1\}$ and $\mathcal{N}_{fringe}:= \{N_1+1,\dots,N\}$ such that $\mathcal{N}_{dominant}\cup \mathcal{N}_{fringe}:=\mathcal{N}_{full}$. Let $S_i=\frac{\mathscr{S}_i}{\sum_{j=1}^N\mathscr{S}_j}$ denote the normalized shares such that $\sum_i S_i =1$. We assume $\forall \, i \,\in \mathcal{N}_{dominant}$, $\,S_i=\Theta_p(1)$, and $\forall \, i \in \mathcal{N}_{fringe},$ $S_{i}=\mathcal{O}_p\left(\frac{1}{N}\right)$. We further assume that the cardinality of the dominant units is fixed as $N \rightarrow \infty$, that is, $|\mathcal{N}_{dominant}|=N_1$ and $N_1$ does not rise with $N$ while the cardinality of the fringe grows with $N$, $|\mathcal{N}_{fringe}|=N-N_1 \rightarrow \infty$ as $N \rightarrow \infty$. \newline
\end{assumpB} 
\begin{assumpR} \label{remark1}
	 The first condition gives us instrumental exogeneity for the FGIV and efficient GMM estimators. The second condition allows for instrumental relevance in the extension of a large N framework. An important implication of the second condition is that the Herfindahl index, $h_{N,\mu}$, has the following asymptotic property
	\begin{align*}
	\sqrt{h_{N,\mu}}=\left|\big| \bs{S} \big|\right|_2 = \begin{cases}
	\Theta_p\left( 1\right) \hspace{13.4mm} \text{for} \hspace{5mm} \mu \in (0,1), \\
	\mathcal{O}_p\left( g_{N,\mu}\right) \hspace{8mm} \text{for} \hspace{5mm} \mu \in [1,2),
	\end{cases}
	\end{align*}
with $\mathcal{O}_p\left( g_{N,\mu}\right) \gg1/\sqrt{N}$. The variance of the just identified estimators is inversely proportional to the Herfindahl index, that is $\mathbbm{V}(\widehat{\phi}^j_{FGIV})=\mathcal{O}(h_{N,\mu}^{-1})$ for $j=s,d$, reflecting the fact that the more concentrated the market, the more precise the GIV methodology will be and also reflecting the fact that if the Herfindahl converges to zero in the limit, the variance will diverge.\footnote{The derivation of the asymptotic behavior of $h_{N,\mu}$ can be found in Supplementary Appendix \ref{herf}.} However, if $\mu$ is slightly greater than 1, theoretically identification breaks down for large $N$ but in any finite sample the GIV could be relevant (precisely due to $\mathcal{O}_p\left( g_{N,\mu}\right) \gg1\sqrt{N}$). Nevertheless, we rule this case out for the purpose of asymptotic inference.\footnote{For more details on instrumental relevance for large $N$, see Section \ref{weakGIV}.} 

Note, that the third condition is consistent with $\mu \in (0,1)$, but is slightly stronger. The third condition is also a generalization of the so-called "granular" weights in the panel data literature, say $\underset{{N \times 1}}{\bs{w}}$, which are typically assumed to satisfy $||\bs{w}||_2=\mathcal{O}\left(\frac{1}{\sqrt{N}}\right)$ and $\frac{w_i}{||\bs{w}||_2}=\mathcal{O}\left(\frac{1}{\sqrt{N}}\right) \,\, \forall \, i$. The third condition allows the share vector to be partitioned into a dominant part and a fringe part. That is, $\bs{S}=\begin{pmatrix} \bs{S}'_{d} & \bs{S}'_f \end{pmatrix}'$ where $\bs{S}_{d}$ is $N_1 \times 1$, is the dominant part and $\bs{S}_f$ is $N_2 \times 1$, is the fringe part; with $N_1+N_2=N$, the key being that $N_1(N)=N_1$ is fixed while $N_2(N) \rightarrow \infty$ as $N \rightarrow \infty$. This assumption can be empirically justified in concentrated markets, see Section \ref{oil} as an example; as well as mathematically justified, see \cite{logan1973limit}. 
\end{assumpR}
%\color{red}
%\begin{prop} \label{newprop1}
%	\color{red} Under Assumption \textbf{(A4ii.)}, for each $t$ the Herfindahl index, $h_{N,\mu, t}$, satisfies
%	\begin{align*}
%	\sqrt{h_{N,\mu,t}}=\left|\big| \bs{S}_{\cdot t} \big|\right|_2 = \begin{cases}
%	\Theta_p\left( 1\right) & \mu \in (0,1), \\
%	\mathcal{O}_p\left( g_{N,\mu,t}\right) & \mu \geq 1,
%	\end{cases}
%	\end{align*}
%	with $\mathcal{O}_p\left( g_{N,\mu,t}\right) \gg1/\sqrt{N}$. Moreover, the number of dominant units, $N_1$ remains fixed while $N \rightarrow \infty$.
%\end{prop}\color{black}

\begin{assumpR}
	Taking the variance of the equilibrium price process (assuming the covariances to be zero for simplicity) we obtain $\mathbbm{V}(p_t) =\frac{1}{(\phi^d-\phi^s)^2}(\mathbbm{V}(u_{St})+\mathbbm{V}(\bs{\lambda}_S'\bs{\eta}_t)+\mathbbm{V}(\varepsilon_t))= \Theta(1),$ where the last equality follows by the second and the third conditions in \ref{A4}, details can be found in Lemma \ref{lemmaC1} in the Appendix. Without these conditions, one would obtain the unsatisfactory result that $\mathbbm{V}(p_t)=\mathcal{O}(N)$, that is, the variance of the price process is unbounded for each $t$ as $N\rightarrow \infty$. Effectively, \ref{A4} allows the coexistence of a finite number of dominant units, in terms of size, whose cardinality can not grow with $N$, while at the same time allowing for a bounded variance for the aggregate endogenous variable $p_t$. 
\end{assumpR}
\section{Limiting Distributions}\label{suppest} 
In this section, we first present the limiting distributions of the FGIV elasticity estimators, corresponding to \eqref{justidentifiedGIV} and \eqref{iterphid} with Algorithm \ref{alg:alg1}. We then move on to the limiting distributions of the efficient GMM elasticity estimators, corresponding to \eqref{gmmdemand} and \eqref{gmmsupply} with Algorithm \ref{alg:alg4}. 

\textbf{\large Just identified demand elasticity.} The just identified demand elasticity estimator in \eqref{justidentifiedGIV} is given by
\begin{align*}
\widehat{\phi}^d(\bs{\widehat{z}})=\dfrac{\sum_{t=1}^T\widehat{z}_td_t}{\sum_{t=1}^T\widehat{z}_tp_t} = \dfrac{\sum_{t=1}^T \sum_{i,j}S_i \widehat{Q}_{ij}\tilde{y}_{jt}d_t}{\sum_{t=1}^T\sum_{i,j}S_i \widehat{Q}_{ij}\tilde{y}_{jt}p_t}.
\end{align*}
Hence,
\begin{align*}
\widehat{\phi}^d-\phi^d &= \left( \sum_t \widehat{z}_tp_t\right)^{-1} \left(\sum_t \widehat{z}_t \varepsilon_t \right), \\
&= \left( T^{-1}\sum_t z_tp_t+T^{-1}\sum_t (\widehat{z}_t-z_t)p_t\right)^{-1} \left(T^{-1}\sum_t z_t \varepsilon_t+T^{-1}\sum_t (\widehat{z}_t-z_t) \varepsilon_t \right).
\end{align*}
From above, it is apparent we need to show $\frac{1}{T}\sum_{t=1}^T(\widehat{z}_t-z_t)\varepsilon_t
=\frac{1}{T}\sum_{t=1}^T \bs{S}'(\bs{\widehat{Q}}-\bs{Q})\bs{\tilde{y}_{\cdot t}}  \varepsilon_t=o_p(1)$ and
$\frac{1}{T}\sum_{t=1}^T(\widehat{z}_t-z_t)p_t =\frac{1}{T}\sum_{t=1}^T \bs{S}'(\bs{\widehat{Q}}-\bs{Q})\bs{\tilde{y}}_{\cdot t}  p_t =o_p(1)$. Indeed, we show in Lemma \ref{lemmaC2}, in the Appendix, that
\begin{align}
T^{-1}\sum_{t=1}^T \bs{S}'(\bs{\widehat{Q}}-\bs{Q})\bs{\tilde{y}}_{\cdot t}  \varepsilon_t =\mathcal{O}_p(C^{-2}_{NT})+\mathcal{O}_p(C^{-2}_{NT})\cdot \mathcal{O}_p\left(\frac{N}{T}\right)+\mathcal{O}_p\left( \frac{1}{\sqrt{N}}\cdot C^{-1}_{NT}\right), \label{one}\\
T^{-1}\sum_{t=1}^T \bs{S}'(\bs{\widehat{Q}}-\bs{Q})\bs{\tilde{y}}_{\cdot t}  p_t =\mathcal{O}_p(C^{-2}_{NT})+\mathcal{O}_p(C^{-2}_{NT})\cdot \mathcal{O}_p\left(\frac{N}{T}\right)+\mathcal{O}_p\left( \frac{1}{\sqrt{N}}\cdot C^{-1}_{NT}\right), \label{two}
\end{align}
where $C_{NT} := \text{min} \{\sqrt{N},\sqrt{T}\}$. The terms in \eqref{one} and \eqref{two} are $o_p(1)$ when $N<T$ without any restrictions; however, when $T < N$ we require the mild restriction that $N/T^2 \rightarrow 0$, i.e., if $T<N$, $T^2$ does not grow too slowly relative to $N$. Thus, making use of \eqref{one} and \eqref{two} we obtain
\begin{align}
\widehat{\phi}^d-\phi^d &= \left( \sum_t \widehat{z}_tp_t\right)^{-1} \left(\sum_t \widehat{z}_t \varepsilon_t \right)= \left(T^{-1}\sum_t z_tp_t\right)^{-1} T^{-1}\sum_t z_t \varepsilon_t  +o_p(1).\label{consisref}
\end{align} 
The order of the sampling error generally relies, in part, on the order of the Herfindahl. The order of the Herfindahl, in turn, critically depends on $\mu$, the tail index of the size distribution (\color{black} see \ref{remark1}\color{black}). Results on the order of the Herfindahl as a function of the tail index parameter $\mu$ entails a total of six possible cases. The results can be found in Table~\ref{regimes} of Supplementary Appendix \ref{herf}. However, for inference, we require $\mu \in (0,1)$ (regularly varying tails) or $\mu \rightarrow 0$ (slowly varying tails) as discussed in detail in the previous section's remarks. Given this, even after pinning down the order of the Herfindahl, the panel dimensions can distinguish more cases as seen above. Nevertheless, as  \eqref{one}, \eqref{two} and \eqref{consisref} indicate, for consistency we have the following result:
\setcounter{theorem}{0}
\begin{theorem}[Consistency of $\widehat{\phi}^d$]\label{sconsis}
	Under Assumptions 1-4, as $(N,T) \overset{j}{\rightarrow} \infty$, we have that when $N\geq T$ and $N/T^2 \rightarrow 0$ or when $N<T$
	\begin{align}
	\widehat{\phi}^d-\phi^d \overset{p}{\rightarrow} 0.
	\end{align}
\end{theorem}
All proofs are deferred tot the appendix. Now, multiplying \eqref{consisref} by $\sqrt{T}$
\begin{equation}
\begin{aligned}
 \sqrt{T}(\widehat{\phi}^d-\phi^d) &= \left( T^{-1}\sum_t z_tp_t\right)^{-1} \left(\frac{1}{\sqrt{T}}\sum_t z_t \varepsilon_t +\mathcal{O}_p\left(\frac{\sqrt{T}}{C^2_{NT}}\right)  +\mathcal{O}_p\left(\frac{N}{\sqrt{T}\cdot C^2_{NT}}\right) +  \mathcal{O}_p\left(\frac{\sqrt{T}}{\sqrt{N} \cdot C_{NT}}\right)\right).
\end{aligned}  
\end{equation}
We can state the following result for the limiting distribution:
\begin{theorem}[Limiting distribution for $\widehat{\phi}^d$]\label{sdist}
Under Assumptions 1-4 as $(N,T) \overset{j}{\rightarrow} \infty$, we have that when $N\geq T$, $N/T^{3/2} \rightarrow 0$ and $\sqrt{T}/N \rightarrow 0$; or when $N<T$ only $\sqrt{T}/N \rightarrow 0$
\begin{align}
\sqrt{T} (\widehat{\phi}^d-\phi^d) \overset{d}{\rightarrow} \mathcal{N}\left(0, \mathbbm{v}_d\right),
\end{align} 
where $\mathbbm{v}_d :=\mathbbm{m}_{zp}^{-2}\,\mathbbm{v}_{z\varepsilon}$, $\mathbbm{v}_{z\varepsilon}:=\mathbbm{E}(z^2_t\varepsilon^2_t)$ and $\mathbbm{m}_{zp} :=\mathbbm{E}(z_tp_t)$.
\end{theorem}
$\mathbbm{v}_{z\varepsilon}$ can be consistently estimated with
\begin{align}
\widehat{\mathbbm{v}}_{z\varepsilon} =\begin{cases} T^{-1}\sum_{t=1}^T \widehat{z}_t^{\,2}\, \widehat{\varepsilon}_t^{\,2} \hspace{88.7mm} \text{HC}, \vspace{3mm} \\
T^{-1}\sum_{t=1}^T \widehat{z}_t^{\,2} \, \widehat{\varepsilon}_t^{\,2} + 2 \cdot T^{-1}\sum_{j=1}^m \left(1-\dfrac{j}{m+1}\right)\sum_{t=j+1}^T \widehat{z}_{t}\widehat{\varepsilon}_{t}\widehat{z}_{t-j} \widehat{\varepsilon}_{t-j} \hspace{5mm} \text{HAC},
\end{cases}
\end{align}
where HC and HAC denote heteroskedasticity-consistent and heteroskedasticity and autocorrelation consistent estimators, respectively. Hence $\frac{\sqrt{T}(\widehat{\phi}^d-\phi^d)}{\widehat{\mathbbm{v}}_d^{1/2}} \sim \text{t}_{df} \overset{d}{\rightarrow} \mathcal{N}(0,1)$, where $\widehat{\mathbbm{v}}_d^{1/2}=\widehat{\mathbbm{m}}_{zp}^{-2}\widehat{\mathbbm{v}}_{z\varepsilon}^{1/2}$, with $\widehat{\mathbbm{m}}_{zp}=T^{-1}\sum_{t=1}^T \widehat{z}_t p_t$ also consistent for $\mathbbm{m}_{zp}$. We will see in Section \ref{MCstudy} that the asymptotic theory provides good approximations to the finite sample distribution. 
\begin{assumpR}\label{R4}
	As in GK, we express $\mathbbm{v}_d$ as inversely related to the Herfindahl, $h_{N,\mu}$, as claimed in \ref{remark1}, for insights on the role of market concentration on precision of the GIV. Assuming conditional homoskedasticity of $\varepsilon_t$ and homoskedasticity of $u_{it}$, we have that
	\begin{align*}
	\mathbbm{v}_{z\varepsilon} &=\mathbbm{E}(z_t^2\varepsilon_t^2)=\sigma^2_{\varepsilon}\cdot\sigma^2_{\tilde{u}} \cdot \mathbbm{E}(\bs{S}'\bs{Q}\bs{S}). \numberthis \label{varherf}
	\end{align*}
If $\bs{\lambda}_i=\bs{\lambda} \,\, \forall i$, then there is no need to purge the factor structure through the loading space. That is, a simple cross-sectional demeaning transformation will suffice,  $\bs{Q}=(\bs{I}_N-\bs{\tilde{\Lambda}(\tilde{\Lambda}'\tilde{\Lambda})^{-1}\tilde{\Lambda'}})=(\bs{I}_N-\bs{\frac{\iota\iota'}{N}})$. We can simplify equation \eqref{varherf} to (where we make use of the normalization that $\bs{S}'\bs{\iota}=1$)
\begin{align*}
\mathbbm{v}_{z\varepsilon} &=\sigma^2_{\varepsilon}\cdot\sigma^2_{\tilde{u}} \cdot \left(\mathbbm{E}(\bs{S'S})-\frac{1}{N}\right) =\sigma^2_{\varepsilon}\cdot \underbrace{\sigma^2_{\tilde{u}} \cdot \left(\mathbbm{E}(h_{N,\mu})-\frac{1}{N}\right)}_{\text{$\mathbbm{E}(z_t^2)$}},
\end{align*}
whereas, $\mathbbm{m}_{zp}=\mathbbm{E}(p_tz_t) \propto \mathbbm{E}(z_t^2)$. Hence,
\begin{align}
\mathbbm{v}_d &\propto \dfrac{\sigma^2_{\varepsilon} \cdot \sigma^2_{\tilde{u}} \cdot \left(\mathbbm{E}(h_{N,\mu})-\frac{1}{N}\right) }{\left[\sigma^2_{\tilde{u}} \cdot \left(\mathbbm{E}(h_{N,\mu})-\frac{1}{N}\right)\right]^2} =\dfrac{\sigma^2_{\varepsilon} }{\sigma^2_{\tilde{u}} \cdot \left(\mathbbm{E}(h_{N,\mu})-\frac{1}{N}\right)}.
\end{align}
Thus, the more concentrated the market, the more precise the estimator. See Section \ref{weakGIV} for a more general treatment. 
\end{assumpR}
\textbf{\large Just identified supply elasticity.}
For the just identified supply elasticity estimator in \eqref{iterphid}, upon convergence of Algorithm \ref{alg:alg1}, we have that
\begin{align}
\widehat{\phi}^s-\phi^s &=\underbrace{\left(T^{-1}\bs{\widehat{z}}^{\,'} \,\bs{M}_{\widehat{\eta}}\,\bs{p}\right)^{-1}}_{\text{$\widehat{A}^{-1}$}} \, \underbrace{T^{-1}\bs{\widehat{z}}^{\,'}\,\bs{M}_{\widehat{\eta}}\,\bs{\eta}\cdot \bs{\lambda}_{\widehat{E}}}_{\text{$\widehat{B}$}}+\left(T^{-1}\bs{\widehat{z}}^{\,'} \,\bs{M}_{\widehat{\eta}}\,\bs{p}\right)^{-1}\, \underbrace{T^{-1}\bs{\widehat{z}}^{\,'}\,\bs{M}_{\widehat{\eta}}\,\bs{u}_{\widehat{E}}}_{\text{$\widehat{C}$}}. 
\end{align}
We can write the scalars $\widehat{A}$, $\widehat{B}$ and $\widehat{C}$ as follows
\begin{align*}
\widehat{A} &=T^{-1} \bs{z}'\,\bs{M}_{\eta}\,\bs{p} +\underbrace{T^{-1}(\bs{\widehat{z}}-\bs{z})'\,\bs{M}_{\widehat{\eta}} \,\bs{p}}_{\text{$a_1$}} +\underbrace{T^{-1}\bs{z}'\,(\bs{M}_{\widehat{\eta}}-\bs{M}_{\eta} )\,\bs{p}}_{\text{$a_2$}}, \numberthis \label{A}\\
\widehat{B} &= \underbrace{T^{-1}(\bs{\widehat{z}}-\bs{z})'\,\bs{M}_{\widehat{\eta}} \,\bs{\eta} \lambda_{\widehat{E}}}_{\text{$b_1$}} +\underbrace{T^{-1}\bs{z}'\,(\bs{M}_{\widehat{\eta}}-\bs{M}_{\eta} )\,\bs{\eta} \bs{\lambda}_{\widehat{E}}}_{\text{$b_2$}}, \numberthis \label{B} \\
\widehat{C} &= T^{-1}\bs{z}'\,\bs{M}_{\eta}\,\bs{u}_{E} +\underbrace{T^{-1}(\bs{\widehat{z}}-\bs{z})'\,\bs{M}_{\widehat{\eta}} \,\bs{u}_{\widehat{E}}}_{\text{$c_1$}} +\underbrace{T^{-1}\bs{z}'\,(\bs{M}_{\widehat{\eta}}-\bs{M}_{\eta} )\,\bs{u}_{\widehat{E}}}_{\text{$c_2$}}+\underbrace{T^{-1}\bs{z}'\,\bs{M}_{{\eta}}\,(\bs{u}_{\widehat{E}}-\bs{u}_E)}_{\text{$c_3$}}.\numberthis \label{C} 
\end{align*} 
It is shown in Lemma \ref{lemmaC3} of the Appendix that the terms $a_i, b_i, c_j$ are $o_p(1)$ for $i=1,2$; $j=1,2,3$, such that
\begin{align}
\widehat{\phi}^s-\phi^s &=\left(T^{-1}\bs{z}'\,\bs{M}_{\eta}\,\bs{p}\right)^{-1}  \left(T^{-1}\bs{z}'\,\bs{M}_{\eta}\,\bs{u}_{E}\right) + o_p(1). \label{dem}
\end{align}
We can now state the following result:
\begin{theorem}[Consistency of $\widehat{\phi}^s$]\label{dconsis}
	Under Assumptions 1-4, as $(N,T) \overset{j}{\rightarrow} \infty$, we have 
	\begin{align}
	\widehat{\phi}^s - \phi^s \overset{p}{\rightarrow} 0.
	\end{align}
\end{theorem}
Now, multiplying \eqref{dem} by $\sqrt{T}$
\begin{equation}
\begin{split}
\sqrt{T}(\widehat{\phi}^s-\phi^s) &=\left(T^{-1}\bs{z}'\,\bs{M}_{\eta}\,\bs{p}\right)^{-1}  \left(\frac{\bs{z}'\,\bs{M}_{\eta}\,\bs{u}_{E}}{\sqrt{T}} +\mathcal{O}_p\left(\frac{\sqrt{T}}{C^2_{NT}}\right) + \mathcal{O}_p\left(\frac{\sqrt{T}}{\sqrt{NT} \cdot C_{NT}}\right)\right) + \\ &\left(T^{-1}\bs{z}'\,\bs{M}_{\eta}\,\bs{p}\right)^{-1}\left(\mathcal{O}_p\left(\frac{N}{\sqrt{T}\cdot C^2_{NT}}\right) +  \mathcal{O}_p\left(\frac{m_N \omega^{1-q}_{N,T}}{\sqrt{T}}\right)\right),
\end{split}
\end{equation}
we can state the following result for the limiting distribution:
\begin{theorem}[Limiting distribution for $\widehat{\phi}^s$]\label{ddist}
	Under Assumptions 1-4, as $(N,T) \overset{j}{\rightarrow} \infty$, we have that when $N\geq T$,  $N/T^{3/2} \rightarrow 0$ and $\sqrt{T}/N \rightarrow 0$; or when $N < T$ only $\sqrt{T}/N \rightarrow 0$
	\begin{align}
	\sqrt{T} (\widehat{\phi}^s-\phi^s) \overset{d}{\rightarrow}\mathcal{N}\left(0, \mathbbm{v}_s\right),
	\end{align}
	where $\mathbbm{v}_s:=\mathbbm{m}_{z\tilde{p}}^{-2}\,\mathbbm{v}_{zu}$, $\mathbbm{v}_{zu}:=\mathbbm{E}(z_t^{\,2} \,(\bs{M}_{\eta}\,\bs{u}_{E})_{t}^{\,2})$ and $\mathbbm{m}_{z\tilde{p}} =\mathbbm{E}(z_t(\bs{M}_{{\eta}}\,\bs{p})_{t})$.
\end{theorem}
$\mathbbm{v}_{zu}$ can be consistently estimated with
\small\begin{align}
\widehat{\mathbbm{v}}_{zu} =\begin{cases}
T^{-1}\sum_{t=1}^T \widehat{z}_t^{\,2} \,(\bs{M}_{\widehat{\eta}}\,\bs{\widehat{u}}_{E})_{t}^{\,2}\hspace{105.9mm} \text{HC,} \vspace{3mm} \\
T^{-1}\sum_{t=1}^T \widehat{z}_t^{\,2} (\bs{M}_{\widehat{\eta}}\,\bs{\widehat{u}}_{\widehat{E}})_t^{2} + 2 \cdot T^{-1}\sum_{j=1}^m \left(1-\dfrac{j}{m+1}\right)\sum_{t=j+1}^T  \widehat{z}_{t} (\bs{M}_{\widehat{\eta}}\,\bs{\widehat{u}}_{\widehat{E}})_{t}\widehat{z}_{t-j} (\bs{M}_{\widehat{\eta}}\,\bs{\widehat{u}}_{\widehat{E}})_{t-j} \hspace{3.7mm} \text{HAC.}
\end{cases}
\end{align}\normalsize
Hence $ \frac{\sqrt{T}(\widehat{\phi}^{\,s}-\phi^s)}{\widehat{\mathbbm{v}}_{s}^{\,\,1/2}} \sim \text{t}_{df} \overset{d}{\rightarrow} \mathcal{N}(0,1)$, where $\widehat{\mathbbm{v}}_s^{1/2}=\widehat{\mathbbm{m}}_{z\tilde{p}}^{-2}\widehat{\mathbbm{v}}_{zu}$, with $\widehat{\mathbbm{m}}_{z\tilde{p}}=T^{-1}\sum_{t=1}^T\widehat{z}_t(\bs{M}_{\widehat{\eta}}\,\bs{p})_{t}$ consistent for $\mathbbm{m}_{z\tilde{p}}$. We will see in Section \ref{MCstudy} that asymptotic theory provides good approximations to the finite sample distribution.  

\textbf{\large Overidentified demand elasticity.} For the overidentified demand elasticity estimator in \eqref{gmmdemand}, recall the estimated instrument matrix consists of $\bs{\widehat{Z}}_d=\begin{pmatrix} \bs{\widehat{z}}_{GIV} & \bs{\widehat{\eta}} \end{pmatrix}$. For the strong factors, $\bs{\widehat{\eta}}$, estimated via PCA, \citet{bai2010instrumental} showed that the generated regressors problem of \citet{pagan1984econometric} does not arise when both $N$ and $T$ are large. Thus, the sampling error in $\bs{\widehat{\eta}}$ is negligible in consideration of the limiting distribution of the overidentified demand elasticity estimate. In the previous section, we established that estimation of $\bs{\widehat{z}}_{GIV}$ is also negligible under regularity, we can then use standard asymptotic theory to also obtain asymptotic normality of the efficient GMM estimator in the case of demand since 
\begin{align}
\widehat{\phi}^d_{GMM}-\phi^d &=\left(\bs{p}'\,\bs{\widehat{Z}}_d \,\bs{\widehat{\Omega}}_d^{-1}\, \bs{\widehat{Z}}'_d \, \bs{p}\right)^{-1}\bs{p}'\,\bs{\widehat{Z}}_d \,\bs{\widehat{\Omega}}_d^{-1}\, \nonumber  \bs{\widehat{Z}}^{\,'}_d \, \bs{\varepsilon}, \\
&=\left(\bs{p}'\,\bs{Z}_d \,\bs{\Omega}_d^{-1}\, \bs{Z}^{\,'}_d \, \bs{p}\right)^{-1}\bs{p}'\,\bs{Z}_d \,\bs{\Omega}_d^{-1}\, \bs{Z}^{\,'}_d \, \bs{\varepsilon}+o_p(1).
\end{align}  
We can now state the following theorem:
\begin{theorem}[Limiting distribution for $\widehat{\phi}^d_{GMM}$]\label{ddistgmm}
	Under Assumptions 1-4, with $\mathbbm{E}(\varepsilon_t\bs{\eta}_t)=0 \, \forall t$, as $(N,T) \overset{j}{\rightarrow} \infty$, we have that when $N\geq T$,  $N/T^{3/2} \rightarrow 0$ and $\sqrt{T}/N \rightarrow 0$; or when $N < T$ only $\sqrt{T}/N \rightarrow 0$
	\begin{align}
	\sqrt{T} (\widehat{\phi}^d_{GMM}-\phi^d) \overset{d}{\rightarrow} \mathcal{N}\left(0, \mathbbm{V}(\widehat{\phi}^d_{GMM})\right),
	\end{align}
	where 
	\begin{align}
	\mathbbm{V}(\widehat{\phi}^d_{GMM})&=\left(\bs{m}'_{{Z}_{d}p}\, \bs{\Omega}_d^{-1} \bs{m}_{{Z}_{d}p}\right)^{-1}, \label{effic}
	\end{align}
 with $\bs{m}_{{Z}_{d}p}=\mathbbm{E}(\bs{Z}_{dt}\,p_t)$ and $\bs{\Omega}_d=\emph{plim }T^{-1}\sum_{t=1}^T\bs{\widehat{Z}}_{dt}\bs{\widehat{Z}}_{dt}^{'}(d_t-p_t\widehat{\phi}^d_{2SLS})^2$.
\end{theorem}
$\mathbbm{V}(\widehat{\phi}^d_{GMM})$ can be consistently estimated using $2SLS$ residuals with 
\begin{align}
\widehat{\mathbbm{V}}(\widehat{\phi}^d_{GMM})=\left(\frac{\bs{p}^{\,'}\bs{\widehat{Z}}_{d}}{T}\, \bs{\widehat{\Omega}}_d^{-1}\, \frac{\bs{\widehat{Z}}_{d}^{\,'}\,\bs{p}}{T}\right)^{-1}, \label{noWC}
\end{align} 
where $\bs{\widehat{\Omega}}_d=T^{-1}\sum_{t=1}^T\bs{\widehat{Z}}_{dt}\bs{\widehat{Z}}_{dt}^{\,'}\,(d_t-p_t\widehat{\phi}^d_{2SLS})^2$. It is well known that $\mathbbm{V}(\widehat{\phi}^d_{GMM})$ attains the semiparametric efficiency bound, as shown by \citet{chamberlain1987asymptotic}, which reduces to \eqref{effic} in the linear model. Standard overidentification tests can be carried out since 
\begin{align}
J_d=T\cdot \left(T^{-1}\sum_{t=1}^T \bs{Z}_{dt} \,\varepsilon_t(\widehat{\phi}_{GMM}^d)\right)' \, \bs{\widehat{\Omega}}_d^{-1} \, \left(T^{-1}\sum_{t=1}^T\bs{Z}_{dt} \,{\varepsilon}_t(\widehat{\phi}_{GMM}^d)\right) \overset{d}{\rightarrow} \chi^2_{df_d},
\end{align} 
where the degrees of freedom is given by $df_d=(1+r)-k_d=r$ and $k_d=1$ is the number of endogenous regressors. We will see that simulation evidence shows that the size of the $J$-test is near the nominal size when the true $r$ is used and when $rmax > r$ factors are used; which is important in empirical work when $r$ is typically estimated and it is generally known that an overestimate of $r$ is preferred in order to prevent an effect akin to omitted variable bias, see \citet{moon2015linear} who formalize this notion. 
\textbf{\large Overidentified supply elasticity.} The full instrument matrix for the overidentified supply elasticity estimator in \eqref{gmmsupply} consists of $\bs{\widehat{Z}}_s=\begin{pmatrix}\bs{\widehat{z}}_{GIV} &\bs{\widehat{\varepsilon}} & \bs{\widehat{\eta}} \end{pmatrix}$, (recall the factors self instrument here, as they are part of the supply equation). We show that the sampling error in $\widehat{\bs{Z}}_s$ is indeed negligible. This is again due to both large $N$ and $T$. As a result, $\underset{(2+r) \times (2+r)}{\bs{\Omega}_s}=\text{plim }\,T^{-1}\sum_{t=1}^T\bs{\widehat{Z}}_{st}\bs{\widehat{Z}}^{'}_{st} (y_{\widehat{E}t}-\bs{\widehat{f}}_t'\bs{\widehat{\theta}}^s_{GMM})^2$ is sufficient when constructing the efficient weighting matrix, even though it does not take the sampling error in our estimate of ${\phi}^d$ into account (since $\bs{\widehat{\varepsilon}}=\bs{\varepsilon}(\widehat{\phi}_{GMM}^d)$). That is 
\begin{align}
\bs{\widehat{\theta}}^s_{GMM}-\bs{\theta}^s &=\left(\bs{f}'\,\bs{\widehat{Z}}_s \,\bs{\widehat{\Omega}}_s^{-1}\, \bs{\widehat{Z}}^{\,'}_s \bs{f}\right)^{-1}\bs{f}'\,\bs{\widehat{Z}}_s \,\bs{\widehat{\Omega}}_s^{-1}\, \nonumber  \bs{\widehat{Z}}^{\,'}_s \bs{u}_E, \\
&=\left(\bs{f}'\,\bs{Z}_s \,\bs{\Omega}_s^{-1}\,\bs{Z}^{\,'}_s \bs{f}\right)^{-1}\bs{f}'\,\bs{Z}_s \,\bs{\Omega}_s^{-1}\, \bs{Z}^{\,'}_s \bs{u}_E+o_p(1).
\end{align}  
We can now state the following theorem:
\begin{theorem}[Limiting distribution for $\bs{\widehat{\theta}}^s_{GMM}$]\label{sdistgmm}
	Under Assumptions 1-4, as $(N,T) \overset{j}{\rightarrow} \infty$, we have that when $N\geq T$,  $N/T^{3/2} \rightarrow 0$ and $\sqrt{T}/N \rightarrow 0$; or when $N < T$ only $\sqrt{T}/N \rightarrow 0$
	\begin{align}
	\sqrt{T} (\bs{\widehat{\theta}}^s_{GMM}-\bs{\theta}^s)&\overset{d}{\rightarrow} \mathcal{N}\left(\bs{0}, \mathbbm{V}(\bs{\widehat{\theta}}^s_{GMM})\right),
	\end{align}
	where 
	\begin{align}
	\mathbbm{V}(\bs{\widehat{\theta}}^s_{GMM})&=\left(\bs{m}'_{{Z}_{s}f}\, \bs{\Omega}_s^{-1} \bs{m}_{{Z}_{s}f}\right)^{-1},  \label{correctvar} 
\end{align} 
with $\bs{m}_{{Z}_{s}f}=\mathbbm{E}(\bs{Z}_{st}\bs{f}'_t)$ and $\bs{\Omega}_s=\emph{plim }T^{-1}\sum_{t=1}^T\bs{\widehat{Z}}_{st}\bs{\widehat{Z}}_{st}^{'}(y_{\widehat{E}t}-\bs{\widehat{f}}_t'\bs{\widehat{\theta}}^s_{GMM})^2$.
\end{theorem}
$\mathbbm{V}(\bs{\widehat{\theta}}^s_{GMM})$ can be consistently estimated using $GMM$ residuals with 
\begin{align}
\widehat{\mathbbm{V}}(\bs{\widehat{\theta}}^s_{GMM})=\left(\frac{\bs{\widehat{f}}^{\,'}\bs{\widehat{Z}}_{s}}{T}\, \bs{\widehat{\Omega}}_s^{-1}\, \frac{\bs{\widehat{Z}}_{s}^{\,'}\bs{\widehat{f}}}{T}\right)^{-1}, \label{noWC2}
\end{align} 
where $\bs{\widehat{\Omega}}_s=T^{-1}\sum_{t=1}^T\bs{\widehat{Z}}_{st}\bs{\widehat{Z}}_{st}^{\,'}\,(y_{\widehat{E}t}-\widehat{\bs{f}}_t'\bs{\widehat{\theta}}^s_{GMM})^2$. 
Just as in the case of the overidentified demand elasticity estimator, \eqref{correctvar} achieves the semiparametric efficiency bound. Overidentification tests can be carried out since
 \begin{align}
 J_s=T\cdot \left(T^{-1}\sum_{t=1}^T \bs{Z}_{st} \,u_{\widehat{E}t}(\widehat{\bs{\theta}}_{GMM}^s)\right)' \, \bs{\widehat{\Omega}}_s^{-1} \, \left(T^{-1}\sum_{t=1}^T\bs{Z}_{st} \,u_{\widehat{E}t}(\widehat{\bs{\theta}}_{GMM}^s)\right) \overset{d}{\rightarrow} \chi^2_{df_s},
 \end{align} 
 where the degrees of freedom are given by $df_s=2-k_s=2-1=1$ and $k_s=1$ is the number of endogenous regressors for the supply equation.
% \footnote{We use \eqref{noWC2} to estimate the variance in our simulations and don't observe the commonly reported downward bias in our standard errors. But, one can of course make use of the \citet{windmeijer2005finite} finite sample variance correction, which is given by 
% 	\begin{align*}
% 	\widehat{\mathbbm{V}}_c(\bs{\widehat{\theta}^s_{gmm}})&=\widehat{\mathbbm{V}}(\bs{\widehat{\theta}^s_{gmm}})+\widehat{C}_{s1}+\widehat{C}_{s2}, \\
% 	\bs{\widehat{C}_{s1}}&=\frac{1}{T}\left[ \bs{\widehat{D}_s}\left(\bs{G_s}'\bs{\widehat{\Omega}_s^{-1}}(\bs{\widehat{\theta}^s_{2sls}})\bs{G_s} \right)^{-1} + \left(\bs{G_s}'\bs{\widehat{\Omega}_s^{-1}}(\bs{\widehat{\theta}^s_{2sls}})\bs{G_s} \right)^{-1}\bs{\widehat{D}_s}'\right]\\
% 	\bs{\widehat{C}_{s2}}&=\bs{\widehat{D}_s}\cdot\widehat{\mathbbm{V}}(\bs{\widehat{\theta}^s_{2sls}})\cdot \bs{\widehat{D}_s},
% 	\end{align*} 
% 	where 
% 	\begin{align*}
% 	\bs{\widehat{D}_s}[\cdot,j]&=\left(\bs{G_s}'\bs{\widehat{\Omega}_s^{-1}}(\bs{\widehat{\theta}^s_{2sls}})\bs{G_s} \right)^{-1}\bs{G_s}'\bs{\widehat{\Omega}_s^{-1}}(\bs{\widehat{\theta}^s_{2sls}})\frac{\partial \bs{\widehat{\Omega}_s}}{\partial \theta^s_j}\bs{\widehat{\Omega}_s^{-1}}(\bs{\widehat{\theta}^s_{2sls}}) \bs{\bar{g}_s}(\bs{\widehat{\theta}^s_{gmm}}), \\ 
% 	\bs{G_s} &=\frac{\partial \bs{\bar{g}_s}(\bs{\theta}^d)}{\partial \bs{\theta}^s}, \\
% 	\bs{\bar{g}_s}(\bs{\widehat{\theta}^s_{gmm}})&=\frac{1}{T}\sum_{t=1}^T\bs{g_{st}}(\bs{\widehat{\theta}^s_{gmm}}), \\
% 	\bs{g_{st}}({\phi}^d)&=\bs{Z_{st}}u_{Et}(\bs{\theta}^s)
% 	\end{align*}} 
\begin{assumpR}\label{FStheory}
	The asymptotic distribution of the FGIV and efficient GMM estimators of demand and supply elasticities are established. However, the finite sample moments of these estimators, are unbounded to different degrees. The extensive literature on the classic simultaneous equations model has documented this result in many forms, see \citet{mariano1973approximations}, \citet{hatanaka1973existence}, \citet{sawa1972finite}, \citet{takeuchi1970exact}, \citet{ullah1974exact}, \citet{sargan1978existence}, \citet{fuller1977some} and \citet{hillier1981exact}. A complete representation of the above results was given by \citet{kinal1980existence}. Kinal's result for 2SLS states that, if the dependent variable, explanatory variables and instruments are jointly normal, then $\mathbbm{E}||\widehat{\phi}^j_{2SLS}||^m<\infty$ for $m < \ell_j - k_j +1$, $j=d,s$, where $\ell_j$ is the number of instruments and $k_j$ is the number of endogenous regressors. 
	
	Thus, the FGIV estimators for supply and demand exhibit no bounded absolute moments since $\ell_j=k_j=1$, $j=d,s$. Whereas, the efficient GMM estimators exhibits $\ell_d-k_d=(1+r)-1=r$ bounded absolute moments in finite samples for the case of demand and $\ell_s-k_s=2-1=1$ bounded absolute moment in finite samples for the case of supply. Hence, the efficient GMM elasticity estimators (overidentified FGIV estimators) exhibit superior finite sample properties relative to their (just identified) FGIV counterparts. Of course, in general, just identified instrumental variables estimators (with strong instruments) exhibit nice properties \textit{asymptotically}.
\end{assumpR}
%In our framework, in unreported simulations, we find that panels with $\text{min}\{N,T\} < 50$ and $\text{max}\{N,T\}$ $ < 100$ are typically problematic for the FGIV estimators due to the Kinal problem laid out above, whereas the efficient GMM estimators perform quite well.  This negative result is a consequence of the fact that the procedure entails a lot of sampling error, which vanishes only for both $N$ and $T$ large, while the rate of convergence is only at the rate of $\sqrt{T}$. 
\section{Weak Instruments}\label{weakGIV}
The classical weak instruments framework introduced by \citet{staiger1997instrumental} has its analog in this framework. Interestingly, here the "weak" aspect is partially linked to the Herfindahl index without making the usual \textit{local-to-zero} assumption as in \citet{staiger1997instrumental}. Moreover, the traditional notion of local-to-zero with $\frac{1}{\sqrt{T}}$ scaling which matches the rate of convergence of the estimator need not necessarily apply here for weak instruments to arise. More specifically, the locality to zero can be expressed as decaying functions of $N$, except in the case of $\mu \in (0,1)$, which we require for inference under our maintained strong instruments assumption; whereas the rate of convergence is at the $\sqrt{T}$ rate. To make things more clear, it is useful to see the reduced form, equilibrium price equation again. Recall from \eqref{eqprice}, we have that $p_t = \dfrac{1}{\phi^d - \phi^s} \left(u_{St} + \bs{\lambda}'_S \bs{\eta}_t -\varepsilon_t \right)$. Thus, it is clear that for finite $N$, $\mathbbm{Cov}(p_t, z_{t}) > 0$, which automatically renders the GIV as relevant. However, for large $N$, writing $z_{t}=\bs{S}'\bs{Q} \bs{\tilde{u}}_{\cdot t}$ we observe that
\begin{align}
\mathbbm{V}(\bs{S}'\bs{Q} \bs{\tilde{u}}_{\cdot t})& =\mathbbm{E}(\bs{S}'\bs{Q} \mathbbm{E}(\bs{\tilde{u}}_{\cdot t}\bs{\tilde{u}}_{\cdot t}'|\bs{\Gamma})\bs{Q}\bs{S})= \mathbbm{E}(\bs{S}'\bs{Q}\,\bs{\Sigma}_{\tilde{u}} \,\bs{Q}\bs{S}), \label{varz}
\end{align} 
where $\bs{\Sigma}_{\tilde{u}}:=\mathbbm{E}(\bs{\tilde{u}}_{\cdot t}\bs{\tilde{u}}_{\cdot t}')$.  The term inside the expectation can be simplified to
\begin{align*}
\bs{S}'\bs{Q}\,\bs{\Sigma}_{\tilde{u}} \,\bs{Q}\bs{S} &=\bs{S}'\bs{Q}\,\bs{\Sigma}_{u} \,\bs{Q}\bs{S} + \mathcal{O}_p\left(N^{-1}\right)\\
&\leq \bs{S}'\bs{Q}\bs{S} \,\gamma_{max}(\bs{\Sigma}_{u})+ \mathcal{O}_p\left(N^{-1}\right) \leq \bs{S}'\bs{S} \cdot \gamma_{max}(\bs{\Sigma}_{u}) \cdot \gamma_{max}(\bs{Q})+ \mathcal{O}_p\left(N^{-1}\right) \\
&=\mathcal{O}\left(1\right) h_{N, \, \mu}+ \mathcal{O}_p\left(N^{-1}\right), \numberthis \label{weakinstr}
\end{align*}
 where we make use of $\gamma_{max}(\bs{\Sigma}_{u}) =\mathcal{O}\left(1\right)$ and the fact that a symmetric idempotent matrix, such as $\bs{Q}$, has eigenvalues of $0$ or $1$ and so $\gamma_{max}(\bs{Q})=1$.  Taken together, \eqref{varz} and \eqref{weakinstr} imply 
 \begin{align}
\mathbbm{V}(z_t) \leq \mathcal{O}\left(1\right)\mathbbm{E}(h_{N,\mu})+ \mathcal{O}\left(N^{-1}\right).
 \end{align}
 As such, only when we are in a tail regime indexed by $\mu \in (0,1)$ do we avoid the locality to zero. For example, when $\mu > 2$, we have that $\mathbbm{V}(\bs{S}'\bs{Q} \bs{\tilde{u}}_{\cdot t}) \leq \mathcal{O}\left(1\right)\mathbbm{E}(h_{N, \, \mu>2})=\mathcal{O}\left(\frac{1}{N}\right)$.  As a result,  when $\mu >2$, we have that $z_{t}=\bs{S}'\bs{Q}\bs{\tilde{u}}_{\cdot t}=\mathcal{O}_p \left(\frac{1}{\sqrt{N}} \right)$, so our equilibrium price equation simplifies to the following large $N$ representation
\begin{align}
p_t = \dfrac{1}{\phi^d - \phi^s} \left(\bs{\lambda}_S' \bs{\eta}_t -\varepsilon_t \right) + \mathcal{O}_p \left(\dfrac{1}{\sqrt{N}}\right). \numberthis \label{pes}
\end{align}
This would render the GIV as very weak since $\mathbbm{Cov}(p_t, u_{St}) =\mathbbm{Cov}(p_t, z_{t}) = \mathcal{O} \left(\frac{1}{N}\right)$. Note that the $\mathbbm{Cov}(p_t, u_{St})$ and $\mathbbm{Cov}(p_t, z_{t})$ are of the same order precisely because $\gamma_{max}(\bs{Q})=1$. \eqref{pes} is effectively the relationship that was exploited by \citet{mohaddes2016country}, who assumed the so-called "granular" weights of order $\mathcal{O}\left(\frac{1}{N}\right)$ and used this weak correlation for large $N$ to ultimately deduce that prices can be treated as weakly exogenous.\footnote{"Granular" has a different definition in the panel data literature, which is referring to properties of weights and heuristically, rules out the existence of dominant units, see e.g., \citet{mohaddes2016country} and \ref{remark1}. On the contrary, our usage of the term "granular" follows \citet{gabaix2011granular} and is essentially referring to the existence of dominant cross-sectional units, see Section \ref{intro}.\label{granfoot}}

Consider the well documented and empirically relevant case where $\mu$ is just above $1$ (Zipf's law corresponds to $\mu=1$); when $\mu \in (1,2)$ we have $h_{N, \, \mu \in (1,2)}=\mathcal{O}_p\left(1/(N^{2-\frac{2}{\mu}})\right)$.  So,
\begin{align}
p_t = \dfrac{1}{\phi^d - \phi^s} \left(\bs{\lambda}'_S \bs{\eta_t} -\varepsilon_t \right) + \mathcal{O}_p \left(\dfrac{1}{N^{1-\frac{1}{\mu}}}\right).
\end{align}
Therefore, even though $\mathbbm{Cov}(p_t, z_{t}) = \mathcal{O}\left(1/(N^{2-\frac{2}{\mu}})\right)$, it is in fact decaying to zero so slowly for $\mu$ near 1, that this potentially corresponds to a highly relevant instrument in any finite sample. That is, $\mathbbm{Cov}(p_t, z_{t}) = \mathcal{O}\left(1/(N^{2-\frac{2}{\mu}})\right)$, is potentially consistent with $z_t$ accounting for large fractions of aggregate variation, see \citet{gabaix2011granular}. 

However, the case we theoretically entertain, for consistency and valid asymptotic inference, requires $\mu \in (0,1)$, which in conjunction with the additional regularity assumptions, renders $\mathbbm{Cov}(p_t, z_{t}) =\Theta(1)$ even as $N \rightarrow \infty$. 

\textbf{\large Rothemberg Representations.} Moreover, to further assess the likelihood of weak instruments, we can analyze the efficient GMM estimator of the demand elasticity which uses both the GIV and the factors as instruments and for comparison we can analyze the just identified demand elasticity estimator which uses only the GIV as an instrument. We analyze the overidentified case with conditional homoskedasticity (assuming only for remainder of this section). Define the projection matrix $\bs{P}_{Z_d} = \bs{Z}_d(\bs{Z}_d'\bs{Z}_d)^{-1}\bs{Z}_d'$, the 2SLS estimator takes the form
\begin{align}
 \widehat{\phi}^d_{GMM}-\phi^d&=\frac{\bs{p}'\,\bs{P}_{Z_d}\,\bs{\varepsilon}}{\bs{p}'\,\bs{P}_{Z_d}\,\bs{p}}. \label{rothen}
\end{align}
Write the structural and reduced form equations as
\begin{align}
\bs{d}=\bs{p}\,\phi^d +\bs{\varepsilon} \nonumber \\ 
\bs{p}=\bs{z}\bs{\pi}' + \bs{v} \label{RF},
\end{align} 
where $\underset{T\times (1+r)}{\bs{z}}=\begin{pmatrix}\bs{u}_S & \bs{\eta} \end{pmatrix}$, $\underset{(1+r) \times 1}{\bs{\pi}}=\begin{pmatrix}\frac{1}{\phi^d-\phi^s} & \frac{1}{\phi^d-\phi^s}\cdot \bs{\lambda}'_S\end{pmatrix}'$ and $\underset{T \times 1}{\bs{v}}=\frac{1}{\phi^d-\phi^s}\cdot \bs{\varepsilon}$. 
\begin{assumpR}\label{Rz}
	The difference between $\bs{z}$ in \eqref{RF} and our actual instrument, $\bs{Z}_d$, boils down to the difference between their first columns, $\bs{Z}_d[\cdot,1]=\bs{z}_{GIV}=\bs{\tilde{u}}_{\cdot \cdot}\bs{Q}\bs{S}$ and $\bs{z}[\cdot,1]=\bs{u}_S=\bs{u}_{\cdot \cdot}\bs{S}$. In the case of the demand equation, $\bs{u}_S$ is ideal, whereas $\bs{z}_{GIV}$ is a proxy. The reason the proxy is used is simply due to a simpler theoretical exposition than a direct estimate for the ideal. Indeed $\bs{z}_{GIV}$ is in fact a good proxy. For example, the correlation between $\bs{u}_S$ and $\bs{z}_{GIV}$ is over 90\% regardless of the complexity of our DGP in Monte Carlo simulations even for small configurations of $(N,T)$. Moreover, in the case of the supply equation, $\bs{u}_S$ is no longer valid, whereas $\bs{z}_{GIV}$ is; see \eqref{magic}. Mathematically, $\bs{S}'\bs{u}_{\cdot t}-\bs{S}'\bs{Q}\bs{\tilde{u}}_{\cdot t} = \bs{S}'\bs{P}_{\tilde{\Lambda}}\bs{u}_{\cdot t}=\bs{S}'\bs{P}_{\tilde{\Lambda}}\bs{P}_{\tilde{\Lambda}}\bs{u}_{\cdot t}$ for each $t$, where $\bs{P}_{\tilde{\Lambda}}$ is the symmetric and idempotent projection matrix in the demeaned loading space. Hence, $\bs{S}'\bs{P}_{\tilde{\Lambda}}\bs{P}_{\tilde{\Lambda}}\bs{u}_{\cdot t}$ is zero when the loadings and the share vector are asymptotically uncorrelated and/or the loadings and idiosyncratic errors are asymptotically uncorrelated; which explains why our simulations exhibit near perfect correlation. 
\end{assumpR}
Then, it follows from \citet{rothenberg1984approximating} that \eqref{rothen} has the following illustrative representation
\begin{align}
\mu_{d,GMM}(\widehat{\phi}^d_{GMM}-\phi^d)&=\left(\frac{\sigma^2_{\varepsilon}}{\sigma^2_v}\right)^{\frac{1}{2}} \frac{X+(\omega_1/\mu_{d,GMM})}{1+2Y/\mu_{d,GMM} + (\omega_2/\mu_{d,GMM}^2)},
\end{align}
where $X=\bs{\pi}'\bs{z}' \bs{P}_{Z_d}\bs{\varepsilon} /(\sigma^2_{\varepsilon} \bs{\pi}'\bs{z}'\bs{z}\bs{\pi})^{\frac{1}{2}}$ and $Y=\bs{\pi}'\bs{z}' \bs{P}_{Z_d} \bs{v} /(\sigma^2_{v} \bs{\pi}'\bs{z}'\bs{z}\bs{\pi})^{\frac{1}{2}}$ are bivariate standard normal variates with correlation coefficient $\rho$. The random variable $\omega_1=\bs{v}' \bs{P}_{Z_d}\bs{\varepsilon} /(\sigma^2_{\varepsilon}\sigma^2_{v})^{\frac{1}{2}}$ has mean equal to  $\text{rank}(P_{Z_d})\rho=(r+1)\rho$ and variance equal to $(r+1)(1+\rho^2)$. The  random variable $\omega_2=\bs{v}'\bs{P}_{Z_d} \bs{v} /\sigma^2_{v}$ has mean equal to $\text{rank}(\bs{P}_{Z_d})=(r+1)$ and variance equal to $2(r+1)$. Finally, $\mu_{d,GMM}$ is the square root of the so-called concentration parameter $\mu_{d,GMM}^2=\bs{\pi}'\bs{z}'\bs{z}\bs{\pi}/\sigma^2_{v}$ for the demand equation. $\mu_{d,GMM}$ plays the role of $\sqrt{T}$, that is, when $\mu_{d,GMM}$ is large, $\mu_{d,GMM}(\widehat{\phi}^d_{GMM}-\phi^d)$ is well approximated by a $\mathcal{N}(0,1)$ variate. Large values of $\mu_{d,GMM}$ are consistent with large values of $T$, i.e., our typical large sample approximations. However, large values of $\mu_{d,GMM}$ are also consistent with small values of $\sigma^2_v$, regardless of the value of $T$, i.e., small-$\sigma$ asymptotics, as introduced originally by \citet{kadane1971comparison}. More insights can be gained by simplifying the concentration parameter for the demand elasticity
\begin{align*}
\mu_{d,GMM}^2&=\bs{\pi}'\bs{z}'\bs{z}\bs{\pi}/\sigma^2_{v}=\frac{1}{\sigma^2_{v}}\cdot\begin{pmatrix}\pi_1 & \bs{\pi_2}'\end{pmatrix}\begin{pmatrix}
\bs{u}'_S\bs{u}_S & \bs{u}'_S\bs{\eta} \\
\bs{\eta}'\bs{u}_S & \bs{\eta}'\bs{\eta} 
\end{pmatrix} \begin{pmatrix}
\pi_1 \\
\bs{\pi_2}
\end{pmatrix} \approx \frac{\bs{u}'_S\,\bs{u}_S+\bs{\lambda}'_S\bs{\lambda}_S}{\sigma^2_{\varepsilon}}, \numberthis \label{concgmm}
\end{align*}
where the approximation is due to ignoring the terms involving $\bs{\eta}'\bs{u}_S$, which are zero only in expectation. \eqref{concgmm} is very intuitive, if the proportion of the volatility in the GIV \textit{and} size-weighted common components dominate the volatility of the demand shocks, so that the ratio in \eqref{concgmm} is large, then the concentration parameter $\mu_{d,GMM}$ will be large and one should expect good approximations to the finite sampling distributions. 

On the other hand, when only the GIV is used as an instrument, if we redefine $\underset{T\times 1}{\bs{z}}=\bs{u}_S$, $\underset{1 \times 1}{\bs{\pi}}=\frac{1}{\phi^d-\phi^s}$ and $\underset{T \times 1}{\bs{v}}=\frac{1}{\phi^d-\phi^s}\cdot (\bs{\varepsilon}+\bs{\eta}\bs{\lambda}_S)$ from \eqref{RF} and simply follow the logic above through \eqref{concgmm}, we arrive at the following concentration parameter for the FGIV estimator
\begin{align}
\mu^2_{d,FGIV} \approx \frac{\bs{u}'_S\bs{u}_S}{\bs{\lambda}'_S\bs{\lambda}_S+\sigma^2_{\varepsilon}}. \label{concgiv}
\end{align}
Thus, by inspection of \eqref{concgmm} against \eqref{concgiv} we can see that in the case of the just identified FGIV estimator, we would need the volatility of just the GIV to drive up the ratio of the concentration parameter, $\mu^2_{d,GIV}$, and the size-weighted common component would be working against us (in the denominator), in this case, instead of working for us as in $\mu^2_{d,GMM}$. 

Although the literature on granularity has demonstrated that idiosyncratic shocks alone can be quite volatile, in this context, we advocate starting with the efficient GMM estimators, since the $J$-test is well sized as illustrated with simulation evidence, because the efficient GMM estimators can exhibit substantially improved finite sample properties relative to the just identified estimators and are less likely to suffer from weak instrument issues as well. 

\section{Monte Carlo}\label{MCstudy} 
We simulate the following panel simultaneous equations system with latent factor structure that was analyzed in the theoretical sections:
\begin{align*}
d_t &= \phi^d p_t + \varepsilon_t,  \,\, y_{it} = \phi^s p_t + \lambda_{1i}\eta_{1t} + \lambda_{2i}\eta_{2t} + u_{it}, \,\,y_{\bs{S}t}=d_t, \\
 p_t &= \frac{1}{\phi^d-\phi^s} \left( u_{St}+\bs{\lambda}'_S\bs{\eta_t}-\varepsilon_t\right), \,\,\mathscr{S}_i = \left(\frac{i}{N}\right)^{-\frac{1}{\mu}}, \,\, S_i = \frac{\mathscr{S}_i}{\sum_{j=1}^N \mathscr{S}_j}.
\end{align*}
We consider two sets of simulated experiments. In Design 1, we let $u_{it}$ be $i.i.d.$ to establish a set of baseline results. In Design 2, we allow for sparse cross-sectional dependence in $u_{it}$. In addition, in unreported simulations, we simulate $z_{t,GIV}$ to be a weak instrument to illustrate that the efficient GMM estimators are robust to this as they optimally shift their weights away from this point of weakness, whereas the just identified estimators of GK and our FGIV will substantially deteriorate in their performances.

\textbf{\large Design 1 - $u_{it} \,\, \bs{i.i.d.}$ case.} We set 
$\phi^s=0.1$ and $\phi^d=-0.3$. We draw the supply factors and loadings as, $\underset{T\times r}{\bs{\eta}} \overset{i.i.d.}{\sim} \mathcal{N}(0,\bs{I}_r)$ and $\underset{N\times r}{\bs{\Lambda}} \overset{i.i.d.}{\sim} \mathcal{N}(0,\sigma^2_{\Lambda}\bs{I}_N)$, respectively, with $r=2$.\footnote{The results do not change significantly if we draw the loadings from a Uniform distribution with non-zero mean.} We draw the idiosyncratic supply shocks as $\underset{T\times N}{\bs{u}} \overset{i.i.d.}{\sim} \mathcal{N}(0,\sigma^2_u\bs{I}_N\otimes \bs{I}_{T})$ and aggregate demand shocks as $\underset{T\times 1}{\bs{\varepsilon}} \overset{i.i.d.}{\sim} \mathcal{N}(0,\sigma^2_{\varepsilon}\bs{I}_T)$. 

\textbf{\large Design 2 - $u_{it}$ non-$\bs{i.i.d.}$ case.} Everything is identical to Design 1, except that we no longer set $\bs{\Sigma}_u=\sigma^2_u \bs{I}_N$ for each $t=1,\dots, T$. We generate a non-diagonal banded covariance matrix; as such, it satisfies the sparsity requirement from \ref{A3a}.\footnote{In unreported simulations, we find that the results do not change significantly if we generate a dense, non-diagonal $\bs{\Sigma}_u$, such as one arising from a cross-sectional AR(p) process. This is because although the cross-sectional AR(p) generates a dense matrix, it is not too dense since the off-diagonals decay exponentially fast to 0 as $|i-j| \rightarrow \infty$.}
We consider the following banded idiosyncratic covariance matrix with cross-sectional dependence and heteroskedasticity 
\begin{align}
\sigma_{u,ij}=\begin{cases}
\tau^{|i-j|}\sqrt{\sigma_{u,i}\sigma_{u,j}} \hspace{16.5mm}  |i-j| \leq k; k \geq 0, \\
0 \hspace{38.75mm}  |i-j|>k, \\
\end{cases}
\end{align}
with bandwidth $k=3$ and $\sigma^2_{u,i}$ are drawn from $\mathcal{U}[0.5,1]$. 
%As a departure from Gibrat's law, we allow for cross-sectional heteroskedasticity by allowing smaller cross-sectional units to have larger variances than larger cross-sectional units.\footnote{In the context of firms in Compustat, this departure has been documented to be small, see \citet{amaral1997scaling} and \citet{gabaix2011granular}. However, it can also be an appropiate approximation in many contexts, e.g. in the country-level analsyses of global crude oil markets.} That is, suppose the cross-sectional units are ordered from smallest to largest, then we generate
%\begin{align}
%\sigma_{u,ii}=\begin{cases}
%\sigma_u^2  \hspace{20mm}  i=1,\dots,[N/3],  \\
%.85 \sigma_u^2 \hspace{15mm}  i=[N/3]+1,\dots,[2N/3], \\
%.70 \sigma_u^2 \hspace{15mm} i=[2N/3]+1,\dots,N; \\
%\end{cases}
%\end{align}

\textbf{\large Target parameterizations.} The variance of the price process takes the following form $\mathbbm{V}(p_t)=c\cdot  \left( \mathbbm{V}(u_{St}) +\mathbbm{V}(\bs{\lambda}'_S\bs{\eta}_t) +\mathbbm{V}(\varepsilon_t) \right)$,
where $c=\dfrac{1}{(\phi^d-\phi^s)^2}$. This conveniently allows us to parameterize the relative volatilities of the various components of equilibrium prices. We parameterize the individual variances, $\sigma^2_u$ (for Design 1), $\sigma^2_{\Lambda}$ and $\sigma^2_{\varepsilon}$ such that 
$\psi_{u} :=  \dfrac{\mathbbm{V}\left(\sqrt{c}\cdot u_{St}\right)}{\mathbbm{V}(p_t)} \in (0.15, 0.35),$
$\psi_{u+\eta} :=  \dfrac{\mathbbm{V}\left(\sqrt{c}\cdot (u_{St}+\bs{\lambda}'_S\bs{\eta}_t)\right)}{\mathbbm{V}(p_t)} \in (0.45, 0.65),$ and 
$\psi_{u+\varepsilon} :=   \dfrac{\mathbbm{V}\left(\sqrt{c}\cdot (u_{St}+\varepsilon_t)\right)}{\mathbbm{V}(p_t)} \in (0.45, 0.65).$\footnote{The interval for $\psi_u$ is consistent with the literature on granularity, which has documented the proportion of aggregate fluctuations traced back to idiosyncratic shocks falling in this specified range.} In Design 1, we achieve an average across simulations of $\bar{\psi}_u \approx 0.23$, $\bar{\psi}_{u+\eta} \approx 0.58$ and $\bar{\psi}_{u+\varepsilon}  \approx 0.65$ which implies that $\bar{\psi}_{\eta}=0.35$ and $\bar{\psi}_{\varepsilon}=0.42$. That is, the idiosyncratic shocks are not the dominating force in terms of observed price volatility; however, their granular role is still substantial enough to draw inferences from when used as instruments. In Design 2, we achieve an average across simulations of $\bar{\psi}_u \approx 0.27$, $\bar{\psi}_{u+\eta} \approx 0.64$ and $\bar{\psi}_{u+\varepsilon}  \approx 0.63$. 

%\textbf{\large Design 3 - weak GIV case.} In Design 3, we take $\mu \in [0,1)$ but target a significantly smaller $\psi_u \approx 0.08$. Everything else is identical to Design 1, but the modifications simulate an environment where $z_{t,GIV}=\bs{S}'\bs{Q}\bs{\widetilde{y}_{\cdot t}}$ is a weak instrument. It should be noted that there are numerous ways to achieve a weak GIV. One mechanism is for the tail index of the size distribution to be significantly out of the $[0,1)$ interval we require for strong instrument asymptotics. Another mechanism, is to make the proportion of price volatility attributed to idiosyncratic shocks small, even for $\mu \in [0,1)$. Both mechanisms result in a small concentration parameter for the just identified estimators. 
Let $\widehat{\phi}^j(m)$, $j=d,s$, denote the estimate in the $m^{th}$ monte carlo repetition, $m=1,...,M$. We report the monte carlo bias: $\text{Bias}(\widehat{\phi}^j)= \left(\dfrac{1}{M} \sum_{m=1}^M \widehat{\phi}^j(m)-\phi^j \right)$ for $j=d,s;$ and square root of the monte carlo MSE: 
$\text{RMSE}(\widehat{\phi}^j)= \sqrt{\dfrac{1}{M}\sum_{m=1}^M(\widehat{\phi}^j-\phi^j)^2}$ for $j=s,d.$ Additionally we report the size of the $t$-test for all estimators and size of the $J$-test for the efficient GMM estimators. The results are reported in Table \ref{design1all} (Design 1) and Table \ref{design2all} (Design 2). In Table \ref{design1all}, we multiply the bias by 100 because all the estimators perform quite well in this ideal setting. For nearly each configuration of $(N,T)$ the GMM estimators perform the best, in terms of bias and RMSE, as the theory suggests. Importantly, the $t$-test and $J$-test is well sized even when $rmax=3 > r=2$ factors are used.  In Table \ref{design2all}, we report the bias as is, and we find that for a given configuration of $(N,T)$ the bias is two orders of magnitude larger in Design 2 relative to Design 1. Nevertheless, as the theory would suggest, the efficient GMM estimators perform the best in terms of bias and RMSE. There are some size distortions for the supply side estimators but the distortions are decreasing in $N$ for a given $T$. 
\section{Application to Global Crude Oil Markets}\label{oil}
\begin{figure}[!htbp]
		\minipage{\textwidth}
		\includegraphics[width=\linewidth]{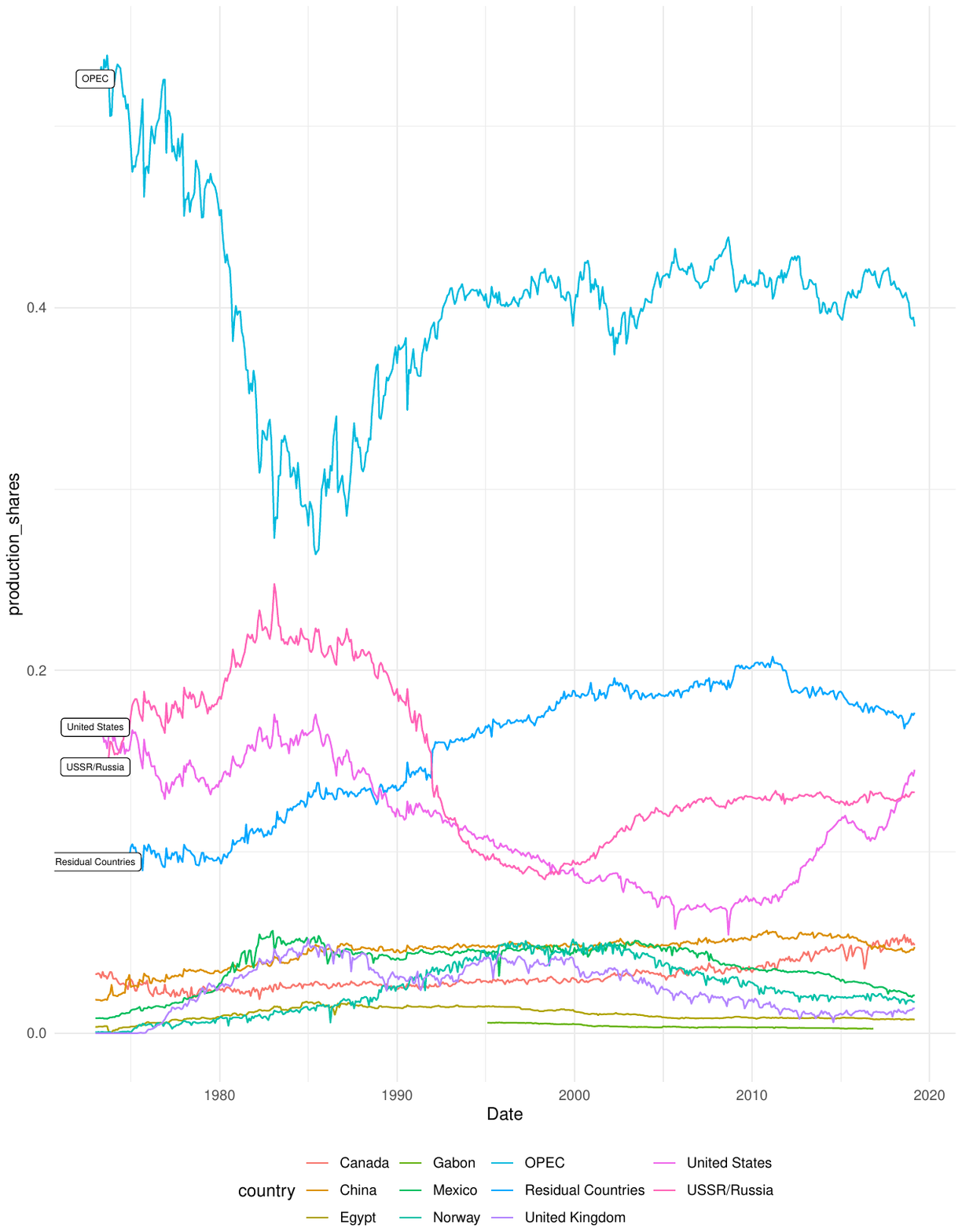}
		\caption{Temporal variation of production shares}\label{temporalshares}
		\endminipage\hfill
	\end{figure}

\begin{figure}[!htbp]
		\minipage{\textwidth}
		\includegraphics[width=\linewidth]{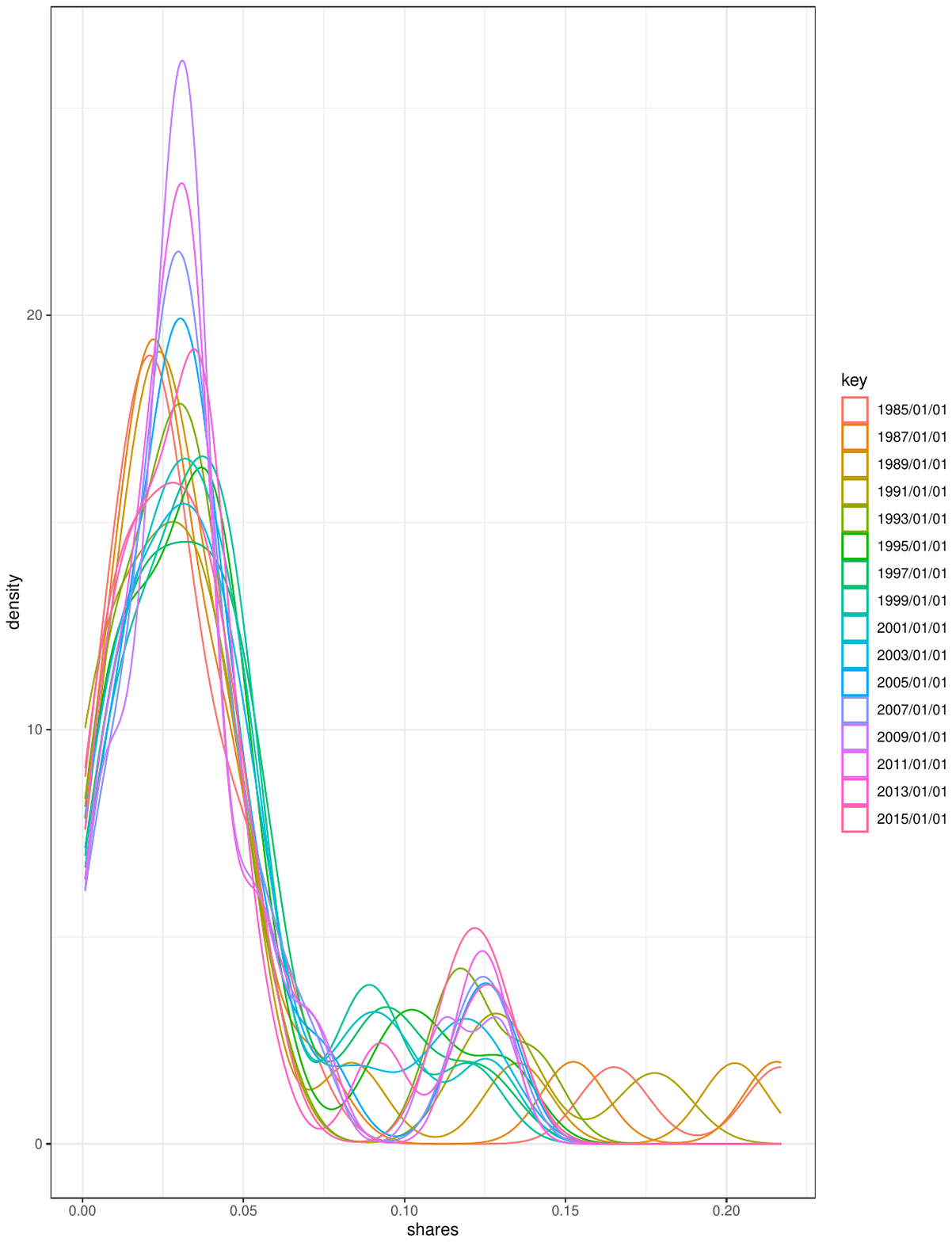}
		\caption{Distribution of production shares}\label{sharesdist}
		\endminipage\hfill
	\end{figure}

\begin{figure}[!htbp]
		\minipage{\textwidth}%
		\includegraphics[width=\linewidth]{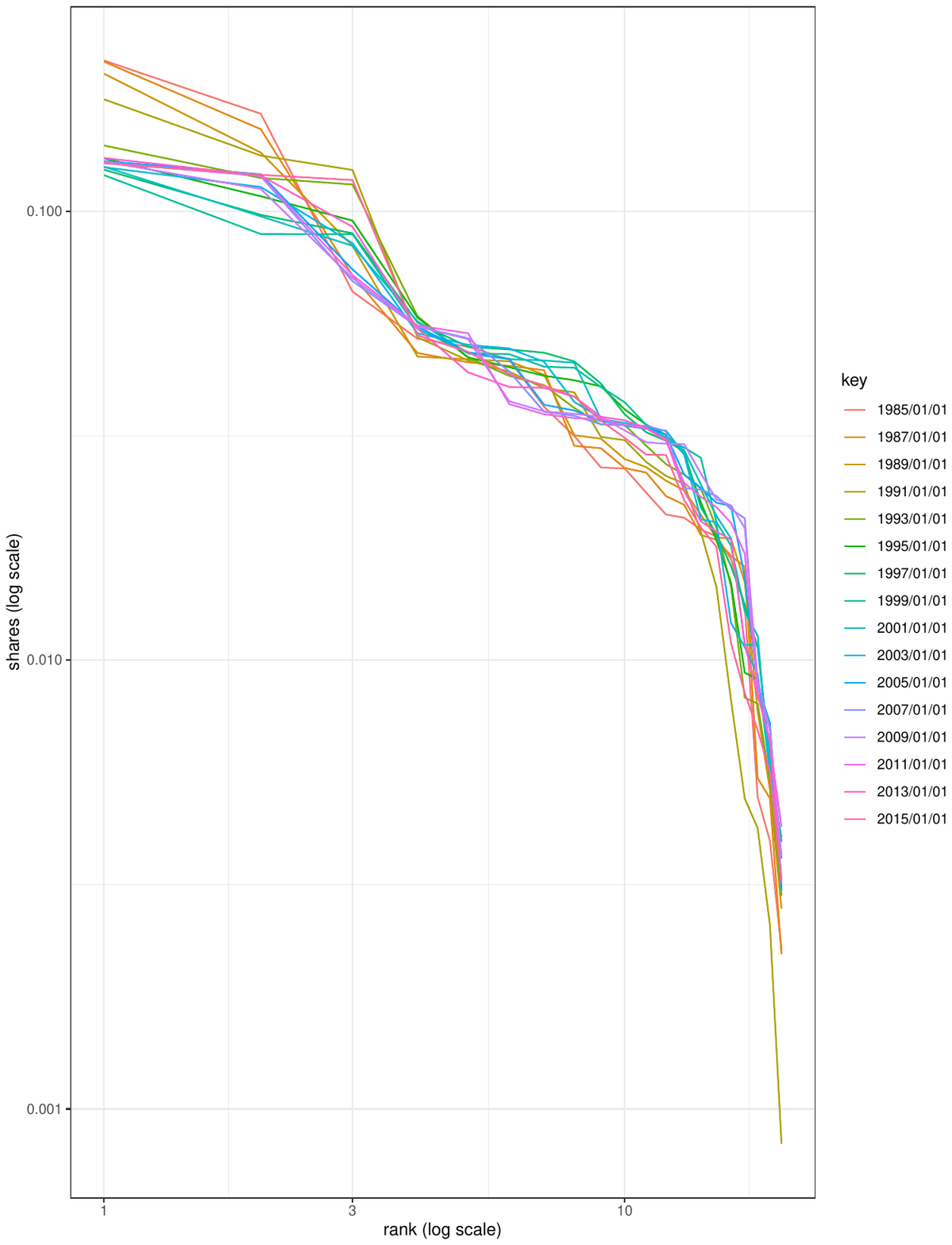}
		\caption{Size-Rank plot (log-log scale)}\label{sizerank}
		\endminipage
	\end{figure}

The data construction follows the recent literature: \citet{kilian2009not}, \citet{caldara2019oil}, \citet{baumeister2019structural} (hereafter BH) and GK.  The following is a breakdown of the raw variables collected for Jan. 1985 - Dec. 2015 $(T=372$ months$)$:  monthly oil production for $N=22$ countries from the U.S. Energy Information Administration (hereafter EIA); world oil production from the U.S. EIA; monthly oil prices based on the refiner acquisition cost of imported crude oil from the U.S. EIA; U.S. CPI from the St. Louis FRED database; monthly change in inventories from BH; monthly industrial production index from BH. The CPI is used to deflate nominal oil prices to arrive at the real price of oil, which is highly non-stationary. Following the aforementioned literature, we take the logarithm of the real price of oil series and then take first differences. We apply the same transformation to the monthly oil production for each country. These transformations render the production and price series stationary as confirmed by a host of Dickey-Fuller tests. For ensuring the tail index of the size-distribution, $\mu$, is in the region the theory requires, we provide visual evidence along with 6 estimates of $\mu$ that all fall beneath 1, see Table \ref{tailindexest}. Also, see Figure \ref{temporalshares}, Figure \ref{sharesdist} \& Figure \ref{sizerank}.

Let $y_{it}$ denote the log difference of the oil supply for country $i$ at time $t$ and $p_t$ denote the log difference of the real price of oil. Following GK, we estimate an OPEC factor using information on the cross-section of countries (i.e., known loadings). To that end, let $o_{it}$ denote a dummy variable equal to 1 if country $i$ is an OPEC member at time $t$ and note that $o_{it}=o_i$ for most $i$, with the exception of Gabon and Ecuador in our sample. Finally, $\bs{c}_{t-1}$ denotes a $4 \times 1$ vector containing: lagged $p_t$, lagged world supply growth, lagged change in inventories, and lagged growth in industrial production. The system is given as follows
\begin{align}
y_{it} &= \phi^s p_t + \bs{\gamma_s}'\bs{c}_{t-1} + o_{it}\eta_{OPEC,t} + \bs{\lambda}_i'\bs{\eta}_t  + u_{it},\\
d_t &= \phi^d p_t + \bs{\gamma}_d'\bs{c}_{t-1} + \lambda_d \eta_{OPEC,t} + \varepsilon_t,\\
\sum_{i=1}^N S_{it}\, y_{it}&=d_t;
\end{align}
where we lose the observation $t=1$ due to differencing. The cross-sectionally demeaned supply equation is given by the approximate factor model,
\begin{align}
\tilde{y}_{it} = y_{it} - \frac{1}{N}\sum_i y_{it}&= \tilde{o}_{it}\eta_{OPEC,t}+\bs{\tilde{\lambda}}_i'\bs{\eta}_t + \tilde{u}_{it} = \tilde{o}_{it}\eta_{OPEC,t}+ \tilde{e}_{it}, \label{oilsup2} 
\end{align}
where $\tilde{e}_{it} :=\bs{\tilde{\lambda}}_i'\bs{\eta}_t + \tilde{u}_{it}$. Note that \eqref{oilsup2} implies we can obtain the OPEC factor, $\eta_{OPEC,t}$, via cross-sectional regression, for each $t>1$, that is $\widehat{\eta}_{OPEC,t}=( \bs{\tilde{o}}_{\cdot t}' \bs{\tilde{o}}_{\cdot t})^{-1} \bs{\tilde{o}}_{\cdot t}'\bs{\tilde{y}}_{\cdot t}$. Hence, in our preliminary stage, we extract $\widehat{\eta}_{OPEC,t}$ and then run PCA on $\widehat{\tilde{e}}_{it}=\tilde{y}_{it}-\widehat{\eta}_{OPEC,t}$ to extract the latent demeaned loadings and latent factors. Define $\bs{y}^*_{\cdot t} :=  \bs{\tilde{y}}_{\cdot t}-\bs{\tilde{x}}_{\cdot t}\widehat{\eta}_{OPEC,t}$, then we purge the latent factors via $\bs{Q}$ as in the main text: $\bs{{Q}}\,\bs{y}^*_{\cdot t}= \bs{{Q}}\,\bs{\tilde{u}}_{\cdot t}$. However, when forming the GIV, there is a minor difference that we have time-varying size-weights, so we no longer construct $\bs{z}_{GIV}$ with a time-invariant share vector $S_i$, but rather we weight each idiosyncratic component at time $t$ with its corresponding share from time $t-1$ to avoid endogeneity issues arising from contemporaneous weighting
\begin{align}
\underset{(T-1) \times 1}{ \bs{z}_{GIV}}=\begin{pmatrix}
 \bs{S}'_{\cdot 1} \bs{Q} \, \bs{\tilde{y}}^*_{\cdot 2} \\
 \bs{S}'_{\cdot 2} \bs{Q} \, \bs{\tilde{y}}^*_{\cdot 3} \\
\vdots \\
 \bs{S}'_{\cdot T-1} \bs{Q} \, \bs{\tilde{y}}^*_{\cdot T} 
\end{pmatrix}.
\end{align}
Besides these modifications from the stylized model in the theory, we estimate the elasticities using the estimators outlined in the main text. The number of factors, $r$, is estimated via the AH procedures (as outlined in the Supplementary Appendix \ref{rhat}). The $ER$ method of AH estimated $\widehat{r}_{ER}=1$, while the $GR$ method estimated $\widehat{r}_{GR}=3$; with $k_{max}=10$. To be safe, we take $\widehat{r}=\widehat{r}_{GR}+1$. 	

\textbf{\large Supply results.} The results for the supply elasticity are presented in Table \ref{oilelassupply}.
%In this particular application, as oil is a storable good, small shocks can be offset by the use of inventories. As such, we construct the instrument with a caveat: shock filtration, such that emphasis is placed on larger size-weighted idiosyncratic shocks by using a hard thresholding procedure given by
%\begin{align}
%\widehat{z}_t(b)=\sum_{i=1}^N \sum_{j=1}^N S_{it-1} \widehat{Q}_{ij} \tilde{y}^*_{jt} \cdot  \mathbbm{1}\left(\bigg|\sum_{j=1}^N S_{it-1} \widehat{Q}_{ij} \tilde{y}_{jt}\bigg|\geq b\right), 
%\end{align}
%We consider $\widehat{z}_t(b)$ with $b \in \{0.4\cdot2^{-m}; m=0,1,\dots,10\}$. We choose $b^*$ which maximizes the first stage $F$-statistic, which corresponds to $b^*=0.0125$. We report the results below for $b=0$, which corresponds to the baseline FGIV, and the results for $b^*$. 
In Table \ref{oilelassupply}, the 2nd column displays GK's results. The instrument GK use is given by \eqref{ZGK} and their dependent variable is simply the cross-sectional average of the log difference of oil supply (i.e., $\bs{E}=\bs{\iota}/N$). Our results are in columns 3 and 4. In contrast to \eqref{ZGK}, the instrument we use in column 3 purges the common factors through the loading space. The instrument we use in column 4 also adds an estimate of the unobserved aggregate demand shocks, $\widehat{\varepsilon}_t$, to our FGIV.  Moreover, the dependent variable we use is weighted using the estimated precision vector $\bs{\widehat{E}}$, which allows for cross-sectional correlations and heteroskedasticity in $u_{it}$. These differences lead to significantly different results. Columns 2 and 3, which attempt to use \textit{only} GIVs as instruments, both lead to weak instruments as indicated by the first-stage $F$-statistics less than the rule of thumb, 10. Nevertheless, the FGIV supply elasticity estimate (0.016) from column 3 (estimated via Algorithm \ref{alg:alg1}) is roughly one third that of GK's (0.044). Whereas, our efficient GMM supply elasticity estimate (0.005) from column 4 (estimated via Algorithm \ref{alg:alg4}) is highly significant at the 1\% level. Additionally, our results reveal that using estimates of unobserved aggregate demand shocks as supply instruments indeed renders a strong instrument as indicated by the first-stage $F$-stat of 14.33 in column 4. Moreover, the $p$-value for the $J$-statistic (0.11) fails to reject the null hypothesis of a valid model. An $F$-stat greater than 10, coupled with a small $J$-statistic provides statistical evidence in favor of our efficient GMM point estimate for the supply elasticity. 

\textbf{\large Demand results.} Turning now to the demand elasticity in Table \ref{oilelasdemand}, the dependent variable GK use in this case is the same as the one we use. However, the instruments are different. Column 2 displays GK's demand elasticity (-0.463), again using the instrument as in \eqref{ZGK}. Column 3 presents our result when using only the FGIV as an instrument (-.0009), which is roughly 400 times smaller than GK's estimate. Columns 4 through 7 sequentially add principal components to the instrument vector for our efficient GMM estimator from column 3 until 4 principal components are used. Here we find that none of the models yield first-stage $F$-statistics greater than 10. It is reassuring, however, that the $J$-statistic for columns 4 through 7 all fail to reject the null of a valid model. Lastly, column 8 presents the \cite{bai2010instrumental} estimator which only includes the four principal components but not the FGIV as instruments, nearly all statistics remain unchanged except that inclusion of the FGIV increases the $t$-stat by about 25\%. 

Taken together, our empirical results suggest that supply shocks, whether they be aggregate or idiosyncratic supply shocks, albeit valid, do not serve as strong instruments for estimation of the demand elasticity. Whereas, aggregate demand shocks indeed seem to be a strong source of exogenous variation to tease out the supply elasticity. 
%To put the efficient GMM point estimate for the supply elasticity into perspective, we find that it falls within 1 standard deviation of the estimated oil elasticities from the vast oil literature. Moreover, \citet{kilian2020understanding} states that \textit{"...evidence in Kilian (2008a,b) and Montiel Olea, Stock and Watson (2020) suggests that all existing oil supply shock instruments are weak in the econometric sense, which explains why this approach has been largely abandoned."} However, the GMM estimators provide instruments which appear to be quite strong.  
%\subsection{Quarterly Data} Although data on monthly exchange rates are available for the countries in our panel, it was not possible to find price indices for our panel at the monthly frequency. As such, we asses our robustness check on the quarterly data of \citet{mohaddes2016country}. 
\section{Concluding Remarks}
In this paper, we have further developed the GIV methodology introduced by \citet{gabaix2020granular}, which takes advantage of panel data to construct instruments for estimation of structural time series regression models that involve endogenous regressors. This paper focuses on the underlying econometric issues involved in developing FGIV in a large $N$ and large $T$ framework where the loadings are treated as unknown parameters to be estimated before constructing the FGIV instrument. We further demonstrate that the sampling error arising from estimating the instrument, factors and a high dimensional precision matrix does not affect the limiting distribution for the structural parameters of interest. We also overidentify the structural parameters, which leads to new and improved results in the crude oil markets application and demonstrate that the $J$-test is well sized with simulation evidence. Our Monte Carlo study illustrates that our estimators and algorithms exhibit desirable performance with the finite sample distributions being well approximated by the asymptotic distributions. 

More fruitful areas of research would be empirical applications of the theoretical results derived in this paper. Interesting theoretical extensions would be to allow for random slope coefficients with correlated heterogeneity, the presence of weak factors and unbalanced panels with data not missing at random.  We are currently pursuing the dynamic panel data extension, as well as adapting the GIV methodology for unit-specific endogenous variables. \newpage  
	\begin{table}[!htbp] 
	\begin{center}
		\caption{Bias$\times 100$, RMSE, Size of $t$-test and Size of $J$-test for Design 1.}
		\label{design1all}
		\resizebox{\textwidth}{!}{
			\begin{threeparttable} 
				\begin{tabular}{|rrrr|rrrr|rrrr|}
					\hline
					&\multicolumn{11}{l}{\textit{Finite sample properties for Design 1.}} \vline\\ 
					&\multicolumn{11}{l}{} \vline \\
					& $N$ & $T$ &$\mu$&$\widehat{\phi}^s_{FGIV}$ & $\widehat{\phi}^s_{GK}$ &$\widehat{\phi}^s_{GMM}$  & $\widehat{\phi}^{\,s,rmax}_{GMM}$ & $\widehat{\phi}^d_{FGIV}$& $\widehat{\phi}^d_{GK}$ & $\widehat{\phi}^d_{GMM}$& $\widehat{\phi}^{d,rmax}_{GMM}$ \\ 
					\hline
					1 & 30 & 400 & 0.92 &0.0612 & 0.0910  & 0.0164  & 0.0273 &-0.2760 &-0.2723   & 0.1100 &0.2100 \\ 
					&  &  & & (0.0344) & (0.0330) & (0.0204) & (0.0204) & (0.0152) & (0.0173) &  (0.0079) &  (0.0080)  \\ 
					&  &  &  & $[0.0635]$& $ [0.1205]$ &  $[0.0830]$ & $[0.0830]$  & $[0.0570]$&  $[0.0510]$ & $[0.0685]$&  $[ 0.0735]$ \\ 
					&  &  &  & $\{\text{N.A.} \}$&  $\{\text{N.A.} \}$&  $ \{0.0540 \}$ & $ \{ 0.0625\}$  &$\{\text{N.A.} \}$ &$\{\text{N.A.} \}$  & $ \{0.0490 \}$&  $ \{ 0.0465\}$ \\ \hline
					2 & 50 & 400 &0.85 & 0.0071 &0.0774  &0.0174  &0.0224  &-0.1803 & -0.1649 & 0.1462 & 0.2306 \\ 
					&  &  & & (0.0313) & (0.0292) &(0.0200) &(0.0200)  &(0.0106) &(0.0186)  & (0.0058)  & (0.0059)\\ 
					&  &  &  & $[0.0650]$& $ [0.2685]$ &  $[0.0710]$ & $[0.0715]$  & $[0.0555]$&  $[0.0555]$ & $[0.0700]$&  $[0.0740]$ \\ 
					&  &  &  &$\{\text{N.A.} \}$ & $\{\text{N.A.} \}$&  $ \{0.0520 \}$ & $ \{ 0.0510 \}$  & $\{\text{N.A.} \}$&  $\{\text{N.A.} \}$& $ \{0.0480 \}$&  $ \{ 0.0550\}$ \\ \hline
					3 & 100 & 400 & 0.80 & 0.0059 & 0.0159  & 0.0070  & 0.0102  & -0.1886&-0.1558  &0.1449  & 0.2330  \\ 
					&  &  &  & (0.0287)&  (0.0271) &  (0.0196) & (0.0197)  & (0.0068) &  (0.0112) &  (0.0039) &  (0.0040) \\ 
					&  &  &  & $[0.0610]$& $ [0.2505]$ &  $[0.0585]$ & $[0.0615]$  & $[0.0515]$&  $[0.0540]$ & $[0.0705]$&  $[0.0790]$ \\ 
					&  &  &  & $\{\text{N.A.} \}$&  $\{\text{N.A.} \}$&  $ \{ 0.0495\}$ & $ \{ 0.0485\}$  & $\{\text{N.A.} \}$& $\{\text{N.A.} \}$& $ \{0.0440 \}$&  $ \{0.0425 \}$ \\ \hline
					4 & 200 & 400 & 0.77& 0.0192 & 0.0246  & -0.006  & 0.0010  & -0.1896&-0.1713  &0.0524  & 0.1409 \\ 
					&  &  &  & (0.0276)&  (0.0263) &  (0.0188) & (0.0188)  & (0.0046) &  (0.0080) &  (0.0027) &  (0.0027) \\ 
					&  &  &  & $[0.0600]$& $ [0.2660]$ &  $[0.0545]$ & $[0.0535]$  & $[0.0410]$&  $[0.0450]$ & $[0.0625]$&  $[0.0635]$ \\ 
					&  &  &  & $\{\text{N.A.} \}$&  $\{\text{N.A.} \}$&  $ \{ 0.0590\}$ & $ \{ 0.0620\}$  & $\{\text{N.A.} \}$& $\{\text{N.A.} \}$& $ \{ 0.0495\}$&  $ \{0.0425 \}$ \\ \hline
					5 & 500 & 400 & 0.75& -0.0013 & -0.0078  & -0.0097  & -0.0102  & 0.2501&0.2255 &-0.0893  & -0.1747 \\ 
					&  &  &  & (0.0287)&  (0.0280) &  (0.0188) & (0.0187)  & (0.0028) &  (0.0030) &  (0.0016) &  (0.0016) \\ 
					&  &  &  & $[0.0545]$& $ [0.0840]$ &  $[0.062]$ & $[0.0625]$  & $[0.0540]$&  $[0.0590]$ & $[0.0680]$&  $[0.0730]$ \\ 
					&  &  &  & $\{\text{N.A.} \}$& $\{\text{N.A.} \}$&  $ \{ 0.0560\}$ & $ \{ 0.0535\}$  & $\{\text{N.A.} \}$& $\{\text{N.A.} \}$& $ \{ 0.054\}$&  $ \{ 0.051\}$ \\ \hline
					\hline
				\end{tabular}
				\begin{tablenotes}
					\item Notes: We report Bias$\times 100$, (RMSE), [$t$-test],and $\{$$J$-test$\}$ (if applicable) with a nominal size of 5\%. $\widehat{\phi}^s_{FGIV}$, $\widehat{\phi}^s_{GMM}$, $\widehat{\phi}^{s,rmax}_{GMM}$ estimated with Algorithm \ref{alg:alg1}, \ref{alg:alg4}, \ref{alg:alg4} with $r =2$ and $rmax=3$, respectively and $\widehat{\phi}^s_{GK}$, $\widehat{\phi}^d_{GK}$ both use \eqref{ZGK} as an instrument. $\widehat{\phi}^d_{FGIV}$, $\widehat{\phi}^d_{GMM}$, $\widehat{\phi}^{d,rmax}_{GMM}$ estimated with \eqref{justidentifiedGIV}, \eqref{gmmdemand}, \eqref{gmmdemand} with $r =2$ and $rmax=3$, respectively. $\mu$ set to maintain $h_{N,\mu}=0.12$ across all configurations of $(N,T)$. $\bar{\psi}_{u}=0.23$, $\bar{\psi}_{u+\eta}=0.58$,  $\bar{\psi}_{u+\varepsilon}=0.65$.
				\end{tablenotes}
		\end{threeparttable}}
	\end{center}
\end{table}

	\begin{table}[!htbp] 
	\begin{center}
		\caption{Bias, RMSE, Size of $t$-test and Size of $J$-test for Design 2.}
		\label{design2all}
		\resizebox{\textwidth}{!}{
			\begin{threeparttable} 
				\begin{tabular}{|rrrr|rrrr|rrrr|}
					\hline
					&\multicolumn{11}{l}{\textit{Finite sample properties for Design 2.}} \vline \\ &\multicolumn{11}{l}{} \vline \\
					& $N$ & $T$ &$\mu$&$\widehat{\phi}^s_{FGIV}$ & $\widehat{\phi}^s_{GK}$ &$\widehat{\phi}^s_{GMM}$  & $\widehat{\phi}^{\,s,rmax}_{GMM}$ & $\widehat{\phi}^d_{FGIV}$& $\widehat{\phi}^d_{GK}$ & $\widehat{\phi}^d_{GMM}$& $\widehat{\phi}^{d,rmax}_{GMM}$ \\ 
					\hline
					1 & 30 & 400 & 0.92 &0.0263 & 0.0225  & 0.0080  & 0.0089 &-0.0021 &-0.0016   & 0.0003 &0.0009 \\ 
					&  &  & & (0.0365) & (0.0259) & (0.0150) & (0.0151) & (0.0371) & (0.0368) &  (0.0148) &  (0.0161)  \\ 
					&  &  &  & $[0.2730]$& $ [0.2895]$ &  $[0.12450]$ & $[0.1410]$  & $[0.0500]$&  $[0.0510]$ & $[0.0560]$&  $[ 0.0600]$ \\ 
					&  &  &  & $ \{ \text{N.A.}\}$& $ \{ \text{N.A.}\}$ &  $ \{0.1895 \}$ & $ \{ 0.3035\}$  & $ \{\text{N.A.} \}$&  $ \{\text{N.A.} \}$ & $ \{0.05200 \}$&  $ \{ 0.0465\}$ \\ \hline
					2 & 50 & 400 &0.85 & 0.0144 & 0.0136  &0.0058 &0.0063  &-0.0009 & -0.0007 & 0.0007 & 0.0013 \\ 
					&  &  & & (0.0245) & (0.0219) &(0.0154) &(0.0155)  &(0.0381) &(0.0327)  & (0.0112)  & (0.0117)\\ 
					&  &  &  & $[0.4030]$& $ [0.4090]$ &  $[0.1245]$ & $[0.1410]$  & $[0.0465]$&  $[0.0490]$ & $[0.0705]$&  $[0.0735]$ \\ 
					&  &  &  & $ \{  \text{N.A.}\}$& $ \{ \text{N.A.}\}$ &  $ \{0.1185 \}$ & $ \{ 0.1615\}$  & $ \{ \text{N.A.}\}$&  $ \{\text{N.A.} \}$ & $ \{0.0425 \}$&  $ \{ 0.0405\}$ \\ \hline
					3 & 100 & 400 & 0.80 & 0.0070 & 0.0065  & 0.0029  & 0.0031  & -0.0006 &-0.0004  &0.0009  & 0.0014  \\ 
					&  &  &  & (0.0221)&  (0.0196) &  (0.0137) & (0.0139)  & (0.0117) &  (0.0196) &  (0.0068) &  (0.0069) \\ 
					&  &  &  & $[0.1365]$& $ [0.3865]$ &  $[0.0925]$ & $[0.0915]$  & $[0.0465]$&  $[0.0455]$ & $[0.0530]$&  $[0.0590]$ \\ 
					&  &  &  & $ \{  \text{N.A.}\}$& $ \{  \text{N.A.}\}$ &  $ \{ 0.0895\}$ & $ \{ 0.0955\}$  & $ \{  \text{N.A.}\}$&  $ \{  \text{N.A.}\}$ & $ \{0.0495 \}$&  $ \{0.0510 \}$ \\ \hline
					4 & 200 & 400 & 0.77& 0.0031 & 0.0029  & 0.0015  & 0.0016  & -0.0011 &-0.0009  &0.0006  & 0.0010 \\ 
					&  &  &  & (0.0216)&  (0.0197) &  (0.0139) & (0.0139)  & (0.0071) &  (0.0130) &  (0.0044) &  (0.0045) \\ 
					&  &  &  & $[0.0950]$& $ [0.3645]$ &  $[0.0745]$ & $[0.0760]$  & $[0.0400]$&  $[0.0455]$ & $[0.0640]$&  $[0.0665]$ \\ 
					&  &  &  & $ \{ \text{N.A.} \}$& $ \{  \text{N.A.}\}$ &  $ \{ 0.0590\}$ & $ \{ 0.0615\}$  & $ \{  \text{N.A.}\}$&  $ \{  \text{N.A.}\}$ & $ \{ 0.0510\}$&  $ \{0.0515 \}$ \\ \hline
					5 & 500 & 400 & 0.75& 0.0014 & 0.0014  & 0.0006  & 0.0006  & -0.0011  &-0.0009 &0.0004  & 0.0008 \\ 
					&  &  &  & (0.0218)&  (0.0209) &  (0.0141) & (0.0141)  & (0.0037) &  (0.0042) &  (0.0024) &  (0.0024) \\ 
					&  &  &  & $[0.0690]$& $ [0.1125]$ &  $[0.0670]$ & $[0.0690]$  & $[0.0515]$&  $[0.0570]$ & $[0.0605]$&  $[0.0675]$ \\ 
					&  &  &  & $ \{ \text{N.A.} \}$& $ \{  \text{N.A.}\}$ &  $ \{ 0.0535\}$ & $ \{ 0.0520\}$  & $ \{  \text{N.A.}\}$&  $ \{  \text{N.A.}\}$ & $ \{ 0.0575\}$&  $ \{ 0.0555\}$ \\ \hline
					\hline
				\end{tabular}
				\begin{tablenotes}
					\item Notes: We report Bias, (RMSE), [$t$-test],and $\{$$J$-test$\}$ (if applicable) with a nominal size of 5\%. $\widehat{\phi}^s_{FGIV}$, $\widehat{\phi}^s_{GMM}$, $\widehat{\phi}^{s,rmax}_{GMM}$ estimated with Algorithm \ref{alg:alg1}, \ref{alg:alg4}, \ref{alg:alg4} with $r =2$ and $rmax=3$, respectively and $\widehat{\phi}^s_{GK}$, $\widehat{\phi}^d_{GK}$ both use \eqref{ZGK} as an instrument. $\widehat{\phi}^d_{FGIV}$, $\widehat{\phi}^d_{GMM}$, $\widehat{\phi}^{d,rmax}_{GMM}$ estimated with \eqref{justidentifiedGIV}, \eqref{gmmdemand}, \eqref{gmmdemand} with $r =2$ and $rmax=3$, respectively. $\mu$ set to maintain $h_{N,\mu}=0.12$ across all configurations of $(N,T)$. $\bar{\psi}_{u}=0.23$, $\bar{\psi}_{u+\eta}=0.58$,  $\bar{\psi}_{u+\varepsilon}=0.65$.
				\end{tablenotes}
		\end{threeparttable}}
	\end{center}
\end{table}

\begin{table}[!htbp]
	\begin{center}
	\resizebox{0.425\textwidth}{!}{
			\begin{threeparttable}
				\centering
				\caption{Tail Index Estimates by Various Methods}
				\label{tailindexest}
				\begin{tabular}{|l|l|}
					\hline
					\textit{Tail index estimator} & $\widehat{\mu}$ \\ &\\[.3cm]
					MLE & 0.4216\\[.3cm]
					OLS & 0.5095\\[.3cm]
					Percentiles Method & 0.8987\\[.3cm]
					Modified Percentiles Method & 0.9000\\[.3cm]
					Geometric Percentiles Method & 0.5208\\[.3cm]
					Weighted Least Squares & 0.3725\\[.3cm]
					\hline 
				\end{tabular}
				\begin{tablenotes}
					\small
					\item Notes: The estimates are for a month selected at random. However, the estimates do not change significantly if we estimate $\widehat{\mu}$ for each month and average across months. 
				\end{tablenotes}
		\end{threeparttable}}
	\end{center}
\end{table}

\begin{table}[!htbp]
	\begin{center}
	\resizebox{0.6\textwidth}{!}{
			\begin{threeparttable}
				\centering
				\caption{Global crude oil market: supply elasticity}
				\label{oilelassupply}
				\begin{tabular}{|c|c|c|c|}
					\hline 
					&$\widehat{\phi}^s_{GK}$&$\widehat{\phi}^s_{FGIV}$ & $\widehat{\phi}^s_{GMM}$ \\ [.3cm]\hline
					Supply instruments&  $\bs{Z}_{GIV}$ &   $\bs{z}_{GIV}$  & $\bs{Z}_s = (\bs{z}_{GIV},\bs{\varepsilon})$\\ [.3cm]
					Dep. variable& $\bs{\bar{y}}$ & $\bs{y}_{\widehat{E}}$& $\bs{y}_{\widehat{E}}$  \\[.3cm]
					$\bs{p}$ &$0.044$  &$0.016$ & $0.005$  \\[.3cm]
					$t$-stat& $(1.43)$ & $(1.35)$ & $( 4.32)$  \\[.3cm]
					$(N,T)$ & $(21,370)$& $(21,370)$& $(21,370)$ \\[.3cm]
					$J$-stat $p$-value & $\{\text{N.A.} \}$&$\{\text{N.A.} \}$ & $0.11$  \\[.3cm]
					First stage $F$-stat &$<10$ & $<10$ & $14.33$ \\[.3cm]
					First stage $R^2$ &$0.26$ & $0.14$  &$0.21$ \\ \hline
				\end{tabular}
				\begin{tablenotes}
					\small
					\item Notes: $\widehat{\phi}^s_{GK}$ is estimated using \eqref{ZGK} as the instrument; whereas $\widehat{\phi}^s_{FGIV}$, and $\widehat{\phi}^s_{GMM}$ are estimated using Algorithm \ref{alg:alg1} and \ref{alg:alg4} respectively. See Section \ref{oil} for more details. The $t$-stat is reported in parenthesis below coefficient estimates. The coefficient estimates on $\bs{\widehat{\eta}_t}$,$\widehat{\eta}_{OPEC,t}$ and $\bs{c}_{t-1}$ are omitted for brevity.
				\end{tablenotes}
		\end{threeparttable}}
	\end{center}
\end{table} 

\begin{table}[!htbp]
	\begin{center}
	\resizebox{1.1\textwidth}{!}{
			\begin{threeparttable}
				\centering
				\caption{Global crude oil market: demand elasticity}
				\label{oilelasdemand}
				\begin{tabular}{|c|c|c|cccc|c|}
					\hline 
					&\multicolumn{1}{c}{$\widehat{\phi}^d_{GK}$} \vline&$\widehat{\phi}^d_{FGIV}$ & \multicolumn{4}{c}{$\widehat{\phi}^d_{GMM}(r)$} \vline&\multicolumn{1}{c}{FGMM, BN}  \\ [.3cm]\hline
					Demand instruments&  $\bs{Z}_{GIV}$ &   $\bs{z}_{GIV}$  & $(\bs{z}_{GIV},\bs{\eta}[,1])$& $(\bs{z}_{GIV},\bs{\eta}[,1:2])$& $(\bs{z}_{GIV},\bs{\eta}[,1:3])$& $(\bs{z}_{GIV},\bs{\eta}[,1:4])$ &  $\bs{\eta}[,1:4]$ \\ [.3cm]
					Dep. variable& $\bs{d}$ & $\bs{d}$& $\bs{d}$ & $\bs{d}$&$\bs{d}$ &$\bs{d}$  &$\bs{d}$ \\ [.3cm]
					$\bs{p}$ &$-0.463$  &$-0.0009$ & $-0.0009$  &$-0.0003$ &$-0.0003$ &$-0.0003$  &$-0.0003$\\[.3cm]
					$t$-stat& $(-3.54)$ & $(-0.88)$ & $(-0.89)$  &$(-0.87)$ &$(-0.93)$ &$(-1.01)$ &$(-0.80)$\\[.3cm]
					$(N,T)$ & $(21,370)$& $(21,370)$& $(21,370)$ & $(21,370)$ & $(21,370)$ &  $(21,370)$ & $(21,370)$ \\[.3cm]
					$J$-stat $p$-value & $\{\text{N.A.} \}$&$\{\text{N.A.} \}$ & $0.83$ &$0.67$ &$0.85$ &$0.93$ &$0.98$ \\[.3cm]
					First stage $F$-stat &$<10$ & $<10$ & $<10$ & $<10$& $<10$& $<10$ &$<10$\\[.3cm]
					First stage $R^2$ &$0.58$ & $0.12$  &$0.12$  &$0.15$ &$0.15$ &$0.16$  &$0.16$\\ \hline
				\end{tabular}
				\begin{tablenotes}
					\small
					\item Note: $\widehat{\phi}^d_{GK}$ is estimated using \eqref{ZGK} as the instrument; whereas $\widehat{\phi}^d_{FGIV}$, and $\widehat{\phi}^d_{GMM}$ are estimated using \eqref{justidentifiedGIV} and \eqref{gmmdemand} respectively. See Section \ref{oil} for more details. The $t$-stat is reported in parenthesis below coefficient estimates. The coefficient estimates on $\widehat{\eta}_{OPEC,t}$ and $\bs{c}_{t-1}$ are omitted for brevity. The final column represents the \citet{bai2010instrumental} factor GMM estimator (FGMM). 
				\end{tablenotes}
		\end{threeparttable}}
	\end{center}
\end{table} 

\newpage  
\begin{appendices}\label{appendix}
	\setlength{\parindent}{0.5cm}
	In this appendix, we prove Theorems 1-6, which require 4 lemmas. Lemmas 1, 2 and 3 are included in this appendix while Lemma 4 is deferred to Appendix \ref{herf} of the Supplementary Appendix. 
	\begin{lemma}\label{lemmaC1} Under Assumptions 1-4, we have that
		\begin{align}
		&(i.) \,\,\,\left(\frac{1}{T}\sum_{t=1}^T \tilde{y}_{i t}\varepsilon_t\right)^2= \mathcal{O}_p\left(\frac{1}{T}\right)  \label{part1} \\
		&(ii.) \,\,\, \mathbbm{V}(p_t)=\Theta(1) \\
		&(iii.) \,\,\,\left(\frac{1}{T}\sum_{t=1}^T \tilde{y}_{i t}p_t\right)^2 = \mathcal{O}_p\left(1\right) \\
		&(iv.)\,\,\,\sum_{i=1}^N\sum_{j=1}^N\left(\frac{1}{T}\sum_{t=1}^TS_i \tilde{y}_{j t}\varepsilon_t\right)^2 =\mathcal{O}_p\left(\frac{N}{T}\right). \label{part5}
		\end{align}
	\end{lemma}
	\textbf{Proof of Lemma \ref{lemmaC1}:} For $(i.)$, we have that for large $T$, the sum is stochastically bounded by the central limit theorem when scaled by $T^{-1/2}$, thus the term inside the square is $\mathcal{O}_p\left(\frac{1}{\sqrt{T}}\right)$.
	For $(ii.)$, note that by \ref{A3a} we can decompose our share vector into a dominant and a fringe part: $\bs{S}=\begin{pmatrix}\bs{S}'_{d} & \bs{S}'_f \end{pmatrix}'$ where $\bs{S}_{d}$ is $N_1 \times 1$, is the dominant part and $\bs{S}_f$ is $N_2 \times 1$, is the fringe part; with $N_1+N_2=N$. The key being that $N_1(N)=N_1$ is fixed while $N_2(N) \rightarrow \infty$ as $N \rightarrow \infty$. Recall that prices are given by $p_t = \dfrac{1}{\phi^d-\phi^s} \left(u_{St} + \bs{\lambda}'_S \bs{\eta}_t -\varepsilon_t\right)$. For simplicity, suppose that supply and demand shocks are uncorrelated, so that (ignoring the squared constant term)
	\begin{align*}
	\mathbbm{V}(p_t) &\propto  \mathbbm{E}(\bs{S}' \bs{\Sigma}_u \bs{S}) + \mathbbm{E}(\bs{S}' \bs{\Lambda}  \bs{\Lambda}' \bs{S}) + \mathbbm{V}(\varepsilon_t)=  \mathbbm{E}(\bs{S}' \bs{\Sigma}_u \bs{S}) + \mathbbm{E}(\bs{S}_{d}' \bs{\Lambda}_{d}  \bs{\Lambda}_{d}' \bs{S}_{d})+ \mathbbm{E}(\bs{S}_{f}' \bs{\Lambda}_{f}  \bs{\Lambda}_{f}' \bs{S}_{f}) + \mathbbm{V}(\varepsilon_t)\\
	&\leq  \mathbbm{E}(||\bs{S}||_2^2 \gamma_{max}(\bs{\Sigma}_u)) +  \mathbbm{E}(|| \bs{S}_{d}||_2^2 \gamma_{max} ( \bs{\Lambda}_{d} \bs{\Lambda}_{d}')) +  \mathbbm{E}(|| \bs{S}_f||_2^2 \gamma_{max} ( \bs{\Lambda}_{f} \bs{\Lambda}_{f}'))+\mathcal{O}(1),
	\end{align*} 
	by \ref{A3a} and \ref{A4}; the first term consists of $|| \bs{S}||^2$, which is $\Theta_p(1)$ for $\mu \in (0,1)$, see Lemma \ref{shareslemma} in Appendix \ref{herf} of the Supplementary Appendix, and $\gamma_{max}( \bs{\Sigma}_u)=\mathcal{O}(1)$ by assumption, the second term is $\mathcal{O}(1)$ by \ref{A4} and the third term is $\mathcal{O}(\frac{1}{N})\cdot\mathcal{O}(N)=\mathcal{O}(1)$ by \ref{A4}. For part $(iii.)$,
	\begin{align*}
	\left(\frac{1}{T}\sum_{t=1}^T \tilde{y}_{i t}p_t \right) &\leq \left(\frac{1}{T}\sum_{t=1}^T\tilde{y}^2_{it}\right)^{\frac{1}{2}} \cdot \left( \frac{1}{T}\sum_{t=1}^T p^2_t\right)^{\frac{1}{2}}=\left( \frac{1}{\sqrt{T}} ||  \bs{\tilde{y}}_{i \cdot}||\right) \cdot \left( \frac{1}{\sqrt{T}}|| \bs{p}||\right) =\mathcal{O}_p(1).
	\end{align*}
	 For part $(iv.)$, we have
	\begin{align*}
	\sum_{i=1}^N\sum_{j=1}^N\left(\frac{1}{T}\sum_{t=1}^TS_i \tilde{y}_{j t}\varepsilon_t\right)^2 &=I+II+III+IV, 
	\end{align*}
	where $I=\mathcal{O}_p\left( \frac{1}{T}\right) = o_p(1)$, $II=\mathcal{O}_p\left( \frac{1}{T}\right) =o_p(1)$, $III=\mathcal{O}_p\left(\frac{1}{NT}\right)=o_p(1)$ and $IV=\mathcal{O}_p(\frac{N}{T})$ which are show below (where we repeatedly make use of part $(i.)$ as well as the decomposition of the share vector into fringe and dominant components)
	\begin{align*}
	I&= \sum_{i=1}^{N_1} S_i^2\sum_{j=1}^{N_1}\left(\frac{1}{T}\sum_{t=1}^T \tilde{y}_{j t}\varepsilon_t\right)^2 \\
	&= || \bs{S}_{d}||^2\cdot \mathcal{O}_p\left( \frac{N_1}{T}\right) = o_p(1)\\
	II&=\sum_{i=N_1+1}^{N}S_i^2\sum_{j=N_1+1}^{N}\left(\frac{1}{T}\sum_{t=1}^T \tilde{y}_{j t}\varepsilon_t\right)^2\\
	&=|| \bs{S}_f||^2\cdot \mathcal{O}_p(N_2)\cdot\mathcal{O}_p\left(\frac{1}{T}\right) = \mathcal{O}_p\left(\frac{1}{N}\right)\mathcal{O}_p\left(\frac{N_2}{T}\right)=o_p(1)\\
	III&= \sum_{i=N_1+1}^N S_i^2\sum_{j=1}^{N_1} \left(\frac{1}{T}\sum_{t=1}^T\tilde{y}_{jt}\varepsilon_t\right)^2\\
	&=||\bs{S}_f||^2\cdot \mathcal{O}_p\left(\frac{N_1}{T}\right) = o_p\left(1\right) \\
	IV&=\sum_{i=1}^{N_1} S_i^2\sum_{j=N_1+1}^{N} \left(\frac{1}{T}\sum_{t=1}^T\tilde{y}_{jt}\varepsilon_t\right)^2\\
	&=\mathcal{O}_p\left(\frac{N}{T}\right)
	\end{align*}
	by \ref{A4} and Lemma \ref{lemmaC1} part (i.) $\hfill \blacksquare$
	
	\noindent
	\begin{lemma}\label{lemmaC2} Under Assumptions 1-4, we have that
		\begin{align}
		\frac{1}{T}\sum_{t=1}^T(\widehat{z}_t-z_t)\varepsilon_t
		&=\frac{1}{T}\sum_{t=1}^T  \bs{S}'( \bs{\widehat{Q}}- \bs{Q}) \bs{\tilde{y}}_{\cdot t}  \varepsilon_t=\mathcal{O}_p(C^{-2}_{NT})+\mathcal{O}_p(C^{-2}_{NT})\cdot \mathcal{O}_p\left(\frac{N}{T}\right)+\mathcal{O}_p\left( \frac{1}{\sqrt{N}}\cdot C^{-1}_{NT}\right) \label{sC2}\\
		\frac{1}{T}\sum_{t=1}^T(\widehat{z}_t-z_t)p_t &=\frac{1}{T}\sum_{t=1}^T  \bs{S}'( \bs{\widehat{Q}}- \bs{Q}) \bs{\tilde{y}}_{\cdot t}  p_t =\mathcal{O}_p(C^{-2}_{NT})+\mathcal{O}_p(C^{-2}_{NT})\cdot \mathcal{O}_p\left(\frac{N}{T}\right)+\mathcal{O}_p\left( \frac{1}{\sqrt{N}}\cdot C^{-1}_{NT}\right),
		\end{align}
		where $C_{NT}=\emph{min}\{\sqrt{N}, \sqrt{T}\}$.
	\end{lemma}
	\textbf{Proof of Lemma \ref{lemmaC2}:} For the first term, it is well known that the loadings are only identified up to scale, so the usual notion of consistency is altered to consider consistency up to a rotation instead. For notational ease we will let $ \bs{\tilde{\Lambda}}$ be denoted by $ \bs{\Lambda}$. Recall, $ \bs{Q}= \bs{I}_N- \bs{P}_{\Lambda H^{-1}}$ is an idempotent matrix spanned by the null space of $ \bs{\Lambda}  \bs{H^{-1}}$ and is invariant to an orthogonal transformation. 
%	that is $ \bs{Q}= \bs{I_N}- \bs{P_{\Lambda H^{-1}}}= \bs{Q}= \bs{I_N}- \bs{P_{\Lambda}}$. 
	Let $ \bs{\widehat{D}}=\dfrac{ \bs{\widehat{\Lambda}}' \bs{\widehat{\Lambda}}}{N}=\dfrac{1}{N}\sum_{i=1}^N \bs{\widehat{\lambda}}_i \bs{\widehat{\lambda}}_i'$ and $ \bs{D}=\dfrac{ \bs{H^{-1'}} ( \bs{\Lambda}' \bs{\Lambda})  \bs{H^{-1}}}{N}=\dfrac{1}{N}  \bs{H^{-1'}}\sum_{i=1}^N \color{black}\bs{{\lambda}}_i \bs{{\lambda}}_i'\color{black} \bs{H^{-1}}$, then we have (omitting subscripts on $ \bs{P}$)
	\begin{align*}
	 \bs{\widehat{Q}}- \bs{Q} &= \bs{\widehat{P}}- \bs{P} \\
	&=N^{-1} \bs{\widehat{\Lambda}}\left(\frac{ \bs{\widehat{\Lambda}}' \bs{\widehat{\Lambda}}}{N}\right)^{-1} \bs{\widehat{\Lambda}}' - N^{-1} \bs{\Lambda}  \bs{H^{-1}}\left(\frac{ \bs{H^{-1'}} ( \bs{\Lambda}' \bs{\Lambda})  \bs{H^{-1}}}{N}\right)^{-1}  \bs{H^{-1'}}  \bs{\Lambda}' \\
	&=N^{-1}\left[ \bs{\widehat{\Lambda}} \bs{\widehat{D}^{-1}} \bs{\widehat{\Lambda}}' -  \bs{\Lambda}  \bs{H^{-1}} \bs{D^{-1}} \bs{H^{-1'}}  \bs{\Lambda}' \right]\\
	&=N^{-1}\left[( \bs{\widehat{\Lambda}}- \bs{\Lambda}  \bs{H^{-1}} + \bs{\Lambda}  \bs{H^{-1}}) \bs{\widehat{D}^{-1}}( \bs{\widehat{\Lambda}}- \bs{\Lambda}  \bs{H^{-1}} + \bs{\Lambda}  \bs{H^{-1}})'-  \bs{\Lambda}  \bs{H^{-1}} \bs{D^{-1}}  \bs{H^{-1'}}  \bs{\Lambda}' \right]\\
	&=N^{-1}\left[ \right.( \bs{\widehat{\Lambda}}- \bs{\Lambda}  \bs{H^{-1}}) \bs{\widehat{D}^{-1}}( \bs{\widehat{\Lambda}}- \bs{\Lambda}  \bs{H^{-1}})'+( \bs{\widehat{\Lambda}}- \bs{\Lambda}  \bs{H^{-1}}) \bs{\widehat{D}^{-1}} \bs{H^{-1'}} \bs{\Lambda}'+ \dots \\
	& \,\,\,\,\,\,\,\,\,\,\,\,\,\,\,\,\,\,\,\,\,\,\,\,\,\,\dots +  \bs{\Lambda}  \bs{H^{-1}}  \bs{\widehat{D}^{-1}} ( \bs{\widehat{\Lambda}}- \bs{\Lambda}  \bs{H^{-1}})+ \bs{\Lambda}  \bs{H^{-1}}( \bs{\widehat{D}^{-1}}- \bs{D^{-1}})  \bs{H^{-1'}}  \bs{\Lambda}' \left.\right]. \nonumber 
	\end{align*}
	
	%	\begin{align}
	%	S'\Lambda \Lambda'S &=\sum_{i=1}^N\sum_{j=1}^N S_i^2 \lambda_i'\lambda_j \\
	%	&\leq \sum_{j=1}^N \lambda_j' \left( \sum_{i=1}^N S^2_i\lambda_i' \right)^2 \lambda_j \\  
	%	&\leq \sum_{j=1}^N \lambda_j' \left( \sum_{i=1}^N S_i^4\right) \left(\sum_{i=1}^N \lambda_i\lambda_i' \right) \lambda_j \\  
	%	&=\left( \sum_{i=1}^N S_i^4\right) \sum_{j=1}^N \lambda_j'  \left( \Lambda'\Lambda \right) \lambda_j \\
	%	&= \frac{\sum_{i=1}^N \mathscr{S}^4_i}{\left(\sum_{j=1}^N \mathscr{S}_j\right)^4} \sum_{j=1}^N \lambda_j'  \left( \Lambda'\Lambda \right) \lambda_j \\ 
	%	&= \frac{1}{N^3}\frac{N^{-1}\sum_{i=1}^N \mathscr{S}^4_i}{\left(N^{-1}\sum_{j=1}^N \mathscr{S}_j\right)^4} \sum_{j=1}^N \lambda_j'  \left( \Lambda'\Lambda \right) \lambda_j
	%	\end{align}
	%	Note that $\mathscr{S}_i^4$ follows a power law with tail index $\frac{\mu}{4}<1$ so it follows that (see Appendix \ref{herf}),
	%	\begin{align}
	%	S'\Lambda \Lambda'S &= \frac{1}{N^3}\frac{N^{-1}\sum_{i=1}^N \left(\frac{i}{N}\right)^{-\frac{4}{\mu}}}{\left(N^{-1}\sum_{j=1}^N \left(\frac{j}{N}\right)^{-\frac{1}{\mu}}\right)^4} \sum_{j=1}^N \lambda_j'  \left( \Lambda'\Lambda \right) \lambda_j \\ 
	%	&= \frac{\frac{1}{N^{4-\frac{4}{\mu}}}\sum_{i=1}^N i^{-\frac{4}{\mu}}}{{\frac{1}{N^{4-\frac{4}{\mu}}}\left(\sum_{j=1}^N j^{-\frac{1}{\mu}}\right)^4}} \sum_{j=1}^N \lambda_j'  \left( \Lambda'\Lambda \right) \lambda_j\\
	%	&= \frac{\zeta(\frac{4}{\mu})}{\left(\zeta(\frac{1}{\mu})\right)^4} \sum_{j=1}^N \lambda_j'  \left( \Lambda'\Lambda \right) \lambda_j	
	%	\end{align}
	Therefore, $\frac{1}{T}\sum_{t=1}^T  \bs{S}'( \bs{\widehat{Q}}- \bs{Q}) \bs{\tilde{y}}_{\cdot t}  \varepsilon_t=I+II+III+IV$. Each term is analyzed below in order.\footnote{In this Appendix, we use the Frobenius norm of a matrix $\bs{A}$ is $|| \bs{A}||_F =\left[\text{tr}( \bs{A}' \bs{A}) \right]^{\frac{1}{2}}=\left[\sum_i \sum_j |a_{ij}|^2\right]^{\frac{1}{2}}$, but omit the subscript $F$ for notational ease.}
	\begin{align*}
	I&=\frac{1}{NT}\sum_{t=1}^T  \bs{S}'( \bs{\widehat{\Lambda}}- \bs{\Lambda}  \bs{H^{-1}}) \bs{\widehat{D}^{-1}}( \bs{\widehat{\Lambda}}- \bs{\Lambda}  \bs{H^{-1}})' \bs{\tilde{y}}_{\cdot t}\varepsilon_t \\
	&=\frac{1}{N}\sum_{i=1}^N\sum_{j=1}^N  ( \bs{\widehat{\lambda}}_i- \bs{H^{-1}} \bs{\lambda}_i  )' \bs{\widehat{D}^{-1}}( \bs{\widehat{\lambda}}_j- \bs{H^{-1}} \bs{\lambda}_j)\cdot\frac{1}{T}\sum_{t=1}^T S_i\tilde{y}_{j t}\varepsilon_t \nonumber\\
	&\leq \left(\frac{1}{N^2}\sum_{i=1}^N\sum_{j=1}^N\left[ ( \bs{\widehat{\lambda}}_i- \bs{H^{-1}} \bs{\lambda}_i  )' \bs{\widehat{D}^{-1}}( \bs{\widehat{\lambda}}_j- \bs{H^{-1}} \bs{\lambda}_j)\right]^2\right)^{\frac{1}{2}}\cdot \left[\sum_{i=1}^N\sum_{j=1}^N\left(\frac{1}{T}\sum_{t=1}^T S_i \tilde{y}_{j t}\varepsilon_t\right)^2\right]^{\frac{1}{2}} \numberthis \label{C7}\\
	&\leq \left(\frac{1}{N^2}\sum_{i=1}^N\sum_{j=1}^N  ||( \bs{\widehat{\lambda}}_i- \bs{H^{-1}} \bs{\lambda}_i  )||^2\cdot || \bs{\widehat{D}^{-1}}||^2\cdot||( \bs{\widehat{\lambda}}_j- \bs{H^{-1}} \bs{\lambda}_j)||^2\right)^{\frac{1}{2}}\cdot \mathcal{O}_p\left(\frac{N}{T}\right) \\
	&= || \bs{\widehat{D}^{-1}}||\cdot \left(\left(\frac{1}{N}\sum_{i=1}^N ||( \bs{\widehat{\lambda}}_i- \bs{H^{-1}} \bs{\lambda}_i  )||^2\right)^2\right)^{\frac{1}{2}} \cdot \mathcal{O}_p\left(\frac{N}{T}\right)= \mathcal{O}_p(1)\cdot \mathcal{O}_p(C^{-2}_{NT})\cdot \mathcal{O}_p\left(\frac{N}{T}\right)=\mathcal{O}_p(C^{-2}_{NT})\cdot \mathcal{O}_p\left(\frac{N}{T}\right),
	\end{align*}
	where $\mathcal{O}_p(C^{-2}_{NT})$ follows from symmetry of Theorem 1 in \citet{bai2002determining}, who show (while proving their Lemma 2) $||\widehat{D}^{-1}||$ is $\mathcal{O}_p(1)$ which again follows symmetrically here. Note, the first inequality follows from Cauchy-Schwarz applied to the summation in $(i,j)$; the second inequality follows from Cauchy-Schwarz applied again but to the left most term's inner term in brackets being squared, in equation \eqref{C7}. 
	\begin{align*}
	II&=\frac{1}{NT}\sum_{t=1}^T  \bs{S}'( \bs{\widehat{\Lambda}}- \bs{\Lambda}  \bs{H^{-1}} ) \bs{\widehat{D}^{-1}} \bs{H^{-1'}} \bs{\Lambda}' \bs{\tilde{y}}_{\cdot t}\varepsilon_t=\frac{1}{N}\sum_{i=1}^N ( \bs{\widehat{\lambda}}_i- \bs{H^{-1}} \bs{\lambda}_i  )'\underbrace{ \bs{\widehat{D}^{-1}}\sum_{j=1}^N  \bs{H^{-1'}}\, \bs{\lambda}_j\cdot \frac{1}{T}\sum_{t=1}^T S_j \tilde{y}_{i t}\varepsilon_t}_{\text{$\underset{r \times 1}{\bs{a_i}}$}} \\
	&=\mathcal{O}_p\left( C^{-2}_{NT}\right),\numberthis \label{b1} 
	%	&\leq \left(\frac{1}{N^2}\sum_{i=1}^N\sum_{j=1}^N\left[ (\widehat{\lambda}_i-H^{-1}\lambda_i  )'\widehat{D}^{-1}H^{-1'}\lambda_j\right]^2\right)^{\frac{1}{2}} \left[\sum_{i=1}^N\sum_{j=1}^N\left(\frac{1}{T}\sum_{t=1}^T S_i \tilde{y}_{j t}\varepsilon_t\right)^2\right]^{\frac{1}{2}} \\
	%	&\leq \left(\frac{1}{N^2}\sum_{i=1}^N\sum_{j=1}^N ||(\widehat{\lambda}_i-H^{-1}\lambda_i  )||^2 \cdot ||\widehat{D}^{-1}||^2 \cdot ||H^{-1'}\lambda_j||^2 \right)^{\frac{1}{2}} \cdot \mathcal{O}_p(1) \\
	%	&= ||\widehat{D}^{-1}||\cdot \left(\frac{1}{N}\sum_{i=1}^N ||(\widehat{\lambda}_i-H^{-1}\lambda_i  )||^2\right)^{\frac{1}{2}} \cdot \left(\frac{1}{N}\sum_{j=1}^N ||H^{-1'}\lambda_j||^2 \right)^{\frac{1}{2}} \cdot \mathcal{O}_p(1)\\	
	\end{align*} 
	which is the same rate as $I$. In \eqref{b1}, we make use of the fact that $N^{-1} \sum_i (\bs{\widehat{\lambda}}_i-\bs{H^{-1}}\bs{\lambda}_i  )'\bs{a}_i =\mathcal{O}_p\left(C^{-2}_{NT}\right)$, which follows symmetrically from Lemma B.1 of \citet{bai2003inferential}. The same logic leads to $III=\mathcal{O}_p\left( C^{-2}_{NT}\right)$.
	\begin{align*}
	IV &=\frac{1}{NT}\sum_{t=1}^T \bs{S}'\bs{\Lambda} \bs{H^{-1}}(\bs{\widehat{D}^{-1}}-\bs{D^{-1}})\bs{H^{-1'}}\bs{\Lambda}'\bs{\tilde{y}}_{\cdot t}\varepsilon_t=\frac{1}{N} \sum_{i=1}^N\sum_{j=1}^N S_i \bs{\lambda}_i' \bs{H^{-1}}(\bs{\widehat{D}^{-1}}-\bs{D^{-1}})\bs{H^{-1'}}\bs{\lambda}_j\cdot \frac{1}{T} \sum_{t=1}^T \tilde{y}_{j t}\varepsilon_t \nonumber \\
	&\leq \left(\frac{1}{N^2} \sum_{i=1}^N\sum_{j=1}^N S^2_i \left[\bs{\lambda}_i' \bs{H^{-1}}(\bs{\widehat{D}^{-1}}-\bs{D^{-1}})\bs{H^{-1'}}\bs{\lambda}_j\right]^2\right)^{\frac{1}{2}} \cdot \mathcal{O}_p\left( 1\right) \nonumber \\
	& \leq \left(\frac{1}{N^2} \sum_{i=1}^N\sum_{j=1}^N S^2_i \cdot ||\bs{\lambda}_i' \bs{H^{-1}}||^2\cdot||\bs{\widehat{D}^{-1}}-\bs{D^{-1}}||^2\cdot ||\bs{H^{-1'}}\bs{\lambda}_j||^2\right)^{\frac{1}{2}}\cdot \mathcal{O}_p\left( 1\right)\\
	&= \mathcal{O}_p(C^{-1}_{NT}) \cdot\left(\frac{1}{N} \sum_{i=1}^N S^2_i \cdot ||\bs{\lambda}_i' \bs{H^{-1}}||^2\right)^{\frac{1}{2}}\cdot\mathcal{O}_p(1) \cdot \mathcal{O}_p\left(1\right) \\
	&= \mathcal{O}_p(C^{-1}_{NT}) \cdot\left(\frac{1}{N} \left[\sum_{i=1}^{N_1} S^2_i \cdot ||\bs{\lambda}_i' \bs{H^{-1}}||^2 +\sum_{i=N_1+1}^N S^2_i \cdot ||\bs{\lambda}_i' \bs{H^{-1}}||^2\right]\right)^{\frac{1}{2}}\cdot\mathcal{O}_p(1) \cdot \mathcal{O}_p\left(1\right) \\
	&\leq \mathcal{O}_p(C^{-1}_{NT}) \cdot\left(\frac{1}{N} \left(\sum_{i=1}^{N_1} S^4_i\right)^{\frac{1}{2}} \cdot \left(\sum_{i=1}^{N_1}  ||\bs{\lambda}_i' \bs{H^{-1}}||^4\right)^{\frac{1}{2}}+\frac{1}{N}\left(\sum_{i=N_1+1}^{N} S^4_i\right)^{\frac{1}{2}} \cdot \left(\sum_{i=N_1+1}^{N}  ||\bs{\lambda}_i' \bs{H^{-1}}||^4\right)^{\frac{1}{2}}\right)^{\frac{1}{2}} \\
	&=\mathcal{O}_p\left( C^{-1}_{NT}\right)\cdot \left( \mathcal{O}_p\left(\frac{N_1}{N}\right) + \mathcal{O}_p\left(\frac{1}{N}\right) \right)^{\frac{1}{2}}=\mathcal{O}_p\left( C^{-1}_{NT}\right)\cdot \mathcal{O}_p\left(\frac{1}{\sqrt{N}}\right),
	\end{align*}
	where $||\bs{\widehat{D}^{-1}} -\bs{D^{-1}}|| = \mathcal{O}_p(C^{-1}_{NT})$ again follows symmetrically from \citet{bai2002determining}. All in all, we have that
	\begin{align*}
	\frac{1}{T}\sum_{t=1}^T \bs{S}'(\bs{\widehat{Q}}-\bs{Q})\bs{\tilde{y}}_{\cdot t}\varepsilon_t &=\mathcal{O}_p(C^{-2}_{NT})+\mathcal{O}_p(C^{-2}_{NT})\cdot \mathcal{O}_p\left(\frac{N}{T}\right)+\mathcal{O}_p\left( \frac{1}{\sqrt{N}}\cdot C^{-1}_{NT}\right),
	\end{align*}
	which is as the Lemma claimed. $\hfill \blacksquare$
	
	\noindent\textbf{Proof of Theorem \ref{sconsis}:} In light of Lemma \ref{lemmaC2}, the result follows immediately from \eqref{consisref} by observing that $\frac{1}{T}\sum_t p_t z_t \overset{p}{\rightarrow} \mathbbm{E}(p_tz_t) > 0$ for $\mu \in (0,1)$. Similarly, $\frac{1}{T}\sum_t z_t \varepsilon_t \overset{p}{\rightarrow} \mathbbm{E}(z_t\varepsilon_t)=0$ by \ref{A4}. $\hfill \blacksquare$
	
	\noindent\textbf{Proof of Theorem \ref{sdist}:} Under Assumptions 1-4 and Lemma \ref{lemmaC2}, the result follows immediately. $\hfill \blacksquare$
	
	As in the case of the demand elasticity, we need Lemma \ref{lemmaC3} before consistency and the limiting distribution of the supply elasticity can be established. 
	\begin{lemma}\label{lemmaC3} Under Assumptions 1-4, we have that the terms $a_i, b_i, c_j$ for $i=1,2$; and $j=1,2$; defined in equations \eqref{A}, \eqref{B} and \eqref{C} are $o_p\left( 1\right)$. While for $j=3$, $c_3$ is $\mathcal{O}_p\left(\frac{m_N \omega^{1-q}_{N,T}}{T}\right)=o_p(1)$.
	\end{lemma}
	\textbf{Proof of Lemma \ref{lemmaC3}:} The terms, $a_1$, $b_1$, and $c_1$ follow very similarly as the proof for Lemma \ref{lemmaC2} and hence are omitted and the terms $a_2$, $b_2$ and $c_2$ follow symmetrically to the proof of Lemma \ref{lemmaC2} and thus, they are also omitted. The term $c_3$ is novel and warrants some further analysis. Recall $c_3$ is given by $c_3=T^{-1} \bs{z}'\,\bs{M}_{\eta}\,(\bs{u}_{\widehat{E}}-\bs{u}_E)$. Let us focus on $(\bs{u}_{\widehat{E}}-\bs{u}_E)=\bs{u_{\cdot \cdot}}(\bs{\widehat{E}}-\bs{E})$ momentarily,
	where $\bs{u_{\cdot \cdot}}$ is the $T \times N$ matrix of idiosyncratic errors and $\bs{\widehat{E}}-\bs{E} = \dfrac{\bs{\widehat{\Sigma}}_u^{-1}\bs{\iota}}{\bs{\iota}'\bs{\widehat{\Sigma}}_u^{-1}\bs{\iota}}-\dfrac{\bs{\Sigma}_u^{-1}\bs{\iota}}{\bs{\iota}'{\bs{\Sigma}}_u^{-1}\bs{\iota}}$. Let $\mathbf{\Theta}_u :=  \bs{\Sigma}_u^{-1}$, $C :=  \frac{\bs{\iota}' \mathbf{\Theta}_u \bs{\iota}}{N}$, then $\bs{E}=\frac{\mathbf{\Theta}_u \bs{\iota} / N}{C}$. We have that 
	\begin{align*}
	\bs{\widehat{E}}-\bs{E}&=\frac{\left[C\widehat{\mathbf{\Theta}}_u \bs{\iota} -\widehat{C}\mathbf{\Theta}_u \bs{\iota} \right]/N}{\widehat{C}C}=\frac{\left[C\widehat{\mathbf{\Theta}}_u \bs{\iota} -C\mathbf{\Theta}_u \bs{\iota}+C\mathbf{\Theta}_u \bs{\iota}-\widehat{C}\mathbf{\Theta}_u \bs{\iota} \right]/N}{\widehat{C}C}, \\ 
	&=\frac{\left[C(\widehat{\mathbf{\Theta}}_u -\mathbf{\Theta}_u )\bs{\iota}+(C -\widehat{C})\mathbf{\Theta}_u \bs{\iota} \right]/N}{\widehat{C}C},  \\
	\implies ||\bs{\widehat{E}}-\bs{E}||_1&\leq \frac{\left[C\cdot ||(\widehat{\mathbf{\Theta}}_u -\mathbf{\Theta}_u )\bs{\iota} ||_1+|C -\widehat{C}|\cdot||\mathbf{\Theta}_u \bs{\iota}||_1 \right]/N}{|\widehat{C}| C}, \numberthis \label{caneretal} 
	\end{align*}	
	where \eqref{caneretal} follows from \citet{callot2021nodewise}. Let $\mathbf{\Theta}_{u,j}$ denote the $j^{th}$ row of $\mathbf{\Theta}_{u}$ written as a column vector. Using H\"{o}lder's inequality we have
	\begin{align*}
	|\widehat{C}-C|&=\bigg|\frac{\bs{\iota}'(\widehat{\mathbf{\Theta}}_u -\mathbf{\Theta}_u)\bs{\iota}}{N} \bigg| \leq \frac{||(\widehat{\mathbf{\Theta}}_u -\mathbf{\Theta}_u)\bs{\iota} ||_1 \cdot||\bs{\iota}||_{max}}{N} \leq \underset{1\leq j \leq N}{\text{max}} ||\widehat{\mathbf{\Theta}}_{u,j} -\mathbf{\Theta}_{u,j}||_1= ||\widehat{\mathbf{\Theta}}_{u} -\mathbf{\Theta}_{u}||_1.
	\end{align*}  
	Thus, 
	\begin{align*}
	||\bs{\widehat{E}}-\bs{E}||_1 &\leq \frac{\left[C\cdot ||(\widehat{\mathbf{\Theta}}_u -\mathbf{\Theta}_u )\bs{\iota} ||_1+|C -\widehat{C}|\cdot||\mathbf{\Theta}_u \bs{\iota}||_1 \right]/N}{|\widehat{C}| C}, \\ 
	&\leq \frac{\left[C\cdot \underset{1\leq j \leq N}{\text{max}}||\widehat{\mathbf{\Theta}}_{u,j} -\mathbf{\Theta}_{u,j}  ||_1+\underset{1\leq j \leq N}{\text{max}}||\widehat{\mathbf{\Theta}}_{u,j} -\mathbf{\Theta}_{u,j}||_1\cdot||\mathbf{\Theta}_u \bs{\iota}||_1/N \right]}{|\widehat{C}| C}, \\ 
	&= \frac{\underset{1\leq j \leq N}{\text{max}}||\widehat{\mathbf{\Theta}}_{u,j} -\mathbf{\Theta}_{u,j}  ||_1\left[C\cdot +||\mathbf{\Theta}_u \bs{\iota}||_1/N \right]}{|\widehat{C}| C} \leq \frac{\underset{1\leq j \leq N}{\text{max}}||\widehat{\mathbf{\Theta}}_{u,j} -\mathbf{\Theta}_{u,j}  ||_1\left[C +\underset{1\leq j \leq N}{\text{max}}||\mathbf{\Theta}_{u,j} ||_1 \right]}{|\widehat{C}| C}, \\ 
	&\leq \frac{ ||\widehat{\mathbf{\Theta}}_u -\mathbf{\Theta}_u||_1 \left[C + ||\mathbf{\Theta}_u||_1 \right]}{|C+o_p(1)| C}=\frac{\mathcal{O}_p(m_N \omega^{1-q}_{N,T}) \left[\mathcal{O}_p(1)+\mathcal{O}_p(1)\right]}{(\mathcal{O}_p(1)+o_p(1))\mathcal{O}_p(1)} =\mathcal{O}_p(m_N \omega^{1-q}_{N,T}), \numberthis \label{precorder}
	\end{align*}
	where $C \geq \gamma_{min}(\mathbf{\Theta}_{u}) > 0$. 
	%	It is clear from \eqref{caneretal} that the choice of how to estimate the precision matrix will directly affect the consistency rate of the precision weights. For example, if we make use of the optimal rate of convergence (under the spectral norm) for sparse covariance matrix estimation, due to \citet{cai2012optimal}, we have that
	%	\begin{align}
	%	||\widehat{\Sigma}_u^{-1}-{\Sigma}_u^{-1}||=\mathcal{O}_p\left(\sqrt{\dfrac{\text{log}(N)}{T}}\right). \label{optcovrate}
	%	\end{align} 
	%	More recently, \citet{callot2021nodewise} obtain a rate of , , by using nodewise regression techniques. While \citet{lee2020optimal} who generalize existing precision matrix estimation by using factor structures and graphical lasso techniques obtain a rate of . 
	Putting it all together for $c_3$, we have that
	\begin{align*}
	c_3 &=T^{-1} \bs{z}' \, \bs{M}_{\eta} (\bs{u}_{\widehat{E}}-\bs{u}_E)=T^{-1} \bs{z}' \, \bs{M}_{\eta} \bs{u_{\cdot \cdot}}(\bs{\widehat{E}}-\bs{E}) \leq \gamma_{max}(\bs{M}_{\eta})\cdot T^{-1}\bs{z}'\bs{u_{\cdot \cdot}}(\bs{\widehat{E}}-\bs{E}), \\
	&=T^{-1} \bs{z}'\bs{u_{\cdot \cdot}}(\bs{\widehat{E}}-\bs{E}) \leq T^{-1} ||\bs{u_{\cdot \cdot}}' \bs{z}||_{1} \cdot ||\bs{\widehat{E}}-\bs{E}||_1 \leq T^{-1} ||\bs{u_{\cdot \cdot}}' \bs{z} ||_{1}\cdot \mathcal{O}_p(m_N \omega^{1-q}_{N,T}), \\
	&=T^{-1} \sum_{i=1}^N |S_i (\bs{u_{\cdot \cdot}}' \bs{y_{\cdot \cdot}} \bs{Q})_i|   \cdot   \mathcal{O}_p(m_N \omega^{1-q}_{N,T})=\mathcal{O}_p\left(\frac{1}{T}\right)  \cdot   \mathcal{O}_p(m_N \omega^{1-q}_{N,T})=o_p(1),
	\end{align*}
	which concludes the proof. 	$\hfill \blacksquare$
	
	\noindent\textbf{Proof of Theorem \ref{dconsis}:} In light of Lemma \ref{lemmaC3}, the result follows immediately from \eqref{dem}. $\hfill \blacksquare$
	
	\noindent\textbf{Proof of Theorem \ref{ddist}:} Under Assumptions 1-4 and Lemma \ref{lemmaC3}, the result follows immediately. $\hfill\blacksquare$
	
	\noindent\textbf{Proof of Theorem \ref{ddistgmm}:} In light of Theorem \ref{sdist} and \citet{bai2010instrumental}, the result follows immediately. $\hfill \blacksquare$
	
	\noindent\textbf{Proof of Theorem \ref{sdistgmm}:} 
	 In light of the theorems in the just identified case and standard GMM theory, we know that $\sqrt{T} (\bs{\widehat{\theta}}^s_{GMM}-\bs{\theta}^s)$ is asymptotically, a normal variate. The question remains whether using $\bs{\widehat{\varepsilon}}=\bs{\varepsilon}(\widehat{\phi}_{GMM}^d)$ introduces sampling error that will effect the standard error of $\bs{\widehat{\theta}}^s_{GMM}$. To that end, let $\underset{(2+r) \times 1}{\bs{g}_{st}} :=\bs{Z}_{st}u_{Et}$, we have that $\bs{\widehat{\theta}}^s_{GMM}$ solves the following first-order condition with probability approaching 1
	\begin{align} 
	0&=\left(\frac{1}{T}\sum_{t=1}^T \frac{\partial}{\partial \bs{\theta}^s}\bs{g}_{st}(\bs{\widehat{\theta}}^s_{GMM};\widehat{\phi}^d)\right)' \, \bs{\widehat{\Omega}}_s^{-1} \,\left( \frac{1}{T}\sum_{t=1}^T  \bs{g}_{st}(\bs{\widehat{\theta}}^s_{GMM};\widehat{\phi}^d)\right) \\ 
	&=\left(\frac{1}{T}\sum_{t=1}^T \frac{\partial}{\partial \bs{\theta}^s}\bs{g}_{st}(\bs{\widehat{\theta}}^s_{GMM};\widehat{\phi}^d)\right)' \, \bs{\widehat{\Omega}}_s^{-1} \,\left( \frac{1}{T}\sum_{t=1}^T  \bs{g}_{st}(\bs{\theta}^s;\widehat{\phi}^d)\right) \nonumber \\
	&+\left(\frac{1}{T}\sum_{t=1}^T \frac{\partial}{\partial \bs{\theta}^s}\bs{g}_{st}(\bs{\widehat{\theta}}^s_{GMM};\widehat{\phi}^d)\right)' \, \bs{\widehat{\Omega}}_s^{-1} \,\left( \frac{1}{T}\sum_{t=1}^T  \frac{\partial}{\partial \bs{\theta}^s}\bs{g}_{st}(\bs{\bar{\theta}}^s;\widehat{\phi}^d)\right)(\bs{\widehat{\theta}}^s_{GMM}-\bs{\theta}^s) \label{mvedem}  \\ 
	&=\bs{G}_s' \, \bs{\widehat{\Omega}}_s^{-1} \,\left( \frac{1}{\sqrt{T}}\sum_{t=1}^T  \bs{g}_{st}(\bs{\theta}^s;\widehat{\phi}^d)\right)+\bs{G}_s' \, \bs{\widehat{\Omega}}_s^{-1}  \,\bs{G}_s \sqrt{T} (\bs{\widehat{\theta}}^s_{GMM}-\bs{\theta}^s) \label{mvedem2}
	\end{align}
	The basic idea of whether the sampling error from estimating $\widehat{\phi}^d$ can be ignored, boils down to whether the following expression holds:
	$\frac{1}{\sqrt{T}} \sum_{t=1}^T \bs{g}_{st}(\bs{\theta}^s; \widehat{\phi}^d) = \frac{1}{\sqrt{T}} \sum_{t=1}^T \bs{g}_{st}(\bs{\theta}^s; \phi^d) +o_p(1)$; when this equation holds, then $\sqrt{T}(\bs{\widehat{\theta}}^s_{GMM}-\bs{\theta}^s)$ will not asymptotically depend on $\sqrt{T} (\widehat{\phi}^d-\phi^d)$. This can be easily seen if we take a mean value expansion of the left hand side of the expression above around $\phi^d$, we obtain
	\begin{align}
	\frac{1}{\sqrt{T}} \sum_{t=1}^T \bs{g}_{st}(\bs{\theta}^s;\widehat{\phi}^d) &= \frac{1}{\sqrt{T}} \sum_{t=1}^T \bs{g}_{st}(\bs{\theta}^s;\phi^d) + \bs{F} \sqrt{T} (\widehat{\phi}^d-\phi^d) +o_p(1), \label{nuisMVE}
	\end{align}  
	where $\underset{(2+r) \times 1}{\bs{F}}:=\mathbbm{E}\left[\nabla_{\phi^d}\bs{g}_{st}(\bs{\theta}^s;\widehat{\phi}^d)\right]$, is generally different from zero, but here we have $\bs{F}=\mathcal{O}_p(\frac{1}{\sqrt{N}})$. This implies the asymptotic variance of $\sqrt{T}(\bs{\widehat{\theta}}^s_{GMM}-\bs{\theta}^s)$ need not take into account the sampling error induced by $\widehat{\phi}^d$. To see why, we need to get an expression for $\sqrt{T}(\widehat{\phi}^d-\phi^d)$, let $\underset{(1+r) \times 1}{\bs{g}_{dt}}=\bs{Z}_{dt}\varepsilon_t$; then taking a similar mean value expansion (as above) of the first-order conditions that $\widehat{\phi}^d$ solves with probability approaching 1 
	\begin{align}
%	&\left(\frac{1}{T}\sum_{t=1}^T \frac{\partial}{\partial \phi^d}\bs{g}_{dt}(\widehat{\phi}^d) \right)'\, \bs{\widehat{\Omega}}_d^{-1} \, \left(\frac{1}{T}\sum_{t=1}^T\bs{g}_{dt}(\widehat{\phi}^d)\right) \\
%	&=\left(\frac{1}{T}\sum_{t=1}^T \frac{\partial}{\partial \phi^d}\bs{g}_{dt}(\widehat{\phi}^d) \right)'\, \bs{\widehat{\Omega}}_d^{-1} \, \left(\frac{1}{T}\sum_{t=1}^T\bs{g}_{dt}(\phi^d)\right) \nonumber\\
%	&+\left(\frac{1}{T}\sum_{t=1}^T \frac{\partial}{\partial \phi^d}\bs{g}_{dt}(\widehat{\phi}^d) \right)'\, \bs{\widehat{\Omega}}_d^{-1} \,\left(\frac{1}{T}\sum_{t=1}^T \frac{\partial}{\partial \phi^d}\bs{g}_{dt}(\widehat{\phi}^d) \right)'(\widehat{\phi}^d-\phi^d) \\
	0&= \bs{G}_d' \bs{\widehat{\Omega}}_d^{-1} \left(\frac{1}{\sqrt{T}}\sum_{t=1}^T\bs{g}_{dt}(\phi^d)\right) + \bs{G}_d'\bs{\widehat{\Omega}}^{-1}_d \bs{G}_d \sqrt{T}(\widehat{\phi}^d-\phi^d),
	\end{align} 
	hence, we obtain the usual influence function representation
	\begin{align}
	\sqrt{T}(\widehat{\phi}^d-\phi^d) &= -\frac{1}{\sqrt{T}}\sum_{t=1}^T (\bs{G}_d'\bs{\Omega}^{-1}_d \bs{G}_d)^{-1}\bs{G}_d' \bs{\Omega}_d^{-1} \bs{g}_{dt}(\phi^d) :=  \frac{1}{\sqrt{T}}\sum_{t=1}^T {r_{dt}}(\phi^d). \label{suppinfl}
	\end{align}
	Making use of \eqref{suppinfl} in \eqref{nuisMVE} we obtain
	\begin{align}
	\frac{1}{\sqrt{T}} \sum_{t=1}^T \bs{g}_{st}(\bs{\theta}^s;\widehat{\phi}^d) &= \frac{1}{\sqrt{T}} \sum_{t=1}^T \bs{\check{g}}_{st}(\bs{\theta}^s;\phi^d)+o_p(1), \label{nuisMVE2}
	\end{align}
	where $\bs{\check{g}}_{st}(\bs{\theta}^s;\phi^d) :=\bs{g}_{st}(\bs{\theta}^s;\phi^d) + \bs{F}\, {r_{dt}}(\phi^d)$. Putting \eqref{nuisMVE2} and \eqref{mvedem2} together and solving for $\sqrt{T}(\bs{\widehat{\theta}}^s_{GMM}-\bs{\theta}^s)$ gives 
	\begin{align}
	\sqrt{T} (\bs{\widehat{\theta}}^s_{GMM}-\bs{\theta}^s)&\overset{d}{\rightarrow} -(\bs{G}_s' \, \bs{\Omega}_s^{-1}  \,\bs{G}_s )^{-1}\bs{G}_s' \, \bs{\Omega}_s^{-1} \,\left( \frac{1}{\sqrt{T}}\sum_{t=1}^T  \bs{g}_{st}(\bs{\theta}^s;\phi^d) + \bs{F}\, {r_{dt}}(\phi^d)\right)+o_p(1), \\ 
	&=-(\bs{G}_s' \, \bs{\Omega}_s^{-1}  \,\bs{G}_s )^{-1}\bs{G}_s' \, \bs{\Omega}_s^{-1} \,\left( \frac{1}{\sqrt{T}}\sum_{t=1}^T  \bs{g}_{st}(\bs{\theta}^s;\phi^d) + \mathcal{O}_p\left(\frac{1}{\sqrt{N}}\right)\,  \mathcal{O}_p\left(1\right)\right)+o_p(1),
	\end{align}
	which gives the result. $\hfill \blacksquare$

\end{appendices}
\end{refsegment}
\section*{References} \pdfbookmark{References}{sec: MainRef}
\printbibliography[heading=none,segment=1]
%\nocite{*}

%\bibliographystyle{chicago}
%\bibliography[segment=1]{econbib}
\newpage
\begin{refsegment}
\appendixpageoff
\appendixtitleoff
\section*{\LARGE Supplementary Appendices} \currentpdfbookmark{Supplementary Appendices}{sec: SupApp}
\setlength{\parindent}{0.5cm}
This is the Supplementary Appendix to the paper, "Inferential Theory for Granular Instrumental Variables in High Dimensions" by Saman Banafti and Tae-Hwy Lee. Section \ref{figures} contain figures pertaining to the empirical work from Section \ref{oil}. Section \ref{herf} contains theoretical results for Herfindahl's in large $N$ markets along with Lemma \ref{shareslemma}. Section \ref{rhat} contains the estimation methods we use when $r$ is unknown. Section \ref{glasso} contains Algorithm \ref{alg:alg4prime} which employs an alternative estimator for the precision matrix, which is a hybrid of the factor approach and graphical models. 
\begin{appendices}\label{SuppApp}
	\section{Figures}\label{figures}
 \newpage 

	\section{Herfindahl's in Large \textit{N} Markets}\label{herf} 
%	 \currentpdfbookmark{Herfindahl's in large \textit{N} markets}{sec: herf}
	In this appendix we provide some basic information about properties of random variables that follow a power law and then we conclude with the statement of Lemma \ref{shareslemma} and its proof. The following draws on  \citet{darling1952influence}, \citet{gnedenkokolmogorov}, \citet{feller1971introduction}, \citet{logan1973limit},  \citet{jakubowski1993minimal}, \citet{jakubowski1997minimal}, \citet{davis1995point}, \citet{malevergne2009professor}, \citet{gabaix2011granular} and \citet{durrett2019probability}. Recall, the sizes $\mathscr{S}_1,\dots,\mathscr{S}_N$ are drawn $i.i.d.$ from a distribution for which the tail follows a power law, with tail index, $\mu >0$. Note that the first and second moments can potentially diverge
	\begin{align}
	\mathbbm{E}(\mathscr{S}) &=\int_{1}^{\infty} s \mu s^{-\mu-1} \text{d}s= \int_{1}^{\infty}  \mu s^{-\mu} \text{d}s  =\dfrac{\mu}{1-\mu} s^{1-\mu} \bigg|_1^{\infty} = \begin{cases} \,\,\,\,\,\,\,\,\,\infty \,\,\,\,\,\,\,\,\,\,\,\,\, \text{for} \,\,\, \mu \in (0,1]\\  -\dfrac{\mu}{1-\mu} \,\,\,\,\,\,\,\, \text{for} \,\,\, \mu \in (1,\infty)\end{cases} \label{appen0}
	\end{align}
	
	\begin{align}
	\mathbbm{E}(\mathscr{S}^2) &=\int_{1}^{\infty} s^2 \mu s^{-\mu-1} \text{d}s= \int_{1}^{\infty}  \mu s^{1-\mu} \text{d}s =\dfrac{\mu}{2-\mu} s^{2-\mu} \bigg|_1^{\infty} = \begin{cases} \,\,\,\,\,\,\,\,\,\infty \,\,\,\,\,\,\,\,\,\,\,\,\, \text{for} \,\,\, \mu \in (0,2]\\  -\dfrac{\mu}{2-\mu} \,\,\,\,\,\,\,\, \text{for} \,\,\, \mu \in (2,\infty)\end{cases} \label{appen01}
	\end{align}
	as a result of  \eqref{appen0} and \eqref{appen01}, $\mathbbm{E}(\mathscr{S})$ is bounded for $\mu >1$, while $\mathbbm{V}(\mathscr{S})$ is bounded for $\mu >2$. The literature refers to the cases $\mu \leq 2$ as thick tail regimes since the variance is infinite, rendering extreme tail events more likely. In light of this, there are some important cases to distinguish from one another when considering the limiting behavior of $h_{N,\mu}$, which are outlined in Table~\ref{regimes} below.
	
		\begin{table}[!hbtp]
		\begin{center}
			\caption{Limiting Behavior of the Asymptotic Herfindahl Index}
			\label{regimes}
			\resizebox{0.6\textwidth}{!}{
				\begin{tabular}{l |c |c | c| c|c|}
					\hline
					& \makecell{Tail \\index \\regime}& Tail variation & \makecell{First\\ moment} & Variance &$\mathcal{O}_p\left( g_{N,\mu}\right)$  \\[.3cm]
					\hline
					\textbf{Case I}&  $\mu>2$& Exponential & $\mathbbm{E}(\mathscr{S}) < \infty$   & $\mathbbm{V}(\mathscr{S}) < \infty$ & $\dfrac{1}{\sqrt{N}}$\\[.3cm]
					\textbf{Case II}& $\mu=2$ &\makecell{Regularly \\varying}& $\mathbbm{E}(\mathscr{S}) < \infty$ & $\mathbbm{V}(\mathscr{S})=\infty$ & $\sqrt{\dfrac{\text{log}(N)}{N}}$  \\[.3cm]
					\textbf{Case III}&$\mu\in (1,2)$ &\makecell{Regularly \\varying}&$\mathbbm{E}(\mathscr{S}) < \infty$ & $\mathbbm{V}(\mathscr{S})=\infty$ & $\dfrac{1}{N^{1-\frac{1}{\mu}}}$ \\[.3cm]
					\textbf{Case IV}& $\mu=1$ & \makecell{Regularly \\varying}&$\mathbbm{E}(\mathscr{S})=\infty$ & $\mathbbm{V}(\mathscr{S})=\infty$ &$\dfrac{1}{\text{log}(N)}$ \\[.3cm]
					\textbf{Case V}&$\mu \in (0,1)$ & \makecell{Regularly \\varying}& $\mathbbm{E}(\mathscr{S}) = \infty$& $\mathbbm{V}(\mathscr{S})=\infty$& $\Theta_p(1)$\\[.3cm]
					\textbf{Case VI}& \makecell{$\mu \rightarrow 0$} & \makecell{Slowly \\varying}& $\mathbbm{E}(\mathscr{S}) = \infty$& $\mathbbm{V}(\mathscr{S})=\infty$&$\Theta_p(1)$ \\
					\hline
			\end{tabular}}
		\end{center}
	\end{table}
The Herfindahl is given by
	\begin{align}
	h_{N,\mu} &= \sum_{i=1}^N S_i^{\,2} =\sum_{i=1}^N \left( \frac{\mathscr{S}_i}{\sum_{j=1}^N \mathscr{S}_j}\right)^2,\label{appen1}
	\end{align}
	and the object of interest is the asymptotic Herfindahl and from \eqref{appen1}, it can readily be written as
	\begin{align}
	h_{\mu} &:= \lim\limits_{N\to \infty} h_{N,\mu} =\lim\limits_{N\to \infty} \sum_{i=1}^N \left( \frac{\mathscr{S}_i}{\sum_{j=1}^N \mathscr{S}_j}\right)^2 =\lim\limits_{N\to \infty}  \dfrac{1}{N} \,\frac{N^{-1}\sum_{i=1}^N \mathscr{S}_i^2}{\left(N^{-1}\sum_{j=1}^N \mathscr{S}_j\right)^2} :=  \lim\limits_{N\to \infty} \frac{1}{N} \frac{a_N}{b_N}. \label{appen2} 
	\end{align}
	
	In \textbf{Case I}, when $\mathbbm{E}(\mathscr{S}),\mathbbm{E}(\mathscr{S}^2) < \infty$, the usual LLN and continuous mapping theorem gives us $a_N \overset{p}{\rightarrow}  a= \mathbbm{E}(\mathscr{S}^2)$ and $b_N \overset{p}{\rightarrow} b= \left(\mathbbm{E}(\mathscr{S})\right)^2$. Therefore, $h_{N,\mu} \rightarrow h_{\mu}= 0$, for thin tailed regimes.\footnote{As we saw in (unreported) simulation evidence, a Herfindahl converging to zero in the large $N$ limit, should not rule out identification by GIV; although theoretically it does. As illustrated in \ref{R4}, the variance of the elasticities diverges.} We will skip further details regarding \textbf{Cases II-IV} and state Lemma \ref{shareslemma} which is pertaining to \textbf{Cases V} and \textbf{VI}.
	
	\begin{lemma}\label{shareslemma}
	Under \ref{A4} (ii.), we have that
	\begin{align}
	\sqrt{h_{N,\mu}} = ||\bs{S}||_2 = \Theta_p(1)\hspace{5mm} \emph{for} \hspace{5mm}\mu \in [0,1).
	\end{align}
	\end{lemma}
	\textbf{Proof of Lemma \ref{shareslemma}:} Note that $\mathbbm{P} (\mathscr{S} >s) = s^{-{\mu}}$ and hence, $\mathbbm{P} (\mathscr{S}^{-{\mu}} >s) =s$. Or, put differently $\mathbbm{P} (\mathscr{S}^{-{\mu}} >s) \sim \text{U}[0,1]$; which is equivalently denoted as $U_i := 1-F_{\mathscr{S}}(s_i)=s_i^{-{\mu}}$, where $F_{\mathscr{S}}(s_i)$ denotes the CDF of $s_i$. As a result, $U_1,\dots,U_N$ are an $i.i.d.$ sample from $\text{U}[0,1]$. It is well known that order statistics, denoted as $U_{(i)} = 1-F_{\mathscr{S}}(s_{(i)})$, of the uniform distribution on the unit interval have marginal distributions belonging to the Beta distribution family. Hence, the PDF of the $i^{th}$ order statistic, $U_{(i),N} \sim \text{Beta}(\alpha,\beta)$, with $\alpha=i$ and $\beta=N-i+1$. Finally, the size of the $i^{th}$ largest unit out of $N$ can be found by manipulating the expected value of the $i^{th}$ order statistic, given by
	\begin{align}\label{appen3}
	\mathbbm{E}(U_{(i),N}) =\mathbbm{E}(\mathscr{S}^{-{\mu}}_{(i),N})=\frac{\alpha}{\alpha+\beta}=\frac{i}{N+1},
	\end{align}
	hence, the size of the $i^{th}$ largest unit is
	\begin{align}\label{appen4}
	\mathscr{S}_{(i),N}=\left(\dfrac{i}{N+1}\right)^{-\frac{1}{\mu}} \simeq \left(\dfrac{i}{N}\right)^{-\frac{1}{\mu}},
	\end{align}  
	where the approximation is negligible for large $N$. Furthermore, $\mathscr{S}_i^2$ has tail index $\frac{\mu}{2} \leq 1$, since $\mathbbm{P}(\mathscr{S}^2>s) = \mathbbm{P}(\mathscr{S}>s^{\frac{1}{2}})=s^{-\frac{\mu}{2}}$. Plugging \eqref{appen4} into \eqref{appen2}, we obtain
	\begin{align*}\label{appen5}
	&\lim\limits_{N\to \infty}  \dfrac{1}{N} \,\frac{N^{-1}\sum_{i=1}^N \mathscr{S}_i^2}{\left(N^{-1}\sum_{j=1}^N \mathscr{S}_j\right)^2} =\lim\limits_{N\to \infty}  \dfrac{1}{N} \,\frac{N^{-1}\sum_{i=1}^N \left(\dfrac{i}{N}\right)^{-\frac{2}{\mu}}}{\left(N^{-1}\sum_{j=1}^N \mathscr{S}_j\right)^2} =\lim\limits_{N\to \infty}  \dfrac{1}{N^2}\dfrac{1}{N^{-\frac{2}{\mu}}} \,\dfrac{ \sum_{i=1}^Ni^{-\frac{2}{\mu}}}{\left(N^{-1}\sum_{j=1}^N \mathscr{S}_j\right)^2} \\
	&=\lim\limits_{N\to \infty}  \dfrac{1}{N^{2-\frac{2}{\mu}}} \,\dfrac{ \sum_{i=1}^{\infty}i^{-\frac{2}{\mu}}}{\left(\frac{1}{N^{1-\frac{1}{\mu}}}\sum_{j=1}^{\infty} j^{-\frac{1}{\mu}}\right)^2} =\lim\limits_{N\to \infty}  \dfrac{1}{N^{2-\frac{2}{\mu}}} \,\dfrac{ \sum_{i=1}^{\infty}i^{-\frac{2}{\mu}}}{\frac{1}{N^{2-\frac{2}{\mu}}}\left(\sum_{j=1}^{\infty} j^{-\frac{1}{\mu}}\right)^2}=\lim\limits_{N\to \infty}  \dfrac{ \sum_{i=1}^{\infty}i^{-\frac{2}{\mu}}}{\left(\sum_{j=1}^{\infty} j^{-\frac{1}{\mu}}\right)^2}  \nonumber \\
	&= \dfrac{\zeta(\frac{2}{\mu}) }{\left(\zeta(\frac{1}{\mu})\right)^2}  \numberthis
	\end{align*}
	Where $\zeta(\cdot)$ denotes the Riemann-zeta function. Therefore, we have just showed that for $\mu \in (0,1)$, $h_{N,\mu} \rightarrow h_{\mu} > 0$. $\hfill \blacksquare$
	
\section{Estimating the Number of Factors}\label{rhat} 
Generally, since the number of factors, $r$, is unknown we must estimate it. There are many estimators for $r$ in static approximate factor models. Some examples are \citet{bai2002determining}, \citet{onatski2010determining} and \citet{ahn2013eigenvalue}. We make use of the $ER(k)$ and $GR(k)$ estimators proposed by \citet{ahn2013eigenvalue} (hereafter AH), which have been shown to outperform the existing estimators in the literature, particularly when the idiosyncratic errors are not $i.i.d.$, which is likely to be the more relevant case. 

The estimators below can be used in \eqref{justidentifiedGIV} for estimation of the demand elasticity without affecting inference, and in Algorithm \ref{alg:alg1}, \ref{alg:alg2}, or \ref{alg:alg4} for estimation of the supply elasticity; again without affecting inference. The AH estimators are given by maximizing the following criteria
\begin{align}
ER(k)&=\dfrac{\tilde{\mu}_{NT,k}}{\tilde{\mu}_{NT,k+1}} \hspace{20mm} k=1,\dots, kmax \\[1.5mm]
GR(k) &= \dfrac{\text{ln}\left[ V(k-1)/V(k)\right]}{\text{ln}\left[V(k)/V(k+1)\right]} \nonumber\\[1.5mm]
&=\dfrac{\text{ln}(1+\tilde{\mu}^*_{NT,k})}{\text{ln}(1+\tilde{\mu}_{NT,k+1})}\hspace{7mm} k=1,\dots, kmax,
\end{align} 
where $\tilde{\mu}_{NT,k} :=  \psi_k \left[ \bs{X}\bs{X}'/(NT)\right]=\psi_k \left[ \bs{X}'\bs{X}/(NT)\right]$, $\bs{X}$ denotes a $T \times N$ matrix and $\psi_k(\bs{A})$ denotes the $k^{th}$ largest eigenvalue of a positive semidefinite matrix $\bs{A}$.  $V(k)=\sum_{j=k+1}^m \tilde{\mu}_{NT,j}$ and $\tilde{\mu}^*_{NT,k} = \tilde{\mu}_{NT,k}/V(k)$. Where $V(k)$ is the sample mean of the squared residuals from the time series regressions of individual response variables on the first $k$ principal components of $\bs{X}\bs{X}'/(TN)$. Hence, the estimators are 
\begin{align}
\widehat{r}_{ER}=\underset{1\leq k \leq kmax}{\text{argmax}}ER\text{(k)}, \\
\widehat{r}_{GR}=\underset{1\leq k \leq kmax}{\text{argmax}}GR\text{(k)}.
\end{align}
The basic idea behind maximizing the $ER(k)$ and $GR(k)$ criteria is that $\dfrac{\tilde{\mu}_{NT,j}}{\tilde{\mu}_{NT,j+1}}=\mathcal{O}_p(1)$ for $j\neq r$, while $\dfrac{\tilde{\mu}_{NT,r}}{\tilde{\mu}_{NT,r+1}}=\mathcal{O}_p(C_{NT})$. This effective idea stems from the seminal paper of \citet{arbitrage1983mean}, who, among other things, demonstrate that only the $r$ eigenvalues arising from the common component remain unbounded as the sample size tends to infinity, while those from the idiosyncratic part remain bounded. In particular, see their Theorem 4. Lastly, some recommendations on the choice of $kmax$ are provided by AH to avoid choosing $\widehat{r}<r$ wpa 1. 

The important fact is that the limiting distribution of the elasticities remain unchanged so long as we use a consistent estimator for $r$. Let $\widehat{\phi}^{\,j}_{\widehat{r}}$, for $j=s,d$, denote the FGIV or efficient GMM estimator, using a consistent estimator for $r$, such as the AH estimator. It is easy to show that $\widehat{\phi}^{\,j}_{\widehat{r}}$ has the same limiting distribution as $\widehat{\phi}^{\,j}_{r}$, for $j=d,s$:
\begin{align}
\mathbbm{P} \left(\sqrt{T}(\widehat{\phi}^{\,j}_{\widehat{r}}-\phi^{\,j}) \leq x \right)&=\mathbbm{P} \left(\sqrt{T}(\widehat{\phi}^{\,j}_{\widehat{r}}-\phi^{\,j}) \leq x |\widehat{r}=r\right) \cdot \mathbbm{P}(\widehat{r}=r) \nonumber \\
&+ \mathbbm{P} \left(\sqrt{T}(\widehat{\phi}^{\,j}_{\widehat{r}}-\phi^{\,j}) \leq x |\widehat{r}\neq r\right) \cdot \mathbbm{P}(\widehat{r}\neq r) \label{rhat1} \\
&\rightarrow \mathbbm{P} \left(\sqrt{T}(\widehat{\phi}^{\,j}_{\widehat{r}}-\phi^{\,j}) \leq x |\widehat{r}=r\right) \label{rhat2}\\
&= \mathbbm{P} \left(\sqrt{T}(\widehat{\phi}^{\,j}_{r}-\phi^{\,j}) \leq x\right) \label{rhat3}.
\end{align}
From \eqref{rhat1} to \eqref{rhat2} we make use of the fact that $\widehat{r}$ is a consistent estimator for $r$, i.e. $\mathbbm{P}(\widehat{r}=r) \rightarrow 1$. From \eqref{rhat2} to \eqref{rhat3} we make use of the fact that conditional on $\widehat{r}=r$, $\widehat{\phi}^{\,j}_{\widehat{r}}=\widehat{\phi}^{\,j}_{r}$. Therefore, 
\begin{align}
\bigg|\mathbbm{P} \left(\sqrt{T}(\widehat{\phi}^{\,j}_{\widehat{r}}-\phi^{\,j}) \leq x \right)-\mathbbm{P} \left(\sqrt{T}(\widehat{\phi}^{\,j}_{r}-\phi^{\,j}) \leq x\right)\bigg| \rightarrow 0 \label{rhat4}.
\end{align}

\section{Estimation of the High-Dimensional Precision Matrix via $\ell_1$-Penalized Bregman Divergence} \label{glasso}
An alternative to using the POET like procedure of \citet{fan2013large} to estimate a high dimensional precision matrix is to use graphical Lasso methods, as in \citet{friedman2008sparse}. One thought would be to directly estimate the precision matrix using graphical models, say by applying the graphical Lasso procedure to the composite error, $v_{it} = \bs{\lambda}_i'\bs{\eta}_t + u_{it}$, or by using a local (nodewise) graphical method as in \citet{callot2021nodewise} and applying to it $v_{it}$. 

However, these approaches rule out the presence of an approximate factor structure, as they assume unconditional sparsity of the composite error term $v_{it}$. It is clear that sparsity of $v_{it}$ fails given our (pervasive) factor structure, as pointed out by \citet{barigozzi2018power}, \citet{brownlees2018realized} and \citet{koike2020biased}. 

An alternative, hybrid, approach to estimation of high dimensional covariance matrices is to adopt an approximate factor structure, thereby decomposing the process into a low rank part (common component), plus a sparse part (idiosyncratic component) like the approach we adopted in the main text of the paper. Except now we will use a graphical model in estimation of the precision matrix, as opposed to the POET estimation procedure. Thus, we can employ a hybrid approach known as the \textit{factor-adjusted graphical lasso} model or simply FGL of \citet{lee2020optimal}. The FGL approach imposes conditional sparsity similar to POET with the exception that sparsity is imposed on the precision matrix, $\bs{\Sigma}_u^{-1}$, of the idiosyncratic term rather than the covariance matrix, $\bs{\Sigma}_u$. That is, once the low dimensional common factors are conditioned on, $\bs{\Sigma}^{-1}_u$ is assumed to be sparse in the sense that many of the off-diagonal elements are zero. Note that with FGL, sparsity is assumed on the precision matrix, $\bs{\Sigma}^{-1}_u$, for the idiosyncratic term and not on the precision matrix of the composite error, $\bs{\Sigma}^{-1}_v$ as in a traditional graphical method. 

Along the POET procedures, \citet{fan2013large}, \citet{fan2018large} and \citet{bai2017inferences} amongst others, estimate the high dimensional covariance matrix via thresholding techniques and then invert the estimate to obtain an estimate of the precision matrix. Whereas, graphical methods directly estimate the precision matrix. The FGL approach is essentially a hybrid of the two approaches. We adopt the FGL approach of \citet{lee2020optimal}, although in their paper endogeneity is not a concern. To that end, suppose momentarily that the idiosyncratic error is observed 
\begin{align}
u_{it} &= y_{it}-\bs{\lambda}_i' \bs{\eta}_t -\phi^s p_t, \nonumber \\
&= y_{it}-\bs{\psi}_{i}' \bs{f}_t, \label{adjdem}
\end{align}
where $\bs{\psi}_{i}:= \begin{pmatrix}\bs{\lambda}_i' & \phi^s \end{pmatrix}'$ and $\bs{f}_t$ is defined as in the main text. We apply the graphical Lasso procedure of \citet{friedman2008sparse} to \eqref{adjdem}, to obtain $\bs{\widehat{\Sigma}}_u^{-1}$ as the solution to the $\ell_1$-penalized Bregman Divergence. Bregman divergence is simply a measure of distance between two objects defined in terms of a strictly convex function, say $f(\cdot)$. Introduce $\mathcal{S}_{++}$ as the set of symmetric positive definite matrices, then, for $\bs{A}_1$, $\bs{A}_2 \in \mathcal{S}_{++}$, the Bregman Divergence in this context, is defined as
\begin{align}
d_f(\bs{A}_1,\bs{A}_2) &:=f(\bs{A}_1)-f(\bs{A}_2)-\langle \nabla f(\bs{A}_2), \bs{A}_1 - \bs{A}_2 \rangle \label{BD}
\end{align}  
where $f(\cdot)$ is strictly convex and continuously differentiable. The Bregman Divergence, $d_f$, can be viewed as the difference of $f(\bs{A}_1)$ from the first-order approximation of $f(\bs{A}_1)$ around $\bs{A}_2$. Moreover, \eqref{BD} nests some important loss functions as special cases for particular choices of $f(\cdot)$, e.g. when $f(\bs{x})=\bs{x}'\bs{B}\bs{x}$, $d_f$ becomes the Mahalanobis distance, which reduces to the squared norm when $\bs{B}=\bs{I}$ and when $f(\bs{x})=\sum_i x_i \text{log} \,x_i$, we obtain $d_f$ as the Kullback-Leibler divergence. When one sets $f(\bs{A})=-\text{log }\text{det}(\bs{A})$, then $\nabla f(\bs{A})=-\bs{A}_2^{-1}$,\footnote{See Section A.4.1 of \citet{boyd2004convex} for an elegant derivation of this gradient.} and the Bregman Divergence takes the following familiar form for $\bs{A}_1=\bs{\Sigma}_u^{-1}$ and $\bs{A}_2=\bs{\widehat{\Sigma}}_u^{-1}$
\begin{align*}
d_f(\bs{\Sigma}_u^{-1},\bs{\widehat{\Sigma}}_u^{-1}) &=-\text{log }\text{det}(\bs{\Sigma}_u^{-1})+\langle \bs{\widehat{\Sigma}}_u, \bs{\Sigma}_u^{-1} - \bs{\widehat{\Sigma}}_u \rangle + c_1 \\
&=-\text{log }\text{det}(\bs{\Sigma}_u^{-1})+\text{tr}(\bs{\widehat{\Sigma}}_u\bs{\Sigma}_u^{-1}) + c_2 \numberthis \label{BD3}
\end{align*}  
where \eqref{BD3} can be viewed as the negative Gaussian log-likelihood of the data, partially maximized with respect to the mean parameter. Adding an $\ell_1$-penalty on the off-diagonal elements of $\bs{\Sigma}_u^{-1}$ to \eqref{BD3} gives us $\bs{\widehat{\Sigma}}_u^{-1}$ as the solution to the $\ell_1$-penalized Bregman Divergence
\begin{align}
\bs{\widehat{\Sigma}}_u^{-1}(\rho)=\underset{\bs{\Sigma}_u^{-1}\in\mathcal{S}_{++}}{\text{arg min }}\,\{ -\text{log} \, \text{det}(\bs{\Sigma}_u^{-1}) + \text{tr}(\bs{\widehat{\Sigma}}_{u}\bs{\Sigma}_u^{-1}) + \rho ||\bs{\Sigma}_u^{-1}||_1\} , \label{loglik}
\end{align}
where $\rho$ is the tuning hyperparameter and only here $||\bs{\Sigma}_u^{-1}||_1  := \sum_{i \neq j} |\Sigma_{u,ij}^{-1}|$ is defined to not penalize the diagonal elements. The routine can be easily implemented in the \texttt{R} package \texttt{glassoFast} or \texttt{CVglasso}. 

However, as noted in \citet{jankova2018inference}, there are theoretical and practical benefits to modify \eqref{loglik} to the so-called weighted FGL (effectively just adaptive Lasso)
\begin{align}
\bs{\widehat{\Sigma}}_u^{-1}(\rho)=\underset{\bs{\Sigma}_u^{-1}\in\mathcal{S}_{++}}{\text{arg min }}\,\{ -\text{log} \, \text{det}(\bs{\Sigma}_u^{-1}) + \text{tr}(\bs{\widehat{\Sigma}}_{u}\bs{\Sigma}_u^{-1}) + \rho \sum_{i \neq j} \widehat{W}_{ii}\widehat{W}_{jj} |\Sigma_{u,ij}^{-1}|\} , \label{loglikw}
\end{align} 
where $\bs{\widehat{W}}^2=\text{diag}(\bs{\widehat{\Sigma}}_u)$. We suggest iterating between estimation of $\bs{\Sigma}_u^{-1}$ by optimizing \eqref{loglikw} with the graphical Lasso algorithm and estimation of $\phi^s(\bs{{z}},\bs{\Sigma}_u^{-1})$ as in \eqref{iterphid} in Algorithm \ref{alg:alg4prime} below. For each iteration, we optimally select the penalty hyperparameter, $\rho$, via cross-validation. 

We do not explore the theoretical properties of the sampling error induced by this weighted FGL estimation procedure, but in some unreported Monte Carlo evidence we find that it performs well. The algorithm below details the overidentified estimation procedure for the case when $k_x=0$.
\begin{center}
	\begin{algorithm}[!htbp]
		\renewcommand{\thealgorithm}{$3^{\prime}$}
		\caption{\textbf{Efficient GMM-FGL for $\phi^s$ (when $k_x=0$):}}
		\label{alg:alg4prime}
		\begin{algorithmic}[1]
			\begin{itemize}
				\item \textit{Step 1:} Run PCA on \eqref{runPCA} and obtain $\widehat{z}_t=\bs{S}'\bs{\widehat{Q}}\bs{\widetilde{y}}_{\cdot t}$ as the sample counterpart of \eqref{GIV}.
				\item  \textit{Step 2:} Initialize $\bs{\widehat{\Sigma}}_u^{-1}=\bs{I}_N$.
				\item  \textit{Step 3:} Estimate \eqref{gmmdemand} to obtain $\bs{\widehat{\varepsilon}}$, initialize $\bs{\widehat{W}}_s=(\bs{\widehat{Z}}_s'\bs{\widehat{Z}}_s)^{-1}$ and obtain $\bs{\widehat{\theta}}_{2SLS}^s(\bs{\widehat{Z}}_s,\bs{\widehat{\Sigma}}_{u}^{-1})$. 
				\item  \textit{Step 4:} Obtain  $\bs{y}_{\widehat{E}}(\bs{\widehat{\Sigma}}_u^{-1})$. 				
				\item \textit{Step 5:} Update $\bs{\widehat{W}}_s= \left(\frac{1}{T}\sum_{t=1}^T\bs{\widehat{Z}}_{st} \bs{\widehat{Z}}'_{st} \widehat{u}^2_{\widehat{E}t}\right)^{-1}$, where $\widehat{u}_{\widehat{E}t}=y_{\widehat{E}t}-\bs{\widehat{\theta}}_{GMM}^s(\bs{\widehat{Z}}_s,\bs{\widehat{\Sigma}}_{u}^{-1})'\bs{f}_t$ and construct $\bs{\widehat{\theta}}_{GMM}^s(\bs{\widehat{Z}}_s,\bs{\widehat{\Sigma}}_{u}^{-1})$ as in \eqref{gmmsupply}.
				\item \textit{Step 6:} Construct the sample counterpart of \eqref{adjdem} to update $\bs{\widehat{\Sigma}}_u^{-1}$ via \eqref{loglikw} and update $\bs{y}_{\widehat{E}}(\bs{\widehat{\Sigma}}_u^{-1})$. 
				\item \textit{Step 7:} Iterate \textit{Step 4} through \textit{Step 6} until convergence.
			\end{itemize} 						
		\end{algorithmic}
	\end{algorithm}
\end{center} 
Note, to obtain $\bs{\widehat{\psi}}_i$ in the sample counterpart of \eqref{adjdem}, we have $\bs{\widehat{\Lambda}}=T^{-1} \bs{y}'_{..}\bs{\widehat{\eta}}$ and $\widehat{\phi}^s_{GMM}$ is an element of $\bs{\widehat{\theta}}_{GMM}^s$. In view of Algorithm \ref{alg:alg2}, Algorithm \ref{alg:alg4prime} can be further extended to the case when $k_x >0$. However, for brevity we omit the details. 
\end{appendices}

\end{refsegment} 
%
%%\nocite{*}
%\newpage
\subsection*{References for Supplemental Appendices}  \currentpdfbookmark{References for Supplemental Appendices}{sec: SupRef}
\printbibliography[heading=none, segment=2]
%\bibliographystyle{chicago}
%\bibliography{bib_appendix}
\end{document}